\documentclass[transmag]{IEEEtran}
\usepackage{latexsym}
\usepackage{graphicx}
\usepackage{amsfonts,amssymb,amsmath}
\usepackage[bookmarks=true,hidelinks,pdfpagelabels=true]{hyperref}
\def\BibTeX{{\rm B\kern-.05em{\sc i\kern-.025em b}\kern-.08em T\kern-.1667em\lower.7ex\hbox{E}\kern-.125emX}}

\usepackage{setspace}
\usepackage{amsthm}
\usepackage{color}
\usepackage{cite}
\usepackage{bm}
\usepackage{multirow}
\usepackage[yyyymmdd,hhmmss]{datetime}
\usepackage{pst-node}
\usepackage{fancyhdr,bbm}
\usepackage{xcolor}
\usepackage{float}
\usepackage{comment}
\usepackage{mathtools}
\usepackage{cuted}
\usepackage[caption=false,font=footnotesize]{subfig}
\usepackage{pdfcomment}
\usepackage{enumitem,kantlipsum}

\usepackage[bottom]{footmisc}

\usepackage[crop=pdfcrop,cleanup={.tex, .dvi, .ps, .pdf, .log, .bbl, .out, .upa, .upb}]{pstool}

\usepackage{dblfloatfix}

\usepackage{balance}

\newcounter{MYtempeqncnt}

\makeatletter
\def\underbracex#1#2{\mathop{\vtop{\m@th\ialign{##\crcr
   $\hfil\displaystyle{#2}\hfil$\crcr
   \noalign{\kern3\p@\nointerlineskip}   #1\crcr\noalign{\kern3\p@}}}}\limits}

\def\underbracea{\underbracex\upbracefilla}

\def\upbracefilla{$\m@th \setbox\z@\hbox{$\braceld$}  \bracelu\leaders\vrule \@height\ht\z@ \@depth\z@\hfill 
\kern\p@\vrule \@width\p@\kern\p@\vrule \@width\p@\kern\p@\vrule \@width\p@
$}

\def\upbracefillb{$\m@th \setbox\z@\hbox{$\braceld$}\vrule \@width\p@\kern\p@\vrule \@width\p@\kern\p@\vrule \@width\p@\kern\p@
 \leaders\vrule \@height\ht\z@ \@depth\z@\hfill\bracerd
  \braceld\leaders\vrule \@height\ht\z@ \@depth\z@\hfill
\kern\p@\vrule \@width\p@\kern\p@\vrule \@width\p@\kern\p@\vrule \@width\p@
$}

\def\upbracefillc{$\m@th \setbox\z@\hbox{$\braceld$}\vrule \@width\p@\kern\p@\vrule \@width\p@\kern\p@\vrule \@width\p@\kern\p@
\leaders\vrule \@height\ht\z@ \@depth\z@\hfill
\kern\p@\vrule \@width\p@\kern\p@\vrule \@width\p@\kern\p@\vrule \@width\p@
$}

\def\upbracefilld{$\m@th \setbox\z@\hbox{$\braceld$}\vrule \@width\p@\kern\p@\vrule \@width\p@\kern\p@\vrule \@width\p@\kern\p@
 \leaders\vrule \@height\ht\z@ \@depth\z@\hfill\braceru$}

\def\underbracebd{\underbracex\upbracefillbd}
\def\upbracefillbd{$\m@th \setbox\z@\hbox{$\braceld$}\vrule \@width\p@\kern\p@\vrule \@width\p@\kern\p@\vrule \@width\p@\kern\p@
\bracerd\braceld
 \leaders\vrule \@height\ht\z@ \@depth\z@\hfill\braceru$}

\DeclareRobustCommand{\mcal}[1]{  \ifcat\noexpand#1\relax\mathnormal{#1}\else\cal{#1}\fi
}

\newcommand{\V}[1]{\bm{#1}}
\newcommand{\Set}[1]{{\mcal{#1}}}
\DeclareMathOperator{\diag}{\mathrm{diag}}
\newcommand{\VG}[1]{\bm{#1}}

\newcommand{\ist}{\hspace*{.3mm}}
\newcommand{\rmv}{\hspace*{-.3mm}}
\newcommand{\T}{\text{T}}
\DeclarePairedDelimiterX{\norm}[1]{\lVert}{\rVert}{#1}

\let\geq\geqslant

\usepackage{accents}

\usepackage{bm,upgreek}

\usepackage{amssymb}
\usepackage{xifthen}
\newcommand{\deq}{\triangleq}

\mathtoolsset{showonlyrefs = true}

\allowdisplaybreaks

\definecolor{dgcolour}{RGB}{200,0,200}

\definecolor{mbcolour}{RGB}{255,0,0}

\allowdisplaybreaks
 
\begin{document}

\title{Cooperative Localization and Multitarget Tracking in \\ Agent Networks with the Sum-Product Algorithm}

\author{
	Mattia Brambilla,~\IEEEmembership{Member,~IEEE}, Domenico~Gaglione,
	Giovanni Soldi,
	Rico Mendrzik, \\
	Gabriele Ferri,~\IEEEmembership{Senior Member,~IEEE},
	Kevin D. LePage,
	Monica Nicoli,~\IEEEmembership{Member,~IEEE}, \\
	Peter Willett,~\IEEEmembership{Fellow,~IEEE},
	Paolo~Braca,~\IEEEmembership{Senior Member,~IEEE},
	and Moe Z. Win,~\IEEEmembership{Fellow,~IEEE}
	\thanks{This work was supported in part by the NATO Allied Command Transformation (ACT) under the DKOE project and by the U.S. Army Research Office through the MIT Institute for Soldier Nanotechnologies under Contract W911NF-13-D-0001. Parts of this paper were previously presented at ICASSP 2020, Barcelona, Spain, May 2020.}
	\thanks{M.\ Brambilla is with Dipartimento di Elettronica, Informazione e Bioingegneria (DEIB), Politecnico di Milano, 20133 Milan, Italy (e-mail: mattia.brambilla@polimi.it).}
	\thanks{ M.\ Nicoli is with
		Dipartimento di Ingegneria Gestionale (DIG), Politecnico di Milano, 20156 Milan, Italy (e-mail:  monica.nicoli@polimi.it).}
	\thanks{D.\ Gaglione, G.\ Soldi, G.\ Ferri, K.\ D.\ LePage, and P.\ Braca are with the NATO STO Centre for Maritime Research and Experimentation (CMRE), 19126 La~Spezia, Italy (e-mail: [domenico.gaglione, giovanni.soldi, gabriele.ferri, kevin.lepage, paolo.braca]@cmre.nato.int).}
	\thanks{R.\ Mendrzik is with Ibeo Automotive Systems GmbH, 22143 Hamburg, Germany (e-mail: rico.mendrzik@tuhh.de).}
	\thanks{P.\ Willett is with the University of Connecticut, Storrs, CT 06269, USA (e-mail: peter.willett@uconn.edu).}
	\thanks{M.\ Z.\ Win is with the Laboratory for Information and Decision Systems (LIDS), Massachusetts Institute of Technology (MIT), Cambridge, MA, USA (e-mail: moewin@mit.edu).}
}

\IEEEtitleabstractindextext{\begin{abstract}
This paper addresses the problem of multitarget tracking using a network of
sensing \textit{agents} with unknown positions.
Agents have to both localize themselves in the sensor network and, at the same time, perform multitarget tracking in the presence of clutter and
miss detection.
These two problems are jointly resolved
using a holistic and centralized approach where
graph theory is used to describe the statistical relationships among
agent states, target states, and observations.
A scalable message passing
scheme, based on the sum-product algorithm,
enables to efficiently approximate the marginal posterior distributions of both agent and target states.
The proposed
method is general enough to accommodate a full multistatic network configuration, with multiple transmitters and receivers.
Numerical simulations show superior performance of the proposed joint
approach with respect to the case
in which cooperative self-localization and multitarget tracking are performed separately, as
the former manages to extract valuable information from targets.
Lastly,
data acquired in 2018 by the NATO Science and Technology Organization (STO) Centre for Maritime Research and Experimentation (CMRE) through a network of autonomous underwater vehicles
demonstrates the effectiveness of the approach in a practical application.
\end{abstract}

\begin{IEEEkeywords}
Belief propagation,
factor graph,
maritime surveillance,
message passing,
probabilistic data association.
\end{IEEEkeywords}

}

\maketitle

\section{Introduction}
\label{sec:intro}

\subsection{Background and motivation}
\label{subsec:background}
Detecting unknown
\textit{targets},
understanding their intentions, and taking reactive countermeasures are common tasks in situational awareness (SA) applications
\cite{WinConMaxSheGifDarChi:M11,GagSolMeyHlaBra:J20,BraGolGerLeP:J16,WinSheDai:J18,WinDaiSheChrPoo:J18,SolGagForDisDafBotQuaMil:J21_1,SolGagForDisDafBotQuaMil:J21_2}.
Depending on the specific use case, different types of sensors
(acoustic, radio frequency, optical, etc. \cite{ZafGkeLeu:J19}) may be used to
sense the environment and
provide the desired information.
Most of SA applications use multiple cooperative sensors, rather than a single
one, to infer the presence and
kinematics of
targets.
Indeed, 
cooperation dramatically increases the perception capabilities of an SA system, as
it relies on a larger dataset of
observations (or \textit{measurements}) of the targets \cite{PatAshKypHerMosCor:05,BarConGioWin:J14,ZheCaoWimVar:J15,CaoChoMasVar:J16,SiaNicGarDenRauWym:J18}.
Examples can be found in several domains such as underwater surveillance networks \cite{FerMunTesBra:J17,BraWilLePMarMat:J14,
AkyPomMel:J05,LiWanYuGua:J19
},
connected vehicles \cite{BraNicSoaDef:J20,ZhaStaJosWanGenDamWyeHoe:J20,
SriEsk:J19}, and internet of things (IoT)
\cite{MaTiJi:J19,SafKhaKarMou:J18,ConMazBarLinWin:J19,SauWin:J20}.
Mobility of sensors can further improve the performance of target
detection and localization
by fusing spatial sensing under different geometries, also enabling
the design of optimized sensor trajectories
\cite{BraGolLepMarMatWil:C14}.
However, this requires the sensors to localize themselves continuously.
Cooperative self-localization techniques
based on belief propagation,
also known as the sum-product algorithm (SPA) \cite{KscFreLoe:01,LoeDauHuKorPinKsc:J07},
have been recently proposed,
with computational complexity that linearly scales with the number of cooperative sensors \cite{IhlFisMosWil:J05,WymLieWin:J09,ConGueDarDecWin:J12,LiHedCol:J15,WinMeyLiuDaiBarCon:J18,MenBau:J19,TeaLiuMeyConWin:J21}.
Additional advantages of SPA methods include the ability to address non-linear and non-Gaussian models and to cope with unknown and time-varying hyperparameters \cite{SolMeyBraHla:J19_67}.

The advanced capability of a surveillance system to firstly detect and localize, and then track over time a number of hypothesized targets which behave as non-cooperative entities (i.e., that do not deliberately share
information with the surveillance system) is referred to as multitarget tracking.
Usually, the presence of these targets represents a dangerous situation or a potential threat; e.g., targets can be vulnerable road users in a vehicular scene,
intruding ships in the maritime domain, intruding aircraft in the aerospace domain,
or
thieves in an IoT surveillance system.
It follows that the development of robust, reliable, scalable and efficient multitarget tracking
algorithms becomes of paramount importance, as safety issues are involved. 
Abundant literature on multitarget tracking is available, starting from the pioneering works in~\cite{Sit:J64,SinSeaHou:J74,Sha:J78,
BarGioWinCon:J15} through very recent
studies such
as \cite{JiaWeiRezLin:J15,SauCoaRab:J17,GaoBatChi:J19,YuLia:J19,LiZhaLuWei:J19,FroLinGraWym:J19,
BarVahSad:J20,HeShiTso:J20,GosRatTenBabAliBatChiHos:J21}.
Approaches based on SPA have been proposed as well, both with
stationary sensors
--- whose location is either known \cite{SolMeyBraHla:J19_67,MeyKroWilLauHlaBraWin:J18}
or unknown \cite{SavWyeLar:J16} --- and with mobile ones \cite{BraNicSoaDef:J20,MeyWin:C18,ShaSauBucVar:J19,MeyHliWye:J16}.
However, 
not all of them handle typical multitarget tracking
challenges like the presence of clutter-generated measurements (i.e., \textit{false alarms}),
missed detections, and measurement origin uncertainty \cite{BarWilTia:B11},
i.e., the problem
of unknown association between targets and measurements.
Focusing
on multitarget tracking algorithms with mobile sensors, the cited works are affected by the following limitations:
in \cite{MeyWin:C18} sensors do not localize themselves cooperatively;
in \cite{ShaSauBucVar:J19} the maximum number of targets that can be tracked simultaneously
is limited and needs to be set a priori;
in \cite{BraNicSoaDef:J20} and \cite{MeyHliWye:J16} the number of targets is
time-invariant and known, and, in addition,
in \cite{BraNicSoaDef:J20} neither false alarms nor missed detections are considered, and in \cite{MeyHliWye:J16} the association between targets and measurements is assumed known.
Random finite sets (RFSs) constitute an alternative framework for the development of multitarget tracking methods\footnote{For the interested reader, similarities and differences between the SPA-based and the RFS-based derivation of multitarget tracking algorithms are described in  \cite{MeyKroWilLauHlaBraWin:J18}.} both with stationary~\cite{JiaWeiRezLin:J15,SauCoaRab:J17,GosRatTenBabAliBatChiHos:J21} and mobile sensors~\cite{FroLinGraWym:J19}.
In particular, in \cite{FroLinGraWym:J19} the authors develop a Poisson multi-Bernoulli multitarget tracking filter
that jointly
estimates the uncertain
mobile sensor states and target states using two types of measurements:
sensor state measurements, e.g., global navigation satellite system (GNSS)
measurements, and
target measurements.
However, this algorithm, as well as those cited above,
are not suitable for a full multistatic network configuration,
and do not consider the case of signal reflections from
mobile sensors, thus easing the data association problem.

The above issues have been
partially
addressed in \cite{MenBraAllNicKocBauLepBra:c20},
that indeed represents the preliminary study at the basis of this research;
it is purpose of this paper to further extend that work
as detailed in the next subsection.

\vspace{-2mm}
\subsection{Contributions and paper organization}
\label{subsec:contrib}
The
method we present here is based on a general framework where the concept of \textit{agents}, rather than sensors, is introduced to address the SA task.
An agent is a device with sensing and communication functionalities (i.e., transmitting and/or receiving acoustic, radio, or optical signals), along with motion and navigation capabilities. The connectivity is used to set up a cooperative and centralized processing platform.
We propose a SPA-based technique that extends the state-of-the-art methods by combining
cooperative self-localization and multitarget tracking in a unified centralized framework. In particular, moving agents, whose states are unknown, are capable of jointly localizing themselves by continuously estimating
their states and, at the same time, detecting and tracking an unknown, arbitrary, and time-varying number of targets by exploiting multiple types of measurements and in presence of clutter, miss detection and association uncertainty.
A fully distributed approach based on consensus strategies \cite{OlfFaxMux:J07,DasMou:J15,DasMou:J17,BatChiFanFarGra:J13,PapaRepMeyBraHla:J18,ShaSauBucVar:J19} can be adopted and customized for the proposed scheme; this study
is not included here and left to future work. We focus the attention on the holistic and centralized approach for cooperative self-localization and multitarget tracking, and on the real world experimentation.
The main contributions of this paper, which advances the work in \cite{MenBraAllNicKocBauLepBra:c20} where
an ad-hoc scenario with defined roles
for transmitter and receiver agents is considered, are the following:
\begin{itemize}
	\item a more general formulation is provided in which moving agents can both sense the environment, thus producing multiple types of measurements, and communicate with each other;
	
	\item all \textit{objects}, i.e., both agents and targets, whose states are unknown and need to be estimated, can reflect signals transmitted by a certain agent and thus produce measurements.
		Therefore, the data association problem is
		not limited to targets only as in \cite{MenBraAllNicKocBauLepBra:c20, ShaSauBucVar:J19, MeyWin:C18, BraNicSoaDef:J20,MeyHliWye:J16}, but it involves agents as well;
	
	\item the factor graph underlying the stochastic problem formulation is carefully derived and all the SPA messages are detailed;
	
	\item the proposed algorithm is validated in a real underwater scenario using data acquired by a network of autonomous vehicles.
\end{itemize}
The proposed SPA-based cooperative self-localization and multitarget tracking algorithm inherits the low computational complexity of the SPA-based multitarget tracking algorithm developed in \cite{MeyKroWilLauHlaBraWin:J18}, that scales linearly in the number of sensors and quadratically in the number of targets; the difference with \cite{MeyKroWilLauHlaBraWin:J18} is that the number of sensors is here the number of transmitter-receiver pairs.

The remainder of the paper is organized as follows.
Section~\ref{sec:problem-description-system-model} describes the
scenario we are considering,
the related mathematical representation, and the connection with practical use cases.
The joint cooperative self-localization and multitarget tracking problem is formulated
in Section~\ref{sec:stochastic-problem-formulation}. Section~\ref{sec:the_prop_algorithm} details the proposed SPA-based
algorithm, which is assessed using simulated and real data in Section~\ref{sec:sim_res}. Finally, concluding remarks are drawn in Section
~\ref{sec:conclusions}.

\subsection{Notation}
\label{subsec:notation}
Throughout this paper, column vectors are denoted by boldface lower-case letters (e.g., $\V{a}$) and matrices by boldface upper-case letters (e.g., $\V{A}$).
${\bf I}$
and $\V{0}$ denote the identity matrix
and the vector of all zeros, respectively,
with the size determined by the subscript or from the context.
We write $\diag(a_{1},\ldots,a_{N})$ for an $N \rmv \times \rmv N$ diagonal matrix with diagonal entries $a_{1},\ldots,a_{N}$.
The transpose of a matrix $\V{A}$ is written as $\V{A}^{\T}$.
The Euclidean norm of vector $\V{a}$ is denoted by $\norm{\V{a}}$.
For a two-dimensional (2D) vector $\V{a}$, $\angle\V{a}$ is the angle 
defined clockwise and 
such that $\angle\V{a} \!=\! 0$
for $\V{a} \rmv=\rmv [0 \,\, 1]^{\T}$.
Sets are denoted by calligraphic letters (e.g., $\Set{A}$) and $|\Set{A}|$ indicates the cardinality of the set.
The symbol $\propto$ denotes equality up to a constant factor.
The Dirac delta function is denoted with $\delta(\cdot)$;
the Kronecker delta is denoted with $\delta_{a,b}$, and is equal to $1$ if $a = b$, and $0$ otherwise.

\section{Problem Description and System Model}
\label{sec:problem-description-system-model}

Hereafter, we provide a high-level description of the scenario under consideration.
Let us suppose to have a set of agents and that each agent is equipped with an on-board device that provides noisy (and possibly
incomplete) observations of the agent's own state, referred to as \textit{navigation data}, and with a transmitter and/or a receiver.
For convenience, we will refer to an agent equipped with a transmitter as a \textit{Tx-agent}, and to an agent equipped with a receiver as an \textit{Rx-agent}; if an agent is equipped with both a transmitter and a receiver (i.e., a transceiver), we will use Tx-agent or Rx-agent depending on its role in each specific context.
The transmitter is used to broadcast a signal as, for example,
an acoustic signal used in sonar, or an electromagnetic signal used in radar; we assume that each agent is aware of the signal transmitted by any other agent and that all these signals are orthogonal in some domain (time, frequency, or code) so that interferences can be neglected.

The signal broadcast by a Tx-agent and received by an Rx-agent can be used by the latter to extract a noisy and
incomplete observation of the state of the Tx-agent; this type of measurement is referred to as \textit{inter-agent} measurement.
It is out of the scope of this paper to describe how this measurement is obtained, however we here provide few examples: the information about the Tx-agent state can be directly encoded into the transmitted signal and retrieved by the Rx-agent; or the Rx-agent can compute the time-of-arrival (ToA) or the angle-of-arrival (AoA) of the received signal, thus obtaining an observation of the Tx-agent state.

The signal broadcast by a Tx-agent may also reach an Rx-agent
after being reflected by a target or another agent present in the scene; in this case, the received signal can be used by the Rx-agent to extract a noisy and
incomplete observation of the object (i.e., either target or agent) that caused the reflection.
This type of measurement is referred to as \textit{multi-object tracking} (MOT) measurement; as before, this can be obtained by computing, for example, the ToA or the AoA of the received signal.
If Tx-agent and Rx-agent coincide, then this MOT measurement is acquired in a \textit{monostatic} configuration; otherwise, if Tx-agent and Rx-agent are different, the MOT measurement is acquired in a \textit{bistatic} configuration.
Note that an MOT measurement can be clutter-generated if not caused by the reflection from an object;
finally, we here assume that an Rx-agent is able to distinguish between inter-agent and MOT measurements.

\begin{figure}[!b]
	\centering
	\vspace{-2mm}
	\includegraphics{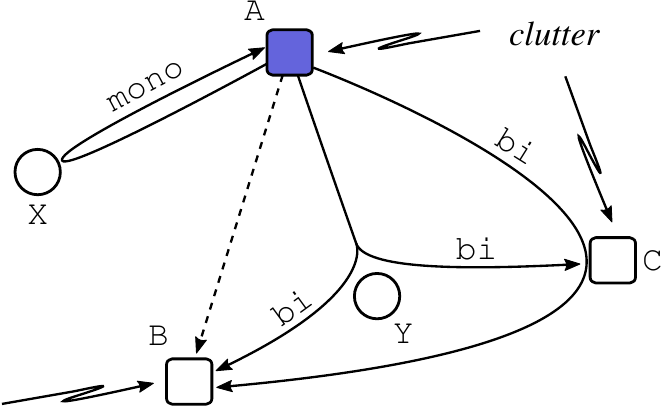}
	\caption{Exemplary illustration of the considered scenario.	Agents are depicted as squares, targets as circles. Agent \texttt{A} (purple square) is equipped with a transceiver; agents \texttt{B} and \texttt{C} (white squares) are equipped with receivers only. Dashed lines represent inter-agent measurements;
		solid lines represent MOT measurements, either \texttt{mono}static or \texttt{bi}static; solid zig-zag lines represent clutter-generated MOT measurements.}
	\label{fig:types-of-observations}
\end{figure}

Fig.~\ref{fig:types-of-observations} illustrates an exemplary scenario with three agents \texttt{A}, \texttt{B}, and \texttt{C}, depicted as
squares, and two targets \texttt{X} and \texttt{Y}, depicted as circles.
Agent \texttt{A} is equipped with a transceiver, and agents \texttt{B} and \texttt{C} are equipped with a receiver only.
Tx-agent
\texttt{A} informs Rx-agent \texttt{B} of its own location by an inter-agent measurement, represented by means of a dashed line; as mentioned before, the information on \texttt{A}'s position might be encoded within the signal, or extracted from the signal by \texttt{B}.
Moreover, the signal transmitted by Tx-agent \texttt{A} bounces off target \texttt{Y} and agent \texttt{C}, and is received by Rx-agent \texttt{B}.
Therefore, \texttt{B} has available an inter-agent and two bistatic MOT measurements --- each sketched as a solid line labeled as ``\texttt{bi}'' --- due to the signal transmitted by Tx-agent \texttt{A}.
However, agent \texttt{B} is unable to a priori associate the two MOT measurements to \texttt{Y} and \texttt{C}, respectively; this
measurement origin uncertainty needs to be handled.
Finally, we observe that the signal reflected by target \texttt{Y} also generates a bistatic MOT measurement at Rx-agent \texttt{C},
and that the signal that bounces off target \texttt{X} is reflected back to Rx-agent \texttt{A}, thus generating a monostatic MOT measurement, pictured as a solid line labeled as ``\texttt{mono}''.
Note that Fig.~\ref{fig:types-of-observations} depicts one possible instance of inter-agent and MOT measurements produced in this configuration of agents and targets, and that many others are possible.

A first objective of this paper is to show that the challenging bistatic
geometry can be fruitfully exploited to enhance the localization of agents along with the detection and tracking of targets.
This requires a mathematical representation of
the agents, the targets, and the available observations, as detailed in the following subsections.

\subsection{Agent state vector, agent pairs, and potential targets}
\label{subsec:agent_state_space_model}
Let $\Set{A} \deq \{ 1, \ldots, A \}$ be the set of agents, whose cardinality $A$ is known and time-invariant\footnote{Note that the number $A$ of agents may also be modeled as being time-variant --- yet, known ---, however this would not change the overall theoretical framework.
}.
The state (e.g., position, velocity, heading) of agent $a \! \in \! \Set{A}$ at time step $t = 1,2,\ldots$
is represented by the vector $\V{s}_{a,t}$
whose evolution in time is
given by the following kinematic model
\begin{align}
	\V{s}_{a,t} = \VG{\varepsilon}_{a} \big( \V{s}_{a,t-1}, \V{u}_{a,t} \big) \ist ,
	\label{eq:agent-dynamic-model}
\end{align}
where $\V{u}_{a,t}$ is a
process noise independent across $a$ and $t$ that accounts for the motion uncertainty of the agent \cite[Sec. 1.5]{BarWilTia:B11}.
The function $\VG{\varepsilon}_{a}(\cdot)$ and the statistics of $\V{u}_{a,t}$ define the agent state transition pdf $\tau_{a} (\V{s}_{a,t} | \V{s}_{a,t-1})$.
We denote the joint agent state vector at time $t$ as $\V{s}_{t} \deq [ \V{s}^{\T}_{1,t} , \ldots , \V{s}^{\T}_{A,t} ]^{\T}$, and the joint agent state vector at all times as $\V{s}_{1:t} \deq [ \V{s}^{\T}_{1}, \ldots, \V{s}^{\T}_{t} ]^{\T}$.

We indicate with $\Set{T} \subseteq \Set{A}$ the set of Tx-agents, and
with $\Set{R} \subseteq \Set{A}$ the set of Rx-agents.
Note that $\Set{T} \cup \Set{R} \! = \! \Set{A}$, and that $\Set{T} \cap \Set{R}$ represents the set of agents equipped with both a
transmitter and a receiver.
When $|\Set{T}|=1$ and $|\Set{R}| \geq 1$ the \textit{network} configuration is
referred to as bistatic, otherwise when $|\Set{T}| > 1$ and $|\Set{R}| \geq 1$ the \textit{network} is referred to as multistatic.
We formally consider the Cartesian product set $\Set{R} \times \Set{T}$ of all the possible pairs $(j_{1},j_{2})$ such that $j_{1} \! \in \! \Set{R}$ and $j_{2} \! \in \! \Set{T}$.
We observe that agents $j_1$ and $j_2$ might also coincide.
To establish an (arbitrary) order among the agent pairs, we introduce the index set $\Set{J} \deq \{ 1, \ldots, J \}$, with $J = |\Set{R}||\Set{T}|$, and define an indexing function $\phi : \mathcal{J} \to \mathcal{R} \times \mathcal{T}$, such that $\phi(j)$ represents the $j$-th agent pair $(j_{1},j_{2})$.
This order, though arbitrary, is used to sequentially process the
MOT measurements collected by the agent pairs as described later.
 
Furthermore, as done in \cite{MeyKroWilLauHlaBraWin:J18}, we account for a time-varying unknown number of targets by introducing the concept of \textit{potential target} (PT). The set of PTs at time $t$ is $\Set{K}_{t} \deq \{ 1, \linebreak \ldots, K_t \}$; the existence of PT $k \in \Set{K}_{t}$ at time $t$ is
indicated by the binary variable $r_{k,t} \in \{0,1\}$, i.e., $r_{k,t} \!=\! 1$ if the PT exists and $r_{k,t} \!=\! 0$ otherwise, and
the state (e.g., position and velocity)
of PT $k \in \Set{K}_{t}$ is denoted as $\V{x}_{k,t}$, and is formally considered also if $r_{k,t} \! = \! 0$.
We combine the state and existence variables of PT $k$ into the \textit{augmented} state vector $\V{y}_{k,t} \deq [ \V{x}_{k,t}^{\T}, r_{k,t} ]^{\T}$, and define the joint vector of all the PTs at time $t$ as $\V{y}_{t} \deq [ \V{y}_{1,t}^{\T}, \ldots, \V{y}_{K_{t},t}^{\T} ]^{\T}$.
We observe that the states $\V{x}_{k,t}$ of nonexisting PTs (i.e., for which $r_{k,t} = 0$) are obviously irrelevant;
thus, all the pdfs defined for the PT augmented states, i.e.,  $f (\V{y}_{k,t}) = f (\V{x}_{k,t},r_{k,t})$, are such that
\begin{align}
	f (\V{x}_{k,t},r_{k,t} = 0) = f_{k,t} \, f_{\text{D}} (\V{x}_{k,t}),
\end{align}
where $f_{k,t} \in [ 0,1 ]$ is a constant, and $f_{\text{D}} (\V{x}_{k,t})$ is an arbitrary ``dummy pdf''.

\subsection{Observations}

\label{subsec:measurements}
As stated above, at any time $t$ an agent $a \in \Set{A}$ might collect navigation data from an on-board device, and
produce two kinds of measurements: inter-agent
and MOT measurements. 

\subsubsection{Navigation data}
The navigation data $\V{g}_{a,t}$ collected by agent $a \! \in \! \Set{A}$ at time $t$ is an observation made by $a$ of its own state, e.g., acquired with an on-board system, such as
GNSS or inertial navigation system (INS).
It is modeled as 
\begin{align}
	\V{g}_{a,t}  = \VG{\theta}_{a} \big( \V{s}_{a,t}, \V{n}_{a,t} \big),
	\label{eq:nav-data}
\end{align}
where $\V{n}_{a,t}$ is a noise term,
independent across $a$ and $t$, modeling the finite accuracy of the on-board system.
Note that also the measurement model $\VG{\theta}_{a} (\cdot)$ depends on the agent $a$, since agents may be equipped with different types of on-board systems (e.g., GNSS or INS).
The function $\VG{\theta}_{a}(\cdot)$ and the statistics of $\V{n}_{a,t}$ define the likelihood function $\mathfrak{g}_{a}( \V{g}_{a,t} | \V{s}_{a,t} )$.
We indicate with $\Set{A}^{\V{g}}_{t} \subseteq	\Set{A}$ the set of agents that  have navigation data available at time $t$,
and we define the stacked vector of all navigation data from all agents at time $t$ as $\V{g}_{t}$, and the stacked vector of all
navigation data from all agents at all times as $\V{g}_{1:t} \deq [ \V{g}^{\T}_{1}, \ldots, \V{g}^{\T}_{t} ]^{\T}$.

\subsubsection{Inter-agent measurements}
\label{subsubsec:inter-agent-measurements}
The inter-agent measurement
produced at time $t$ by Rx-agent $a \! \in \! \Set{R}$
using the signal transmitted by Tx-agent $a' \! \in \! \Set{T} \setminus \{ a \}$ is modeled as
\begin{align}
	\VG{\rho}^{(a,a')}_{t}  = \VG{\vartheta} \big( \V{s}_{a,t},\V{s}_{a'\rmv,t},\V{w}^{(a,a')}_{t} \big) \ist ,
	\label{eq:inter-agent-meas}
\end{align}
where $\V{w}^{(a,a')}_{t}$ is an inter-agent measurement noise term
independent across $a$, $a'$, and $t$.
The function $\VG{\vartheta}(\cdot)$ and the statistics of $\V{w}^{(a,a')}_{t}$ define the likelihood function $\mathfrak{d}( \VG{\rho}^{(a,a')}_{t} | \V{s}_{a,t},\V{s}_{a'\rmv,t} )$.
We indicate with
$\Set{R}^{(a)}_{t} \! \subseteq \! \Set{R} \setminus \{ a \}$ the set of Rx-agents that
produce an inter-agent measurement using the signal transmitted
by Tx-agent $a$ at time $t$, and
with $\Set{T}^{(a)}_{t} \! \subseteq \! \Set{T} \setminus \{ a \}$ the set of Tx-agents that provide an inter-agent measurement to Rx-agent $a$ at time $t$.
Moreover, we define the stacked vector of all inter-agent measurements acquired by Rx-agent $a$ at time $t$ as $\VG{\rho}^{(a)}_{t}$,
the stacked vector of all inter-agent measurements acquired by all Rx-agents at time $t$ as $\VG{\rho}_{t}$,
and the vector of all inter-agent measurements acquired by all
Rx-agents at all times as $\VG{\rho}_{1:t} \deq [ \VG{\rho}^{\T}_{1}, \ldots, \VG{\rho}^{\T}_{t} ]^{\T}$.
We remark that, since the Rx-agent $a$ is aware of the signal transmitted by the Tx-agent $a'$, there is no uncertainty on the origin of the inter-agent measurement $\VG{\rho}^{(a,a')}_{t}$, thus no data association is required.

\subsubsection{MOT measurements}
\label{subsubsec:MOT}
Let us consider the $j$-th agent pair $(j_1,j_2)$, with $j \in \Set{J}$, $j_1 \in \Set{R}$, and $j_2 \in \Set{T}$, such that\footnote{Note that, in the remainder of the paper, we will be referring to a specific agent pair with ``$j$'' and ``$(j_{1},j_{2})$'' interchangeably, without using the indexing function $\phi (\cdot)$.} $\phi(j) = (j_{1},j_{2})$, and assume that Rx-agent $j_1$ produces $M_{t}^{(j)}$ MOT measurements from the signal broadcast by the Tx-agent $j_2$; we indicate with $\V{z}_{m,t}^{(j)}$, $m \in \Set{M}_{t}^{(j)} \deq \{ 1, \ldots, M_{t}^{(j)} \}$, the $m$-th MOT measurement produced by the $j$-th agent pair at time $t$.
We recall that an MOT measurement can be generated by the signal transmitted by Tx-agent $j_2$ reflecting off either an existing PT (i.e., for which $r_{k,t} = 1$) or another agent (i.e., other than $j_1$ and $j_2$), or it can be generated by clutter;
and that it can be either bistatic, if $j_1 \neq j_2$, or monostatic, if $j_1 = j_2$.
An MOT measurement $\V{z}_{m,t}^{(j)}$ generated by PT $k$ is modeled as
\begin{align}
	\V{z}^{(j)}_{m,t}  = \V{\gamma}_{\texttt{mono}} \big(\V{x}_{k,t}, \V{s}_{j_1,t},\V{v}^{(j)}_{m,t} \big)
	\label{eq:inter-agent-meas1}
\end{align}
if acquired in a monostatic configuration; and modeled as
\begin{align}
	\V{z}^{(j)}_{m,t}  = \V{\gamma}_{\texttt{bi}} \big(\V{x}_{k,t}, \V{s}_{j_1,t},\V{s}_{j_2,t},\V{v}^{(j)}_{m,t} \big)
	\label{eq:inter-agent-meas2}
\end{align}
if acquired in a bistatic configuration, where $\V{v}^{(j)}_{m,t}$ is a noise term independent across $j$, $m$ and $t$. 
The functions $\V{\gamma}_{\texttt{mono}} (\cdot)$ and $\V{\gamma}_{\texttt{bi}} (\cdot)$, and the statistics of $\V{v}^{(j)}_{m,t}$ define the likelihood functions 
$f_{\texttt{mono}} \big(\V{z}_{m,t}^{(j)} \big| \V{x}_{k,t}, \V{s}_{j_{1}\rmv,t} \big)$ 
and 
$f_{\texttt{bi}} \big(\V{z}_{m,t}^{(j)} \big| \V{x}_{k,t}, \V{s}_{j_{1}\rmv,t}, \V{s}_{j_{2}\rmv,t} \big) $, respectively.
In case the MOT measurement $\V{z}_{m,t}^{(j)}$ is generated by another agent $j' \neq j_1, j_2$, the models in \eqref{eq:inter-agent-meas1}-\eqref{eq:inter-agent-meas2} still apply as long as $\V{x}_{k,t}$ is replaced by $\V{s}_{j'\rmv,t}$.

For convenience, we stack
all MOT measurements produced
at agent pair $j$ at time $t$
into the vector $\V{z}^{(j)}_{t} \deq \big[ \V{z}^{(j)\T}_{1,t}, \ldots, \linebreak \V{z}^{(j)\T}_{M^{(j)}_{t}\rmv\rmv,t} \big]^{\T}$,
all MOT measurements produced at all agent pairs
at time $t$
into the vector $\V{z}_{t} \deq [ \V{z}_{t}^{(1)\T}, \ldots, \V{z}_{t}^{(J)\T} ]^{\T}$, and
all MOT measurements produced at all
agent pairs at all times
into the vector $\V{z}_{1:t} \deq [ \V{z}^{\T}_{1}, \ldots, \V{z}^{\T}_{t} ]^{\T}$.
Furthermore, we
define the vector of numbers of MOT measurements produced at all agent pairs at time $t$ as $\V{m}_{t} = [ M^{(1)}_{t}, \ldots, M^{(J)}_{t} ]^{\T}$, and the vector of numbers of MOT measurements produced
at all agent pairs at all times as $\V{m}_{1:t} \deq [ \V{m}^{\T}_{1}, \ldots, \V{m}^{\T}_{t} ]^{\T}$.
Finally, we recall that the MOT measurements, unlike inter-agent measurements, have \textit{unknown} origins, i.e., it is unknown if a given MOT measurement is generated by an object --- either target or agent --- and by which object.

\subsection{Legacy PTs and new PTs}
\label{subsec:target_state_space_model}
Following \cite[Sec. VIII-B]{MeyKroWilLauHlaBraWin:J18}, each PT at time $t$ and agent pair $j$ is either a ``legacy'' PT or a ``new'' PT.
A legacy PT is a PT that
has already been introduced in the past,
either at current time $t$ at any previous agent pair $j' < j$, or at any previous time $t' < t$.
We denote with $\Set{L}_{t}^{(j)} \deq \{ 1, \ldots, L_{t}^{(j)} \}$ the set of $L_{t}^{(j)}$ legacy PTs at time $t$
at agent pair $j$,
and indicate with
$\underline{\V{y}}_{\ell,t}^{(j)} \! \deq \! [ \underline{\V{x}}^{(j)\T}_{\ell,t}, \underline{r}_{\ell,t}^{(j)} ]^{\T}$
the augmented state of legacy PT $\ell \! \in \! \Set{L}_{t}^{(j)}$, and with
$\underline{\V{y}}_{t}^{(j)} \! \deq \! [ \underline{\V{y}}^{(j)\T}_{1,t}, \ldots, \underline{\V{y}}^{(j)\T}_{L_{t}^{(j)}\rmv\rmv,t} ]^{\T}$
the joint legacy PT augmented state vector.

New PTs
model those targets that are detected for the first time by agent pair $j$ at time $t$.
Each new PT corresponds to an MOT measurement $\V{z}^{(j)}_{m,t}$; therefore, the number of new PTs at time $t$
at agent pair $j$ is $M_{t}^{(j)}$.
The augmented state of a new PT is denoted by $\overline{\V{y}}^{(j)}_{m,t} \! \deq \! [ \overline{\V{x}}^{(j)\T}_{m,t}, \overline{r}^{(j)}_{m,t} ]^{\T}$, $m \! \in \! \Set{M}^{(j)}_{t}$, and $\overline{r}^{(j)}_{m,t} = 1$ thus means that MOT measurement $m$ was generated by a target that was never detected before, namely, a \textit{newly detected} target.
We define the joint augmented state vector of all
new PTs at time $t$
at agent pair $j$ as $\overline{\V{y}}^{(j)}_{t} \! \deq \! \big[ \overline{\V{y}}^{(j)\T}_{1,t}, \ldots, \overline{\V{y}}^{(j)\T}_{M^{(j)}_{t}\rmv\rmv,t} \big]^{\T}$, and the joint augmented state vector of all
new PTs introduced at time $t$ as $\overline{\V{y}}_{t} \! \deq \! [ \overline{\V{y}}^{(1)\T}_{t}, \ldots, \overline{\V{y}}^{(J)\T}_{t} ]^{\T}$.

Legacy PTs and new PTs at time $t$
at agent pair $j$, become legacy PTs at the next agent pair $j + 1$, if $j < J$, or at the first agent pair at the next time step $t + 1$, if $j = J$;
in the latter case, this operation also implies performing the prediction from $t-1$ to $t$ of the PT states. It
then follows that the number of legacy PTs grows as $L_{t}^{(j)} = L_{t}^{(j - 1)} + M_{t}^{(j - 1)}$, where
$L_{t}^{(1)} = K_{t - 1}$, i.e., the number of legacy PTs at time $t$ at the first agent pair $j = 1$ is equal to the number of PTs at time $t - 1$.
Analogously, we can
reinterpret the vector $\underline{\V{y}}_{t}^{(j)}$ of all the 
legacy PT augmented states at time $t$
at agent pair $j$, as the vector stacking all the legacy PT augmented states at time $t$ at the previous agent pair $j-1$, and the new PT augmented states introduced at time $t$ at the previous agent pair $j-1$, that is, $\underline{\V{y}}_{t}^{(j)} = [\underline{\V{y}}_{t}^{(j - 1)\T}, \overline{\V{y}}_{t}^{(j-1)\T}]^{\T}$.
This
correspondence between legacy and new PTs at agent pair $j-1$, and legacy PTs at agent pair $j$, will hereafter be referred to as ``PT mapping''.
Given the sequential construction of the joint legacy PT augmented state vector $\underline{\V{y}}_{t}^{(j)}$ shown above, the vector $\V{y}_{t}$ of all the PT augmented states at time $t$ introduced in Section \ref{subsec:agent_state_space_model} can now be formally defined as
$\V{y}_{t}
\deq [ \underline{\V{y}}_{t}^{(1)\T}, \overline{\V{y}}_{t}^{(1)\T}, \overline{\V{y}}_{t}^{(2)\T}, \ldots, \overline{\V{y}}_{t}^{(J)\T} ]^{\T}
= [ \underline{\V{y}}_{t}^{(2)\T}, \linebreak \overline{\V{y}}_{t}^{(2)\T}, \ldots, \overline{\V{y}}_{t}^{(J)\T} ]^{\T}
= \cdots
= [ \underline{\V{y}}_{t}^{(J)\T}, \overline{\V{y}}_{t}^{(J)\T} ]^{\T}$.
The number of PTs at time $t$, after all the MOT measurements are incorporated, is therefore
$K_{t} = L_{t}^{(1)} + \sum_{j = 1}^{J} M_{t}^{(j)} = L_{t}^{(J)} + M_{t}^{(J)}$.
Note that the set of PTs at time $t = 0$ is assumed to be empty, i.e., $K_{0} = 0$, and so is the set of legacy PTs at time $t = 1$ at the first agent pair $j = 1$, i.e., $L_{1}^{(1)} = K_{0} = 0$.

\begin{figure}[!tb]
	\centering
	\includegraphics[width=1\linewidth]{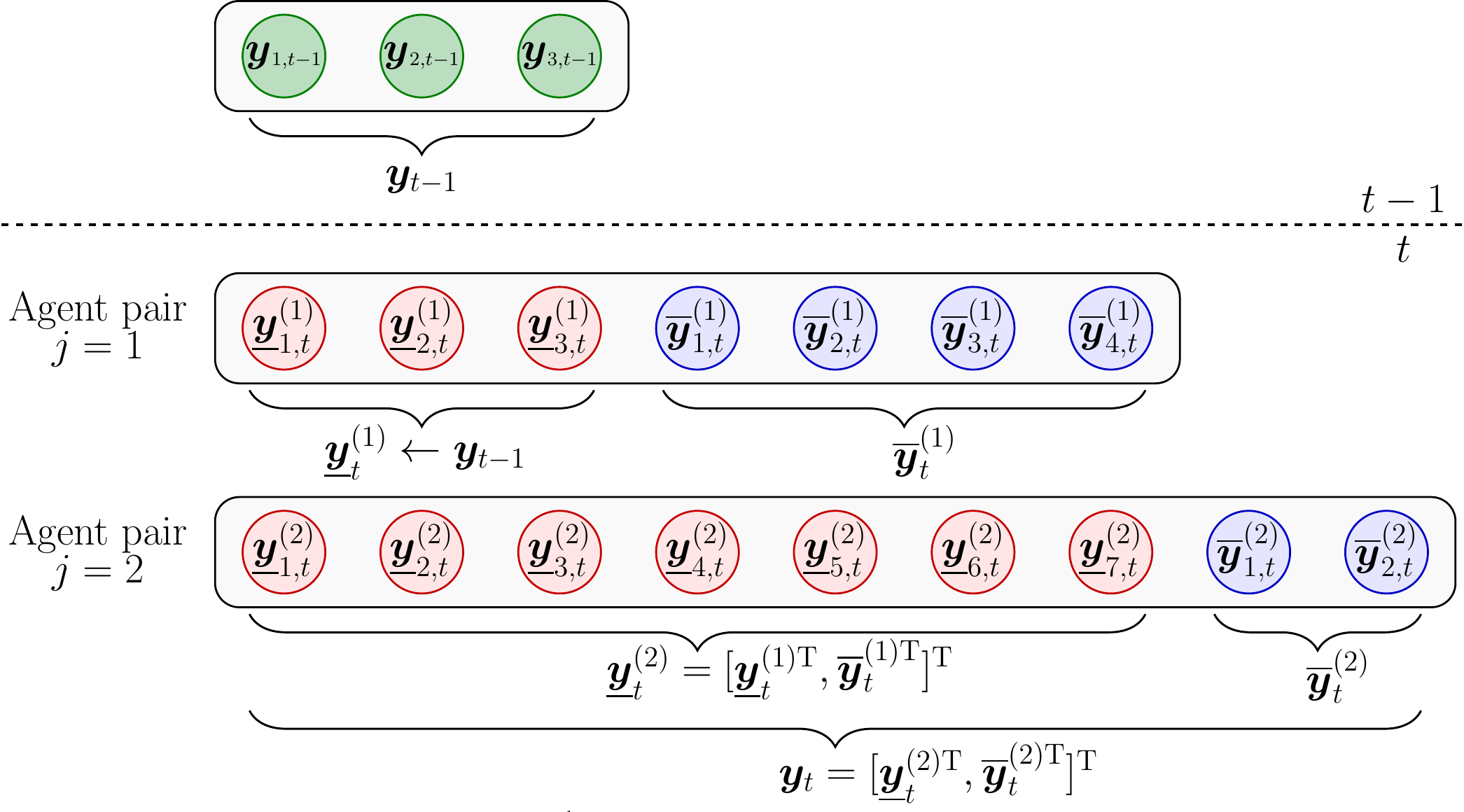}
	\vspace{-4mm}
	\caption{Illustration that describes the introduction of new PTs and the PT mapping between times $t - 1$ and $t$, and between agent pairs $j - 1$ and $j$. Green circles represent PTs at previous time $t-1$; red circles represent legacy PTs at time $t$ at agent pair $j$; blue circles represent new PTs introduced at time $t$ at agent pair $j$.
		    }
	\label{fig:PTevolution}
	\vspace{-2mm}
\end{figure}

We now provide a brief example that, with the support of Fig.~\ref{fig:PTevolution}, shows how new PTs are introduced and how the PT mapping works.
Let us consider one Tx-agent $a'$, i.e., $\Set{T}=\{a'\}$, and two Rx-agents, i.e., $\Set{R}=\{1,2\}$; thus we have $\Set{J}=\{1,2\}$, and we assume that the first agent pair $j=1$ is $(1,a')$, and the second agent pair $j=2$ is $(2,a')$. Note that this choice is arbitrary.
Furthermore, we assume that at time $t - 1$ the number of PTs is $K_{t - 1} = 3$, i.e., $\V{y}_{t - 1} = [ \V{y}_{1,t-1}^{\T}, \V{y}_{2,t-1}^{\T}, \linebreak \V{y}_{3,t-1}^{\T} ]^{\T}$, represented as green circles in the first row of Fig.~\ref{fig:PTevolution}, and that at time $t$ the number of MOT measurements at the agent pair $j = 1$ is $M_{t}^{(1)} = 4$, and the number of MOT measurements at agent pair $j = 2$ is $M_{t}^{(2)} = 2$.
The PTs at previous time $t - 1$ become --- once PTs' states prediction is performed --- legacy PTs at time $t$ at agent pair $j = 1$, and are represented as red circles in the second row of Fig.~\ref{fig:PTevolution}; therefore, $L_{t}^{(1)} = K_{t - 1} = 3$ and we formally have that the first PT at time $t-1$ becomes the first legacy PT at time $t$ at agent pair $j =1$, i.e., $\underline{\V{y}}_{1,t}^{(1)}
\gets \V{y}_{1,t-1}$, and so forth for the other PTs, i.e., $\underline{\V{y}}_{2,t}^{(1)}
\gets \V{y}_{2,t-1}$ and $\underline{\V{y}}_{3,t}^{(1)}
\gets \V{y}_{3,t-1}$.
Moreover, since the number of MOT measurements at agent pair $j = 1$ is $M_{t}^{(1)} = 4$, four new PTs are introduced at this stage, i.e., $\overline{\V{y}}_{1,t}^{(1)}$, $\overline{\V{y}}_{2,t}^{(1)}$, $\overline{\V{y}}_{3,t}^{(1)}$, and $\overline{\V{y}}_{4,t}^{(1)}$, represented as blue circles in the second row of Fig. \ref{fig:PTevolution}.
Then, legacy PTs and new PTs at agent pair $j = 1$ become legacy PTs at agent pair $j = 2$, represented as red circles in the third row of Fig. \ref{fig:PTevolution}.
Therefore, $L_{t}^{(2)} = L_{t}^{(1)} + M_{t}^{(1)} = 3 + 4 = 7$, and we formally have that the first legacy PT at agent pair $j = 1$ becomes the first legacy PT at agent pair $j = 2$, i.e., $\underline{\V{y}}_{1,t}^{(2)}
\gets \underline{\V{y}}_{1,t}^{(1)}$, and so forth for the other legacy PTs at agent pair $j = 1$, and that the first new PT at agent pair $j = 1$ becomes the fourth legacy PT at agent pair $j = 2$, i.e., $\underline{\V{y}}_{4,t}^{(2)}
\gets \overline{\V{y}}_{1,t}^{(1)}$, and so forth for the other new PTs at agent pair $j = 1$.
The vector of all the legacy PT augmented states at time $t$ and agent pair $j = 2$ is thus $\underline{\V{y}}_{t}^{(2)} = [\underline{\V{y}}_{1,t}^{(1)\T}, \ldots, \underline{\V{y}}_{3,t}^{(1)\T}, \overline{\V{y}}_{1,t}^{(1)\T}, \ldots, \overline{\V{y}}_{4,t}^{(1)\T}]^{\T}$.
This is an example of PT mapping from agent pair $j - 1$ to agent pair $j$.
Moreover, since the number of MOT measurements at agent pair $j = 2$ is $M_{t}^{(2)} = 2$, two new PTs are introduced at this stage, i.e., $\overline{\V{y}}_{1,t}^{(2)}$ and $\overline{\V{y}}_{2,t}^{(2)}$, represented as blue circles in the third row of Fig. \ref{fig:PTevolution}.
The number of PTs at time $t$ is thus $K_{t} = L_{t}^{(2)} + M_{t}^{(2)} = 7 + 2 = 9$; these nine PTs will then become legacy PTs at time $t + 1$ at agent pair $j = 1$.

Note that using this mechanism the number of PTs grows indefinitely over time. To
keep a tractable number of PTs,
a sub-optimal pruning step is performed once all the MOT measurements at time $t$ are processed; details are provided in Section \ref{subsubsec:particle_impl}.
Finally, for the reader's convenience, the main sets introduced in this section are summarized in Table \ref{tab:set}.

\begin{table}[!t]
\centering
    	\renewcommand{\arraystretch}{1.2}
	\small	
	\caption{Summary of the sets introduced in Section \ref{sec:problem-description-system-model}}
	\label{tab:set}
	\centering
	\begin{tabular}{ l | c  }
		\textsc{Description}	&	\textsc{Symbol}	\\[.5mm]
		\hline \hline
		Agents & $\Set{A}$ \\
		Tx-agents & $\Set{T}\subseteq \Set{A}$ \\
		Rx-agents & $\Set{R}\subseteq \Set{A}$ \\
		Agent pairs & $\mathcal{J}$ \\
		PTs at time $t$ & $ \Set{K}_t$ \\
		Legacy PTs at time $t$ at agent pair $j$ & $\Set{L}_{t}^{(j)}$  \\[.5mm]
		MOT measurements/new PTs & \multirow{2}{*}{$\Set{M}_{t}^{(j)}$} \\
		\hspace{1mm} at time $t$ at agent pair $j$ &  \\
		\hline
	\end{tabular}
\vspace{-2mm}
\end{table}

\subsection{Practical use cases}
\label{sec:relationToRealCases}
This section links the concepts of agents, targets, and observations introduced so far
to practical use cases. The intention is to
ground the reader in the contextualization and identification of potential applications. In this regard, we provide three examples in popular areas of interest.
\begin{enumerate}[wide]
    \item[$\bullet$] \textit{Maritime situational awareness}: in an acoustic underwater wireless sensor network, agents can be nearly stationary communication gateways (either anchored on the seabed or floating), and underwater or surface mobile vehicles (either manned or unmanned) with communication capabilities that form a
    multistatic active sonar system: one of these vehicles is equipped with a sonar source, while the others act as receivers and use arrays of hydrophones \cite{FerMunTesBra:J17}.
        Agent state $\V{s}_{a,t}$
includes position and possibly other kinematic parameters, such as velocity and acceleration; in case of 2D position $\check{\V{s}}_{a,t}$, velocity $\dot{\check{\V{s}}}_{a,t}$, and acceleration $\ddot{\check{\V{s}}}_{a,t}$, the agent state becomes $\V{s}_{a,t} = [ \check{\V{s}}_{a,t}^{\T} , \dot{\check{\V{s}}}_{a,t}^{\T}, \ddot{\check{\V{s}}}_{a,t}^{\T} ]^{\T} \in \mathbb{R}^6$.
    For surface agents, navigation data $\V{g}_{a,t}$ can be
        obtained through GNSS, in which case the noise-free navigation data can be $\V{g}_{a,t} = \check{\V{s}}_{a,t}$;
        whereas for underwater agents, the noise-free navigation data can be provided in terms of acceleration by the INS, in which case $\V{g}_{a,t} = \ddot{\check{\V{s}}}_{a,t}$.
    Mobile vehicles
        produce inter-agent measurements
        through different
        localization techniques \cite{MaoFidAnd:J07}, with which the range between two agents can be determined; furthermore, using the array of hydrophones the bearing
        can also be obtained. In such a case, the noise-free inter-agent measurement produced by
        agent pair $(a,a')$ is $\VG{\rho}_{t}^{(a,a')} = [\norm{\check{\V{s}}_{a,t} - \check{\V{s}}_{a'\rmv,t}}, \angle(\check{\V{s}}_{a'\rmv,t} - \check{\V{s}}_{a,t})]^{\T}$.
Similarly, the MOT measurement produced by agent $a$ acting as receiver due to the signal transmitted by the sonar source $a'$ that reflects off
a target (e.g., submarine) whose 2D position is $\check{\V{x}}_{k,t}$, generally consists of range and bearing information, that is, $\V{z}_{m,t}^{(a,a')} = [\norm{\check{\V{s}}_{a,t} - \check{\V{x}}_{k,t}} + \norm{\check{\V{s}}_{a'\rmv,t} - \check{\V{x}}_{k,t}}, \angle(\check{\V{x}}_{k,t} - \check{\V{s}}_{a,t})]^{\T}$.
    \item[$\bullet$] \textit{Cooperative intelligent transportation system}: in this scenario, agents are mobile land vehicles and road infrastructure nodes equipped with sensing
        and communication
        systems. Common devices used for sensing include radar, lidar, camera and ultrasound technologies, while the communication among the agents is achieved through ITS-G5/DSRC\footnote{Dedicated Short-Range Communications} or C-V2X\footnote{Cellular-Vehicle-to-Everything} \cite{ZhoXuCheWan:J20}.
GNSS is the most widely adopted technology to obtain navigation data,
often combined with INS for the identification of abrupt kinematic events. 
Inter-agent measurements can rely on V2X communications to collect range and bearing (if antenna arrays are used) information, and multipath signals can be used to detect the presence of scattering
targets (e.g., pedestrians, cyclists, other vehicles) in the environment.
    \item[$\bullet$] \textit{Drone swarm}: agents are flying drones equipped with a GNSS receiver, INS, ultra wideband (UWB) technology, and a camera.
Navigation data from GNSS and INS is paired with UWB ranging
for a cooperative localization of the agents, while targets are detected by cameras with overlapping fields of view. Targets can vary from one application to another, as drone swarms are used in a variety of areas, e.g., smart cities, agriculture, environmental monitoring or mapping, military operations, search and rescue.
\end{enumerate}

These use cases represent only a
subset of
applications in which the proposed cooperative self-localization and multitarget tracking algorithm can be employed.

\section{Stochastic Problem Formulation}
\label{sec:stochastic-problem-formulation}
This section describes the detection and state estimation of a PT at time $t$, the state estimation of an agent, the MOT measurement model and data association problem (caused by the MOT measurement-origin uncertainty), and summarizes the assumptions used in the proposed formulation.
Then, a factorization of the joint posterior pdf of the PT augmented states, agent states, and data association variables (introduced in the next section)
is provided; this factorization is eventually used to compute the marginal posterior pdfs $f(\V{s}_{a,t} | \V{g}_{1:t}, \VG{\rho}_{1:t}, \V{z}_{1:t} )$ and $f(\V{x}_{k,t} , r_{k,t} |  \V{g}_{1:t}, \VG{\rho}_{1:t}, \V{z}_{1:t} )$.

\subsection{Agent self-localization and target tracking}
\label{subsec:target_det_state_est}
The objective of this work is the cooperative self-localization of agents, jointly with the detection and localization of PTs.
This task is performed
with a Bayesian approach based on navigation data $\V{g}_{1:t}$, inter-agent measurements $\VG{\rho}_{1:t}$, and MOT measurements $\V{z}_{1:t}$,
that boils down to the computation of the marginal posterior pdfs
$f(\V{s}_{a,t} | \V{g}_{1:t}, \VG{\rho}_{1:t}, \V{z}_{1:t} )$, $\forall a \in \Set{A}$, and $
f(\V{x}_{k,t} ,  r_{k,t} | \V{g}_{1:t}, \VG{\rho}_{1:t}, \V{z}_{1:t} )$, $\forall k \in \Set{K}_{t}$.

The detection and state estimation of a PT at time $t$ is performed once all the MOT measurements are processed.
The detection of a PT $k \in \Set{K}_{t}$  amounts to calculating the marginal posterior existence probability $f(r_{k,t} \!=\! 1 | \V{g}_{1:t}, \VG{\rho}_{1:t}, \V{z}_{1:t}) \! = \! \int f(\V{x}_{k,t}, r_{k,t} = 1 | \V{g}_{1:t}, \VG{\rho}_{1:t}, \V{z}_{1:t}) \mathrm{d}\V{x}_{k,t}$, and comparing it with a suitably chosen threshold $P_{\text{ex}}$; that is, if $f(r_{k,t} \!=\! 1 | \V{g}_{1:t}, \linebreak  \VG{\rho}_{1:t}, \V{z}_{1:t}) \! > \! P_{\text{ex}}$, the existence of PT $k$ is confirmed.
Then, for each detected PT $k$, an estimate of its state $\V{x}_{k,t}$ is provided by the minimum mean square error (MMSE) estimator
\begin{align}
	\hat{\V{x}}^\text{MMSE}_{k,t} \, \deq \rmv \int \rmv \V{x}_{k,t} \ist f(\V{x}_{k,t} | r_{k,t} \!=\! 1, \V{g}_{1:t}, \VG{\rho}_{1:t}, \V{z}_{1:t} ) \ist \mathrm{d}\V{x}_{k,t} \ist ,
\end{align}
where $f(\V{x}_{k,t} | r_{k,t} = 1, \V{g}_{1:t}, \VG{\rho}_{1:t}, \V{z}_{1:t} ) = f(\V{x}_{k,t}, r_{k,t} = 1 | \V{g}_{1:t}, \linebreak \VG{\rho}_{1:t}, \V{z}_{1:t} ) / f(r_{k,t} = 1 | \V{g}_{1:t}, \VG{\rho}_{1:t}, \V{z}_{1:t} )$.
Likewise, an estimate of the agent state $\V{s}_{a,t}$, $a \in \Set{A}$, is provided by the MMSE estimator
\begin{align}
	\hat{\V{s}}^\text{MMSE}_{a,t} \, \deq \rmv \int \rmv \V{s}_{a,t} \ist f(\V{s}_{a,t} | \V{g}_{1:t}, \VG{\rho}_{1:t}, \V{z}_{1:t} ) \ist \mathrm{d}\V{s}_{a,t} \ist .
\end{align}

\subsection{MOT measurement model and data association}
\label{subsec:meas_model}
As mentioned above (cf. Section~\ref{subsubsec:MOT}), the MOT measurements $\V{z}^{(j)}_{m,t}$, $m \! \in \! \Set{M}^{(j)}_{t}$, produced at time $t$
at agent pair $j$, have unknown origins.
Specifically, we make the assumption --- known as \textit{point-target assumption} --- that each MOT measurement $\V{z}^{(j)}_{m,t}$, $m \in \Set{M}_{t}^{(j)}$, at time $t$ at agent pair $j$, originates either from
a legacy PT or agent, hereafter aggregated under the term \textit{legacy object}, or from a new PT (i.e., a PT never detected before), or from clutter (i.e., a false alarm), and it cannot originate from more than one source
(legacy object, new PT, or clutter) simultaneously.
Conversely, each legacy object or new PT
can generate at most one MOT measurement at time $t$
at agent pair $j$ \cite{BarWilTia:B11}.
To handle this uncertainty, firstly we introduce the joint
legacy object state vector $\V{o}^{(j)}_{t} \deq \big[ \V{o}_{1,t}^{(j)\T}, \ldots, \linebreak \V{o}_{O_{t}^{(j)}\rmv\rmv,t}^{(j)\T} \big]^{\T}$, with $O^{(j)}_{t} \deq L^{(j)}_{t} + A$, as the vector stacking at time $t$ the legacy PT states at agent pair $j$, and the agent states.
That is,
$\V{o}^{(j)}_{i,t} = \underline{\V{x}}^{(j)}_{\ell,t}$ if $i = \ell$ and  $\ell \! \in \! \Set{L}_{t}^{(j)}$, and $\V{o}^{(j)}_{i,t} = \V{s}_{a,t}$ if $i = L_{t}^{(j)} + a$ and $a \! \in \! \Set{A}$.
We observe that the
vector $\V{o}^{(j)}_{t}$ includes the state vectors of the Rx-agent and the Tx-agent
at $i = L_{t}^{(j)} + j_{1}$ and $i = L_{t}^{(j)} + j_{2}$, respectively, which cannot generate any MOT measurement at agent pair $j$.
Therefore, we assume that
a legacy
object $i \in \Set{O}_{t}^{(j)} \deq \{ 1, \ldots, \linebreak O_{t}^{(j)} \}$ is ``detected'' by the
agent pair $j$ --- i.e.,
it generates a measurement $\V{z}_{m,t}^{(j)}$ at the
agent pair $j$ --- with probability
$P_{\text{d}}^{(j)}(\V{o}_{i,t}^{(j)}, \V{s}_{j_{1}\rmv,t}, \V{s}_{j_{2}\rmv,t})\ist$,
defined for $i \neq  L_{t}^{(j)} + j_{1}$ and $i \neq L_{t}^{(j)} + j_{2}$ as
\begin{align}
	&P_{\text{d}}^{(j)} \big( \V{o}_{i,t}^{(j)}, \V{s}_{j_{1}\rmv,t}, \V{s}_{j_{2}\rmv,t} \big) \deq
	\begin{dcases}
		P_{\text{d},\texttt{mono}}^{(j)} \big( \V{o}_{i,t}^{(j)}, \V{s}_{j_{1}\rmv,t} \big) & j_{1} = j_{2} \ist , \\
		P_{\text{d},\texttt{bi}}^{(j)} \big( \V{o}_{i,t}^{(j)}, \V{s}_{j_{1}\rmv,t}, \V{s}_{j_{2}\rmv,t} \big) & j_{1} \neq j_{2} \ist ,
	\end{dcases}
	\nonumber \\[-1mm]
	\label{eq:def-pd1} \\[-1mm]
	\intertext{and for $i = L_{t}^{(j)} + j_{1}$ or $i = L_{t}^{(j)} + j_{2}$ as}
	&P_{\text{d}}^{(j)} \big( \V{o}_{i,t}^{(j)}, \V{s}_{j_{1}\rmv,t}, \V{s}_{j_{2}\rmv,t} \big) = 0 \ist,
	\label{eq:def-pd2}
\end{align}
where $P_{\text{d},\texttt{mono}}^{(j)}(\cdot)$ is the monostatic detection probability of agent $j_{1} = j_{2}$, and $P_{\text{d},\texttt{bi}}^{(j)}(\cdot)$ is the bistatic detection probability of the
agent pair $j$, with $j_{1} \neq j_{2}$.
Secondly, following \cite{MeyKroWilLauHlaBraWin:J18},
we introduce:
\textit{(i)} the set $\Set{N}_{t}^{(j)}$ of
MOT measurements generated by
new PTs at time $t$
at agent pair $j$, that is, $\Set{N}_{t}^{(j)} \deq \{ m \in \Set{M}_{t}^{(j)} : \overline{r}_{m,t}^{(j)} = 1 \}$;
\textit{(ii)} the legacy object-oriented association vector $\VG{\alpha}^{(j)}_{t} \! \deq \! \big[ \alpha^{(j)}_{1,t}, \ldots, \alpha^{(j)}_{O^{(j)}_{t}\rmv\rmv,t} \big]^{\T}$; 
and \textit{(iii)} the MOT measurement-oriented association vector $\VG{\beta}^{(j)}_{t} \! \deq \! \big[ \beta^{(j)}_{1,t}, \ldots, \linebreak \beta^{(j)}_{M^{(j)}_{t}\rmv\rmv,t} \big]^{\T}$.
Here, $\alpha^{(j)}_{i,t}$, $i \in \Set{O}_{t}^{(j)}$, is defined as $m \in \Set{M}^{(j)}_{t}$ if legacy object $i$ generates MOT measurement $m$, and $0$ if legacy object $i$ does not generate any MOT measurement; and $\beta^{(j)}_{m,t}$, $m \in \Set{M}_{t}^{(j)}$, is defined as $i$ if MOT measurement $m$ originates from legacy object $i$, and $0$ if MOT measurement $m$ does not originate from any legacy object.
The point-target assumption can therefore be expressed by the indicator function $\Phi(\VG{\alpha}^{(j)}_{t}, \VG{\beta}^{(j)}_{t})$, defined as
\begin{align}
	\Phi \big( \VG{\alpha}^{(j)}_{t}, \VG{\beta}^{(j)}_{t} \big) \deq \Psi \big( \VG{\alpha}^{(j)}_{t}, \VG{\beta}^{(j)}_{t} \big) \hspace{-2mm} \prod_{m \in \Set{N}_{t}^{(j)}} \hspace{-2mm} \Gamma \big( \beta^{(j)}_{m,t} \big) \ist ,
	\label{eq:Phi-function}
\end{align}
where
\begin{align}
	\Gamma \big( \beta^{(j)}_{m,t} \big) \deq
	\begin{dcases}
		0	& \beta_{m,t}^{(j)} \in \Set{O}_{t}^{(j)} \ist , \\
		1	& \beta_{m,t}^{(j)} = 0 \ist ,
	\end{dcases}
	\label{eq:gamma-func-data-ass}
\end{align}
and
\begin{align}
	\Psi \big( \VG{\alpha}^{(j)}_{t},\VG{\beta}^{(j)}_{t} \big) \deq \prod_{i \in \Set{O}^{(j)}_{t}} \prod_{m \in \Set{M}^{(j)}_{t}} \psi \big( \alpha^{(j)}_{i,t},\beta^{(j)}_{m,t} \big) \ist,
	\label{eq:Psi-function}
\end{align}
with
\begin{align}
	\psi (\alpha^{(j)}_{i,t},\beta^{(j)}_{m,t}) \deq
	\begin{cases}
		0	& \quad \alpha^{(j)}_{i,t} = m \text{ and } \beta^{(j)}_{m,t} \neq i \,, \\
			& \hspace{5mm} \quad  \text{or } \alpha^{(j)}_{i,t} \neq m \text{ and } \beta^{(j)}_{m,t} = i \,, \\
		1	& \quad	\text{otherwise} \, .
\end{cases}
\end{align}
(We observe that, since the product in \eqref{eq:Phi-function} is over the set $\Set{N}_{t}^{(j)}$, the indicator function $\Phi(\VG{\alpha}^{(j)}_{t}, \VG{\beta}^{(j)}_{t})$ formally depends also on the existence variables $\overline{r}_{m,t}^{(j)}$, $m \in \Set{M}_{t}^{(j)}$.)
Stated differently, valid associations described by $\VG{\alpha}_{t}^{(j)}$ and $\VG{\beta}_{t}^{(j)}$ are those for which $\Phi ( \VG{\alpha}^{(j)}_{t}, \VG{\beta}^{(j)}_{t} ) = 1$; and we note that
$\Psi (\VG{\alpha}^{(j)}_{t},\VG{\beta}^{(j)}_{t})$ is $0$ if an MOT measurement is associated with two or more
legacy objects (and, vice versa, if
a legacy object is associated with two or more MOT measurements), and $1$ otherwise;
and that the product over $m \in \Set{N}_{t}^{(j)}\!$ of $\Gamma(\beta^{(j)}_{m,t})$ is $0$ if
any MOT measurement generated by a
new PT is
also associated with a
legacy object,
and $1$ otherwise.
For convenience, we also define the vectors $\VG{\alpha}_{t} \! \deq \! [ \VG{\alpha}^{(1)\T}_{t}, \ldots, \VG{\alpha}^{(J)\T}_{t} ]^{\T}$ and $\VG{\beta}_{t} \! \deq \! [ \VG{\beta}^{(1)\T}_{t}, \ldots, \VG{\beta}^{(J)\T}_{t} ]^{\T}$, as well as $\VG{\alpha}_{1:t} \! \deq \! [ \VG{\alpha}^{\T}_{1}, \ldots, \VG{\alpha}^{\T}_{t} ]^{\T}$ and $\VG{\beta}_{1:t} \! \deq \! [ \VG{\beta}^{\T}_{1}, \ldots, \VG{\beta}^{\T}_{t} ]^{\T}$.

Finally,
if
the MOT measurement $m$ is generated by legacy object $i$ ($i \neq  L_{t}^{(j)} + j_{1}$ and $i \neq L_{t}^{(j)} + j_{2}$), 
the statistical dependence of $\V{z}_{m,t}^{(j)}$ on the legacy object state $\V{o}_{i,t}^{(j)}$, the Rx-agent state $\V{s}_{j_{1}\rmv,t}$, and the Tx-agent state $\V{s}_{j_{2}\rmv,t}$, is given by
\begin{align}
	&\mathfrak{f} \big(\V{o}_{i,t}^{(j)},  \V{s}_{j_{1}\rmv,t}, \V{s}_{j_{2}\rmv,t} ; \V{z}_{m,t}^{(j)} \big) \\
	&\hspace{12mm}\deq 
	\begin{dcases}
		f_{\texttt{mono}} \big(\V{z}_{m,t}^{(j)} \big| \V{o}_{i,t}^{(j)}, \V{s}_{j_{1}\rmv,t} \big) & j_{1} = j_{2} \ist , \\
		f_{\texttt{bi}} \big(\V{z}_{m,t}^{(j)} \big| \V{o}_{i,t}^{(j)}, \V{s}_{j_{1}\rmv,t}, \V{s}_{j_{2}\rmv,t} \big) & j_{1} \neq j_{2} \ist ,
	\end{dcases}
	\label{eq:def-likelihood-f0}
\end{align}
where the likelihoods $f_{\texttt{mono}} (\V{z}_{m,t}^{(j)} \big| \V{o}_{i,t}^{(j)}, \V{s}_{j_{1}\rmv,t} )$ and $f_{\texttt{bi}} (\V{z}_{m,t}^{(j)} \big| \linebreak \V{o}_{i,t}^{(j)}, \V{s}_{j_{1}\rmv,t}, \V{s}_{j_{2}\rmv,t} )$ were introduced in Section \ref{subsubsec:MOT}.
If
the MOT measurement $m$ is generated by a new PT,
the statistical dependence of $\V{z}_{m,t}^{(j)}$ on the new PT state $\overline{\V{x}}_{m,t}^{(j)}$, Rx-agent state $\V{s}_{j_{1}\rmv,t}$, and Tx-agent state $\V{s}_{j_{2}\rmv,t}$, is still described by
the likelihoods in \eqref{eq:def-likelihood-f0} in which the legacy object state $\V{o}_{i,t}^{(j)}$ is replaced by the new PT state $\overline{\V{x}}_{m,t}^{(j)}$.

\subsection{Assumptions}
\label{subsec:assumptions}
The assumptions underlying the proposed stochastic formulation --- besides the point-target assumption --- are here summarized: some
are basic assumptions commonly used in multisensor multitarget tracking \cite{BarWilTia:B11}, while others belong to the specific formulation borrowed from~\cite{MeyHliWye:J16,MeyKroWilLauHlaBraWin:J18}.
\begin{enumerate}

\item[(A1)] The joint agent state vector $\V{s}_{t}$ evolves over time according to a first-order Markov model, and each agent state vector $\V{s}_{a,t}$, evolves independently \cite{BarWilTia:B11};
therefore, the joint agent state transition pdf $f(\V{s}_{t} | {\V{s}}_{t-1})$ factorizes as
\begin{align}
	f(\V{s}_{t} | {\V{s}}_{t-1}) = \prod_{a \in \Set{A}} \tau_{a}(\V{s}_{a,t}  | {\V{s}}_{a,t-1} ) \, ,
	\label{eq:agents-states-transition}
\end{align}
where $\tau_{a}(\V{s}_{a,t}  | {\V{s}}_{a,t-1} )$ is a known state-transition pdf (cf. Section~\ref{subsec:agent_state_space_model}).

\item[(A2)] The joint PT augmented state vector $\V{y}_{t}$ evolves over time according to a first-order Markov model, and each PT augmented state vector $\V{y}_{k,t}$ evolves independently \cite{BarWilTia:B11}.
Recalling that for each PT augmented state $\V{y}_{k,t - 1}$, $k \in \Set{K}_{t-1}$, at time $t-1$, there is one legacy PT augmented state $\underline{\V{y}}_{\ell,t}^{(1)}$, $\ell \in \Set{L}_{t}^{(1)}$, at the first agent pair at time $t$ (in other words, $\Set{L}_{t}^{(1)} = \Set{K}_{t-1}$), the joint PT augmented state transition pdf is given by
\begin{align}
	\hspace{-5mm}f \big( \underline{\V{y}}_{t}^{(1)} \big| {\V{y}}_{t-1} \big) = \hspace{-1.5mm} \prod_{k \in \Set{K}_{t-1}} \hspace{-1.5mm} f \big( \underline{\V{x}}_{k,t}^{(1)} , \underline{r}_{k,t}^{(1)} \big| {\V{x}}_{k,t-1} , {r}_{k,t-1} \big) \ist .
	\label{eq:legacy-PT-transition}
\end{align}
Let us recall from Section \ref{subsec:target_state_space_model} that the set of PTs at time $t = 0$ is empty, i.e., $\Set{K}_{0} = \varnothing$, and so is the set of legacy PTs at time $t = 1$ at the first agent pair $j = 1$, i.e., $\Set{L}_{1}^{(1)} = \varnothing$.
Therefore, the state transition pdf in \eqref{eq:legacy-PT-transition} cannot be performed at time $t = 1$, and we formally
introduce $f (\underline{\V{y}}_{1}^{(1)} | \V{y}_{0}) = 1$ for future use.
An expression of $f(\underline{\V{x}}_{k,t}^{(1)}, \linebreak \underline{r}_{k,t}^{(1)} | {\V{x}}_{k,t-1} , {r}_{k,t-1})$ is provided in \cite[Sec. VIII-C]{MeyKroWilLauHlaBraWin:J18}, and is here
reported for completeness.
If PT $k$ does not exist at time $t-1$, i.e., if $r_{k,t-1} = 0$, it cannot exist as legacy PT at time $t$, i.e, $\underline{r}_{k,t}^{(1)} = 0$, and thus its state pdf is $f_\text{D} (\underline{\V{x}}_{k,t}^{(1)})$. That is,
\begin{align}
	&f \big( \underline{\V{x}}_{k,t}^{(1)} , \underline{r}_{k,t}^{(1)} \big| \V{x}_{k,t-1} , {r}_{k,t-1} = 0 \big) \\
	&\hspace{30mm}=
	\begin{cases}
		f_\text{D} \big( \underline{\V{x}}_{k,t}^{(1)} \big)	&	\underline{r}_{k,t}^{(1)} = 0 \,, \\
		0	&	\underline{r}_{k,t}^{(1)} = 1 \,.
	\end{cases}
\end{align}
Conversely, if PT $k$ exists at time $t-1$, i.e., if $r_{k,t-1} = 1$,\linebreak it survives as legacy PT with probability $p_\text{s}({\V{x}}_{k,t-1})$, and its state $\underline{\V{x}}_{k,t}^{(1)}$ is distributed according to the state transition pdf $f(\underline{\V{x}}_{k,t}^{(1)} | {\V{x}}_{k,t-1})$. Thus,
\begin{align}
	&f \big( \underline{\V{x}}_{k,t}^{(1)} , \underline{r}_{k,t}^{(1)} \big| \V{x}_{k,t-1} , {r}_{k,t-1} = 1 \big) \nonumber \\
	&\hspace{10mm} =
	\begin{cases}
		\big( 1-p_\text{s} ({\V{x}}_{k,t-1}) \big) f_\text{D} \big( \underline{\V{x}}_{k,t}^{(1)} \big)	&	\underline{r}_{k,t}^{(1)} = 0 \,, \\
		p_\text{s} ({\V{x}}_{k,t-1}) f \big( \underline{\V{x}}_{k,t}^{(1)} \big| {\V{x}}_{k,t-1} \big)	&	\underline{r}_{k,t}^{(1)} = 1 \,.
	\end{cases}
\end{align}
The state transition pdf is defined by the PT
kinematic model, that is,
\begin{align}
	\underline{\V{x}}_{k,t}^{(1)} = \VG{\varsigma} \big( \V{x}_{k,t-1}, \V{e}_{k,t} \big) \ist ,
	\label{eq:PT-dynamic-model}
\end{align}
and by the statistics of the
process noise $\V{e}_{k,t}$.

\item[(A3)] The joint agent state vector $\V{s}_{t}$ and the joint PT augmented state vector $\V{y}_{t}$
evolve independently over time \cite{MeyHliWye:J16,BarWilTia:B11}.

\item[(A4)] The
states of legacy PTs and new PTs at time $t$ at agent pair $j$ are independent \cite{MeyKroWilLauHlaBraWin:J18}.

\item[(A5)] The number of new PTs at time $t$
at agent pair $j$ is a priori (i.e., before the MOT measurements are observed) Poisson distributed with mean $\mu_{\text{n}}^{(j)}$. The states of new PTs are independent and identically distributed
according to the prior pdf $f_{\text{n}} (\overline{\V{x}}_{m,t}^{(j)})$ \cite{MeyKroWilLauHlaBraWin:J18}.

\item[(A6)] Given the agent states and PT augmented states at time $t-1$, the observations (navigation data, inter-agent 
and MOT measurements),
association variables, agent states, and PT augmented states at time $t$, are conditionally independent of all the past ($t' < t$) variables \cite{MeyHliWye:J16,MeyKroWilLauHlaBraWin:J18}.

\item[(A7)] Given the agent states and legacy PT augmented states at time $t$, the new PT augmented states, observations, and association variables at time $t$, are conditionally independent of all the past ($t' < t$) agent states and PT augmented states \cite{MeyHliWye:J16,MeyKroWilLauHlaBraWin:J18}.

\item[(A8)] Similarly, given the agent states at time $t$, and the legacy PT augmented states at time $t$ at agent pair $j$, the new PT augmented states, observations, and association variables at time $t$ at current and future agent pairs, are conditionally independent of all the past ($j' < j$) variables \cite{MeyHliWye:J16,MeyKroWilLauHlaBraWin:J18}.

\item[(A9)] The navigation data and the inter-agent measurements are conditionally independent of each other, and of all the other variables, given the joint agent state vector $\V{s}_{t}$ \cite{BarWilTia:B11}.
Then, recalling their measurement models (cf. \eqref{eq:nav-data} and \eqref{eq:inter-agent-meas}), and in particular that the noise terms are independent across agent states and agent pairs, respectively, it follows that the joint navigation data likelihood $f(\V{g}_{t} | \V{s}_{t})$ factorizes as
\begin{align}
	f (\V{g}_{t} | \V{s}_{t}) = \prod_{a \in \Set{A}^{\V{g}}_{t}} \mathfrak g_{a}(\V{g}_{a,t} | \V{s}_{a,t}) \ist ,
	\label{eq:navigation-meas-joint-likelihood}
\end{align}
and that the joint likelihood $f (\VG{\rho}_{t} | \V{s}_{t})$ factorizes as
\begin{align}
	f (\VG{\rho}_{t} | \V{s}_{t}) = \prod_{a \in \Set{R}} \prod_{a' \in \Set{T}^{(a)}_{t}} \mathfrak{d} \big( \VG{\rho}^{(a,a')}_{t} | \V{s}_{a,t}, \V{s}_{a'\rmv,t} \big) \ist.
	\label{eq:direct-meas-joint-likelihood}
\end{align}

\item[(A10)] The number of false alarm MOT measurements at time $t$
at agent pair $j$ is Poisson distributed with mean $\mu_{\text{c}}^{(j)}$. False alarm MOT measurements are
independent and identically distributed according to the pdf $f_{\text{c}}^{(j)}(\V{z}_{m,t}^{(j)})$ \cite{BarWilTia:B11,MeyKroWilLauHlaBraWin:J18}.

\item[(A11)]
The agent-originated and PT-originated MOT measurements at time $t$ at agent pair $j$, are conditionally independent of each other, and conditionally independent of the false alarm MOT measurements, given the agent states $\V{s}_{j_{1},t}$ and $\V{s}_{j_{2},t}$,
PT augmented states, and
association variables \cite{MeyHliWye:J16,MeyKroWilLauHlaBraWin:J18}.

\end{enumerate}

\subsection{Joint posterior pdf}
\label{subsec:joint-posterior-pdf}
The posterior pdfs $f(\V{y}_{k,t} | \V{g}_{1:t}, \VG{\rho}_{1:t}, \V{z}_{1:t} )$, $k \in \Set{K}_{t}$, and $f(\V{s}_{a,t} | \V{g}_{1:t}, \VG{\rho}_{1:t}, \V{z}_{1:t} )$, $a \in \Set{A}$, introduced in Section~\ref{subsec:target_det_state_est}, are marginal densities of the joint posterior pdf $f (\V{y}_{1:t}, \V{s}_{0:t}, \VG{\alpha}_{1:t}, \linebreak \VG{\beta}_{1:t} | \V{g}_{1:t}, \VG{\rho}_{1:t}, \V{z}_{1:t})$. By using assumptions (A3), (A6), (A7), (A8), and (A9), 
this joint posterior pdf can be factorized as (details are provided in the Appendix)
\begin{align}
	&f \big( \V{y}_{1:t}, \V{s}_{0:t}, \VG{\alpha}_{1:t}, \VG{\beta}_{1:t} \big| \V{g}_{1:t}, \VG{\rho}_{1:t}, \V{z}_{1:t} \big) \propto
		f \big( \V{s}_{0} \big) \nonumber \\[0mm] 
	&\hspace{5mm} \times \prod_{t' = 1}^{t} f \big( \V{s}_{t'} \big| \V{s}_{t'-1} \big) f \big( \underline{\V{y}}_{t'}^{(1)} \big| \V{y}_{t'-1} \big) f \big(\V{g}_{t'} \big| \V{s}_{t'} \big) f \big( \VG{\rho}_{t'} \big| \V{s}_{t'} \big) \nonumber \\[0mm]
	&\hspace{5mm} \times \prod_{j = 1}^{J} f \big( \overline{\V{y}}_{t'}^{(j)}, \VG{\alpha}_{t'}^{(j)}, \VG{\beta}_{t'}^{(j)}, \V{z}_{t'}^{(j)}, M_{t'}^{(j)} \big| \underline{\V{y}}_{t'}^{(j)}, \V{s}_{t'} \big) \ist ,
	\label{eq:joint-posterior-pdf}
\end{align}
for some prior
pdf $f ( \V{s}_{0} )$, where we recall that $f (\underline{\V{y}}_{1}^{(1)} | \V{y}_{0}) = 1$ (see assumption (A2)).
Then,
observing that the description of the data association given by $\VG{\alpha}_{t}^{(j)}$ and $\VG{\beta}_{t}^{(j)}$ is redundant once $M_{t}^{(j)}$ is observed --- indeed, $\VG{\alpha}_{t}^{(j)}$ can be derived from $\VG{\beta}_{t}^{(j)}$, and vice versa, when $M_{t}^{(j)}$ is known \cite{MeyKroWilLauHlaBraWin:J18} ---, each factor $ f ( \overline{\V{y}}_{t}^{(j)}, \VG{\alpha}_{t}^{(j)}, \VG{\beta}_{t}^{(j)}, \V{z}_{t}^{(j)}, M_{t}^{(j)} | \underline{\V{y}}_{t}^{(j)}, \V{s}_{t} )$ can be further expressed as
\begin{align}
	f \big( \overline{\V{y}}_{t}^{(j)}, & \, \VG{\alpha}_{t}^{(j)}, \VG{\beta}_{t}^{(j)}, \V{z}_{t}^{(j)}, M_{t}^{(j)} \big| \underline{\V{y}}_{t}^{(j)}, \V{s}_{t} \big) \nonumber \\[0mm] 
	&= f \big( \V{z}_{t}^{(j)} \big| \overline{\V{y}}_{t}^{(j)}, \VG{\alpha}_{t}^{(j)}, \VG{\beta}_{t}^{(j)}, M_{t}^{(j)}, \underline{\V{y}}_{t}^{(j)}, \V{s}_{t} \big) \nonumber \\[0mm]
	&\hspace{10mm} \times f \big( \overline{\V{y}}_{t}^{(j)}, \VG{\alpha}_{t}^{(j)}, \VG{\beta}_{t}^{(j)}, M_{t}^{(j)} \big| \underline{\V{y}}_{t}^{(j)}, \V{s}_{t} \big) \nonumber \\[0mm]
	&= f \big( \V{z}_{t}^{(j)} \big| \overline{\V{y}}_{t}^{(j)}, \VG{\alpha}_{t}^{(j)}, M_{t}^{(j)}, \underline{\V{y}}_{t}^{(j)}, \V{s}_{t} \big) \nonumber \\[0mm] 
	&\hspace{10mm} \times f \big( \overline{\V{y}}_{t}^{(j)}, \VG{\alpha}_{t}^{(j)}, \VG{\beta}_{t}^{(j)}, M_{t}^{(j)} \big| \underline{\V{y}}_{t}^{(j)}, \V{s}_{t} \big) \ist .
	\label{eq:joint-pdf-per-agent-pair}
\end{align}
Hereafter, following the derivations in \cite{MeyKroWilLauHlaBraWin:J18}, we provide expressions for the prior data association pdf $f ( \overline{\V{y}}_{t}^{(j)}, \VG{\alpha}_{t}^{(j)}, \VG{\beta}_{t}^{(j)}, \linebreak M_{t}^{(j)} | \underline{\V{y}}_{t}^{(j)}, \V{s}_{t} )$, and the joint MOT measurements likelihood $f ( \V{z}_{t}^{(j)} | \overline{\V{y}}_{t}^{(j)}, \VG{\alpha}_{t}^{(j)}, M_{t}^{(j)}, \underline{\V{y}}_{t}^{(j)}, \V{s}_{t} )$.

\subsubsection{Prior data association pdf}
\label{subsec:cond_pdf_DA_NewPT_numMeas}
By using assumptions (A4), (A5), (A10),
and the point-target assumption, the pdf $f ( \overline{\V{y}}_{t}^{(j)}, \VG{\alpha}_{t}^{(j)}, \VG{\beta}_{t}^{(j)}, M_{t}^{(j)} | \underline{\V{y}}_{t}^{(j)}, \V{s}_{t} )$ can be expressed as
\begin{align}
	&f \big( \overline{\V{y}}_{t}^{(j)}, \VG{\alpha}_{t}^{(j)}, \VG{\beta}_{t}^{(j)}, M_{t}^{(j)} \big| \underline{\V{y}}_{t}^{(j)}, \V{s}_{t} \big) = C \big( M_{t}^{(j)} \big) \Psi \big( \VG{\alpha}^{(j)}_{t}, \VG{\beta}_t^{(j)} \big) \nonumber \\[1mm]
	&\hspace{10mm} \times \hspace{-1.3mm} \prod_{\ell \in \Set{L}^{(j)}_{t}} \hspace{-1.3mm} q_{1} \big( \underline{\V{y}}^{(j)}_{\ell,t}, \alpha^{(j)}_{\ell,t}, \V{s}_{j_{1},t}, \V{s}_{j_{2},t}; M^{(j)}_{t} \big) \nonumber \\
	&\hspace{10mm} \times \prod_{a \in \Set{A}} h_{1} \big( \V{s}_{a,t}, \alpha^{(j)}_{L + a,t}, \V{s}_{j_{1},t}, \V{s}_{j_{2},t}; M^{(j)}_{t} \big) \nonumber \\
	&\hspace{10mm} \times \hspace{-2.6mm} \prod_{m \in \Set{M}^{(j)}_{t}} \hspace{-2.6mm} \upsilon_{1} \big( \overline{\V{y}}^{(j)}_{m,t}, \beta^{(j)}_{m,t} \big).
	\label{eq:joint-prior-distribution-all}
\end{align}
Here, $C( M_{t}^{(j)} )$ is a normalization factor that depends only on the number of MOT measurements $M_{t}^{(j)}$ (see \cite{MeyKroWilLauHlaBraWin:J18,MeyKroWilLauHlaBraWin:J18-suppl} for details), $\Psi(\VG{\alpha}^{(j)}_{t}, \VG{\beta}_t^{(j)})$ is defined in \eqref{eq:Psi-function}, and the functions $q_{1}(\cdot)$, $h_{1}(\cdot)$, and $\upsilon_{1}(\cdot)$ represent the contributions to the prior data association pdf of the legacy PTs, the agents, and the new PTs, respectively.
The derivation of the pdf in \eqref{eq:joint-prior-distribution-all} closely follows the derivation of the pdf in \cite[Eq. (60)]{MeyKroWilLauHlaBraWin:J18} and is thus omitted. The main difference is given by the product over the agents $a \in \Set{A}$ of the function $h_{1} (\cdot)$, that is a direct consequence of the involvement of the agents in the data association procedure.
Next we provide detailed definitions of the functions $q_{1}(\cdot)$, $h_{1}(\cdot)$, and $\upsilon_{1}(\cdot)$.
The function $q_{1} ( \underline{\V{y}}^{(j)}_{\ell,t}, \alpha^{(j)}_{\ell,t}, \V{s}_{j_{1},t}, \V{s}_{j_{2},t}; M^{(j)}_{t} ) = q_{1} ( \underline{\V{x}}^{(j)}_{\ell,t}, \underline{r}^{(j)}_{\ell,t}, \alpha^{(j)}_{\ell,t}, \linebreak \V{s}_{j_{1},t}, \V{s}_{j_{2},t}; M^{(j)}_{t} )$
is defined for $\underline{r}^{(j)}_{\ell,t} = 1$ as
\begin{align}
	&q_{1} \big( \underline{\V{x}}^{(j)}_{\ell,t}, \underline{r}^{(j)}_{\ell,t} = 1, \alpha^{(j)}_{\ell,t}, \V{s}_{j_{1},t}, \V{s}_{j_{2},t}; M^{(j)}_{t} \big) \nonumber \\ 
	&\hspace{10mm}\deq \begin{cases}
		\dfrac{P_{\text{d}}^{(j)} ( \underline{\V{x}}^{(j)}_{\ell,t}, \V{s}_{j_{1},t}, \V{s}_{j_{2},t} )}{\mu^{(j)}_{\text{c}}}	&	\alpha^{(j)}_{\ell,t} \in \Set{M}^{(j)}_{t} \ist , \\[5mm]
		1 - P_{\text{d}}^{(j)} ( \underline{\V{x}}^{(j)}_{\ell,t}, \V{s}_{j_{1},t}, \V{s}_{j_{2},t} )	&	\alpha^{(j)}_{\ell,t} = 0 \ist , 
	\end{cases} \nonumber \\[-4mm]
	\label{eq:def-q1-r1}
	\intertext{and for $\underline{r}^{(j)}_{\ell,t} = 0$ as}
	&q_{1} \big( \underline{\V{x}}^{(j)}_{\ell,t}, \underline{r}^{(j)}_{\ell,t} = 0, \alpha^{(j)}_{\ell,t}, \V{s}_{j_{1},t}, \V{s}_{j_{2},t}; M^{(j)}_{t} \big) \deq \delta_{\alpha^{(j)}_{\ell,t},0} \ist .
	\label{eq:def-q1-r0}
\end{align}
The function $h_{1} ( \V{s}_{a,t}, \alpha^{(j)}_{L + a,t}, \V{s}_{j_{1},t}, \V{s}_{j_{2},t}; M^{(j)}_{t} )$
is similarly defined as
\begin{align}
	h_{1} \big( \V{s}_{a,t}, & \alpha^{(j)}_{L + a,t}, \V{s}_{j_{1},t}, \V{s}_{j_{2},t}; M^{(j)}_{t} \big) \nonumber \\
	&\deq \begin{cases}
		\dfrac{P_{\text{d}}^{(j)} ( \V{s}_{a,t}, \V{s}_{j_{1},t}, \V{s}_{j_{2},t} )}{\mu^{(j)}_{\text{c}}}	&	\alpha^{(j)}_{L + a,t} \in \Set{M}^{(j)}_{t} \ist , \\[5mm]
		1 - P_{\text{d}}^{(j)} ( \V{s}_{a,t}, \V{s}_{j_{1},t}, \V{s}_{j_{2},t} )	&	\alpha^{(j)}_{L + a,t} = 0 \ist ,
	\end{cases}
	\nonumber \\[-4mm]
	\label{eq:def-h1}
\end{align}
where, with an abuse of notation, the number of legacy PTs at time $t$
at agent pair $j$, i.e., $L_{t}^{(j)}$, is simply referred to as $L$. \linebreak
We observe that if agent $a \in \Set{A}$ is either the Rx-agent $j_{1}$, or the Tx-agent $j_{2}$, that is, $a = j_{1}$ or $a = j_{2}$, according to
\eqref{eq:def-pd2} it follows that $h_{1} ( \V{s}_{a,t}, \alpha^{(j)}_{L + a,t}, \V{s}_{j_{1},t}, \V{s}_{j_{2},t}; M^{(j)}_{t} ) = \delta_{\alpha^{(j)}_{L + a,t},0}$, which intuitively means that no MOT measurements can be associated to the Rx-agent and the Tx-agent.
Finally,
$\upsilon_{1} ( \overline{\V{y}}^{(j)}_{m,t}, \linebreak \beta^{(j)}_{m,t} ) = \upsilon_{1} ( \overline{\V{x}}^{(j)}_{m,t}, \overline{r}^{(j)}_{m,t}, \beta^{(j)}_{m,t} )$
is defined for $\overline{r}^{(j)}_{m,t} = 1$ as
\begin{align}
	&\upsilon_{1} \big( \overline{\V{x}}^{(j)}_{m,t}, \overline{r}^{(j)}_{m,t} = 1, \beta^{(j)}_{m,t} \big) \deq \Gamma \big( \beta_{m,t}^{(j)} \big) \dfrac{\mu^{(j)}_{\text{n}}}{\mu^{(j)}_{\text{c}}} f_{\text{n}} (\overline{\V{x}}^{(j)}_{m,t}) \nonumber \\[1mm] 
	&\hspace{8mm} = \begin{cases}
		0	&		\beta_{m,t}^{(j)} \in \Set{O}_{t}^{(j)} \ist , \\[3mm]
		\dfrac{\mu^{(j)}_{\text{n}}}{\mu^{(j)}_{\text{c}}} f_{\text{n}} (\overline{\V{x}}^{(j)}_{m,t})	&		\beta_{m,t}^{(j)} = 0  \ist ,
	\end{cases}
	\intertext{and for $\overline{r}^{(j)}_{m,t}  = 0$ as}
	&\upsilon^{(j)}_{1} \big( \overline{\V{x}}^{(j)}_{m,t}, \overline{r}^{(j)}_{m,t} = 0, \beta^{(j)}_{m,t} \big) \deq f_{\text{D}} (\overline{\V{x}}^{(j)}_{m,t}) \ist .
\end{align}
Note that the function $\upsilon_{1}(\cdot)$ incorporates the indicator function $\Gamma(\cdot)$ defined in \eqref{eq:gamma-func-data-ass}; and that the combined use in \eqref{eq:joint-prior-distribution-all} of the functions $\Psi(\VG{\alpha}_{t}^{(j)}\rmv,\VG{\beta}_{t}^{(j)})$ and $\upsilon_{1}(\cdot)$
describes the point-target assumption as done by the indicator function $\Phi(\cdot)$ defined in \eqref{eq:Phi-function}. \linebreak

\subsubsection{Joint MOT measurements likelihood}
\label{subsubsec:joint-mot-likelihood}

\begin{figure*}[!b]

\hrulefill

\vspace{-2mm}

\setcounter{MYtempeqncnt}{\value{equation}}
\setcounter{equation}{28}	
\begin{align}
	&f \big( \V{y}_{1:t}, \V{s}_{0:t}, \VG{\alpha}_{1:t}, \VG{\beta}_{1:t} \big| \V{g}_{1:t}, \VG{\rho}_{1:t}, \V{z}_{1:t} \big) \nonumber \\[0mm]
	&\hspace{5mm} \propto
		f \big( \V{s}_{0} \big) \prod_{t' = 1}^{t} \underbrace{\Bigg( \prod_{a \in \Set{A}} \tau_{a} \big( \V{s}_{a,t'} \big| {\V{s}}_{a,t'-1} \big) \Bigg)}_{\textsc{Agents' States Prediction}} \underbrace{\Bigg( \prod_{a \in \Set{A}^{\V{g}}_{t'}} \mathfrak g_{a} \big( \V{g}_{a,t'} \big| \V{s}_{a,t'} \big) \Bigg) \Bigg( \prod_{a \in \Set{R}} \prod_{a' \in \Set{T}^{(a)}_{t'}} \mathfrak{d} \big( \VG{\rho}^{(a,a')}_{t'} \big| \V{s}_{a,t'}, \V{s}_{a'\rmv,t'} \big) \Bigg)}_{\textsc{Agents' Cooperative Self-Localization}} \nonumber \\[0mm]
	&\hspace{10mm} \times \underbrace{\Bigg( \prod_{\ell \in \Set{L}_{t'}^{(1)}} \!\!\! f \big(\underline{\V{y}}_{\ell,t'}^{(1)} \big| {\V{y}}_{\ell,t'-1} \big) \Bigg)}_{\textsc{PTs' Augmented States Prediction}} \prod_{j = 1}^{J} \underbracea{\Bigg( \prod_{\ell \in \Set{L}_{t'}^{(j)}} q \big( \underline{\V{y}}_{\ell,t'}^{(j)}, \alpha_{\ell,t'}^{(j)}, \V{s}_{j_{1},t'}, \V{s}_{j_{2},t'}; \V{z}_{t'}^{(j)} \big) \!\! \prod_{m \in \Set{M}_{t'}^{(j)}} \!\! \psi\big( \alpha_{\ell,t'}^{(j)}, \beta_{m,t'}^{(j)} \big) \Bigg)} \nonumber \\[0mm]
	&\hspace{10mm} \times \underbracebd{\Bigg( \prod_{a \in \Set{A}} h \big( \V{s}_{a,t'}, \alpha_{L + a,t'}^{(j)}, \V{s}_{j_{1},t'}, \V{s}_{j_{2},t'}; \V{z}_{t'}^{(j)} \big) \!\! \prod_{m \in \Set{M}_{t'}^{(j)}} \!\! \psi\big( \alpha_{L + a,t'}^{(j)}, \beta_{m,t'}^{(j)} \big) \Bigg) \! \prod_{m \in \Set{M}_{t'}^{(j)}} \!\! \upsilon \big( \overline{\V{y}}_{m,t'}^{(j)}, \beta_{m,t'}^{(j)}, \V{s}_{j_{1},t'}, \V{s}_{j_{2},t'}; \V{z}_{m,t'}^{(j)} \big)}_{\textsc{MOT Measurements Evaluation and Data Association}} \ist. \nonumber \\[-3mm]
	\label{eq:final-factorization}
\end{align}

\setcounter{equation}{\value{MYtempeqncnt}}

\end{figure*}

By using assumptions
(A10),
(A11), and the point-target assumption, the joint MOT measurements likelihood $f (\V{z}_{t}^{(j)} | \linebreak \overline{\V{y}}_{t}^{(j)}, \VG{\alpha}_{t}^{(j)}, M_{t}^{(j)}, \underline{\V{y}}_{t}^{(j)}, \V{s}_{t})$ can be expressed as
\begin{align}
	& f \big(\V{z}_{t}^{(j)} \big| \overline{\V{y}}_{t}^{(j)}, \VG{\alpha}_{t}^{(j)}, M_{t}^{(j)}, \underline{\V{y}}_{t}^{(j)}, \V{s}_{t} \big) = C \big( \V{z}_{t}^{(j)} \big) \nonumber \\[1mm]
	&\hspace{10mm} \times \hspace{-1.3mm} \prod_{\ell \in \Set{L}^{(j)}_{t}} \hspace{-1.3mm} q_{2} \big( \underline{\V{y}}^{(j)}_{\ell,t}, \alpha^{(j)}_{\ell,t}, \V{s}_{j_{1},t}, \V{s}_{j_{2},t}; \V{z}^{(j)}_{t} \big) \nonumber \\[0mm]
	&\hspace{10mm} \times \prod_{a \in \Set{A}} h_{2} \big( \V{s}_{a,t}, \alpha^{(j)}_{L + a,t}, \V{s}_{j_{1},t}, \V{s}_{j_{2},t}; \V{z}^{(j)}_{t} \big) \nonumber \\[0mm]
	&\hspace{10mm} \times \hspace{-2.6mm} \prod_{m \in \Set{M}^{(j)}_{t}} \hspace{-2.6mm} \upsilon_{2} \big( \overline{\V{y}}^{(j)}_{m,t}, \V{s}_{j_{1},t}, \V{s}_{j_{2},t}; \V{z}^{(j)}_{m,t} \big) \ist .
	\label{eq:joint-likelihood-all}
\end{align}
Here, $C ( \V{z}_{t}^{(j)} )$ is a normalization factor that depends only on the MOT measurements vector $\V{z}_{t}^{(j)}$ (see \cite{MeyKroWilLauHlaBraWin:J18,MeyKroWilLauHlaBraWin:J18-suppl} for details), and the functions $q_{2}(\cdot)$, $h_{2}(\cdot)$, and $\upsilon_{2}(\cdot)$ embed the MOT measurement likelihoods related to the legacy PTs, the agents, and the new PTs, respectively.
The derivation of the likelihood in \eqref{eq:joint-likelihood-all} closely follows the derivation of the likelihood in \cite[Eq. (64)]{MeyKroWilLauHlaBraWin:J18} and is thus omitted. The main difference is given by the product over the agents $a \in \Set{A}$ of the function $h_{2} (\cdot)$, that represents the likelihoods of the MOT measurements when these are generated by the agents.
Next we provide detailed definitions of the functions $q_{2}(\cdot)$, $h_{2}(\cdot)$, and $\upsilon_{2}(\cdot)$.
The function $q_{2} ( \underline{\V{y}}^{(j)}_{\ell,t}, \alpha^{(j)}_{\ell,t}, \V{s}_{j_{1},t}, \V{s}_{j_{2},t}; \V{z}^{(j)}_{t} ) = q_{2} ( \underline{\V{x}}^{(j)}_{\ell,t}, \underline{r}^{(j)}_{\ell,t}, \alpha^{(j)}_{\ell,t}, \V{s}_{j_{1},t}, \linebreak \V{s}_{j_{2},t}; \V{z}^{(j)}_{t} )$
is defined for $\underline{r}^{(j)}_{\ell,t} = 1$ as
\begin{align}
	&q_{2} \big( \underline{\V{x}}^{(j)}_{\ell,t}, \underline{r}^{(j)}_{\ell,t} = 1, \alpha^{(j)}_{\ell,t}, \V{s}_{j_{1},t}, \V{s}_{j_{2},t}; \V{z}^{(j)}_{t} \big) \nonumber \\[0mm]
	&\hspace{10mm}\deq
	\begin{cases}
		\dfrac{
				\mathfrak{f}(\underline{\V{x}}_{\ell,t}^{(j)},  \V{s}_{j_{1}\rmv,t}, \V{s}_{j_{2}\rmv,t} ; \V{z}_{m,t}^{(j)})}{f^{(j)}_{\text{c}} \big( \V{z}^{(j)}_{m,t} \big)}	&	\alpha^{(j)}_{\ell,t} \in \Set{M}^{(j)}_{t} \ist , \\[0mm]
		1	&	\alpha^{(j)}_{\ell,t} = 0 \ist ,
	\end{cases} \nonumber \\[-4mm]
	\label{eq:def-q2-r1}
	\intertext{and for $\underline{r}^{(j)}_{\ell,t} = 0$ as}
	&q_{2} \big( \underline{\V{x}}^{(j)}_{\ell,t}, \underline{r}^{(j)}_{\ell,t} = 0, \alpha^{(j)}_{\ell,t}, \V{s}_{j_{1},t}, \V{s}_{j_{2},t}; \V{z}^{(j)}_{t} \big) \deq 1 \,.
	\label{eq:def-q2-r0}
\end{align}
The function $h_{2} ( \V{s}_{a,t}, \alpha^{(j)}_{L + a,t}, \V{s}_{j_{1},t}, \V{s}_{j_{2},t}; \V{z}^{(j)}_{t} )$ is similarly defined for $a \neq j_{1}$ and $a \neq j_{2}$ as
\begin{align}
	&h_{2} \big( \V{s}_{a,t}, \alpha^{(j)}_{L + a,t}, \V{s}_{j_{1},t}, \V{s}_{j_{2},t}; \V{z}^{(j)}_{t} \big) \nonumber \\[0mm]
	&\hspace{10mm}\deq
	\begin{cases}
		\dfrac{		\mathfrak{f}(\V{s}_{a,t},  \V{s}_{j_{1}\rmv,t}, \V{s}_{j_{2}\rmv,t} ; \V{z}_{m,t}^{(j)})}{f^{(j)}_{\text{c}} \big( \V{z}^{(j)}_{m,t} \big)}	&	\alpha^{(j)}_{L + a,t} \in \Set{M}^{(j)}_{t} \ist,	\\[0mm]
		1	&	\alpha^{(j)}_{L + a,t} = 0 \ist ,
	\end{cases} 	\nonumber \\[-4mm]
	\label{eq:def-h2-1}
	\intertext{and for $a = j_{1}$ or $a = j_{2}$ as}
	&h_{2} \big( \V{s}_{a,t}, \alpha^{(j)}_{L + a,t}, \V{s}_{j_{1},t}, \V{s}_{j_{2},t}; \V{z}^{(j)}_{t} \big) \deq 1 \ist .
	\label{eq:def-h2-2}
\end{align}
Finally, $\!\upsilon_{2} ( \overline{\V{y}}^{(j)}_{m,t}, \V{s}_{j_{1},t}, \V{s}_{j_{2},t}; \V{z}^{(j)}_{m,t} ) \! = \! \upsilon_{2} ( \overline{\V{x}}^{(j)}_{m,t}, \overline{r}_{m,t}^{(j)}, \V{s}_{j_{1},t}, \V{s}_{j_{2},t}; \linebreak \V{z}^{(j)}_{m,t} )$
is defined as
\begin{align}
	&\upsilon_{2} \big( \overline{\V{x}}^{(j)}_{m,t}, \overline{r}^{(j)}_{m,t}, \V{s}_{j_{1},t}, \V{s}_{j_{2},t}; \V{z}^{(j)}_{m,t} \big) \nonumber \\[0mm]
	&\hspace{10mm}\deq
	\begin{cases}
		\dfrac{		\mathfrak{f} \big(\overline{\V{x}}_{m,t}^{(j)},  \V{s}_{j_{1}\rmv,t}, \V{s}_{j_{2}\rmv,t} ; \V{z}_{m,t}^{(j)} \big)}{f^{(j)}_{\text{c}} \big( \V{z}^{(j)}_{m,t} \big)} &	\overline{r}^{(j)}_{m,t} = 1	\,, \\
1	&	\overline{r}^{(j)}_{m,t} = 0 \,.
	\end{cases}
	\label{eq:def-upsilon-2}
\end{align}

The final factorization of the joint posterior pdf $f ( \V{y}_{1:t}, \V{s}_{0:t}, \linebreak \VG{\alpha}_{1:t}, \VG{\beta}_{1:t} | \V{g}_{1:t}, \VG{\rho}_{1:t}, \V{z}_{1:t} )$ is obtained by inserting \eqref{eq:joint-prior-distribution-all} and \eqref{eq:joint-likelihood-all} into \eqref{eq:joint-pdf-per-agent-pair},
and \eqref{eq:agents-states-transition}--\eqref{eq:legacy-PT-transition}, \eqref{eq:navigation-meas-joint-likelihood}--\eqref{eq:direct-meas-joint-likelihood},
\eqref{eq:joint-pdf-per-agent-pair}, into \eqref{eq:joint-posterior-pdf}.
\addtocounter{equation}{1}
Its expression is reported in \eqref{eq:final-factorization}, where the equality $\Set{L}_{t}^{(1)} = \Set{K}_{t-1}$ has been used (see assumption (A2)), and where $q(\cdot) \deq q_{1}(\cdot) \ist q_{2}(\cdot)$, $h(\cdot) \deq h_{1}(\cdot) \ist h_{2}(\cdot)$, and $\upsilon(\cdot) \deq  \upsilon_{1}(\cdot) \ist \upsilon_{2}(\cdot)$.
For the reader's convenience, we summarize in Table~\ref{tab:variables} all the variables involved in this factorization.
\begin{table}[!t]
\centering
	\renewcommand{\arraystretch}{1.3}
	\small	
	\caption{Summary of the variables used in \eqref{eq:final-factorization}}
	\label{tab:variables}
	\centering
	\begin{tabular}{ l | c  }
		\textsc{Description}	&	\textsc{Symbol}		\\[.5mm]
		\hline \hline
		Agent state & $\V{s}_{a,t}$ \\
		Legacy PT augmented state & $\underline{\V{y}}_{\ell,t}^{(j)}$ \\
		New PT augmented state & $\overline{\V{y}}^{(j)}_{m,t} $  \\
		Navigation data & $\V{g}_{a,t}$ \\
		Inter-agent measurement & $\VG{\rho}^{(a,a')}_{t}$ \\
		MOT measurement &$\V{z}^{(j)}_{m,t}$  \\
		Legacy object-oriented DA variable & $\alpha^{(j)}_{\ell,t}$\\
		MOT measurement-oriented DA variable & $\beta^{(j)}_{m,t}$	\\	\hline
	\end{tabular}
\vspace{-2mm}
\end{table}

\section{The Proposed Algorithm}
\label{sec:the_prop_algorithm}

\begin{figure*}
	\centering
	\includegraphics{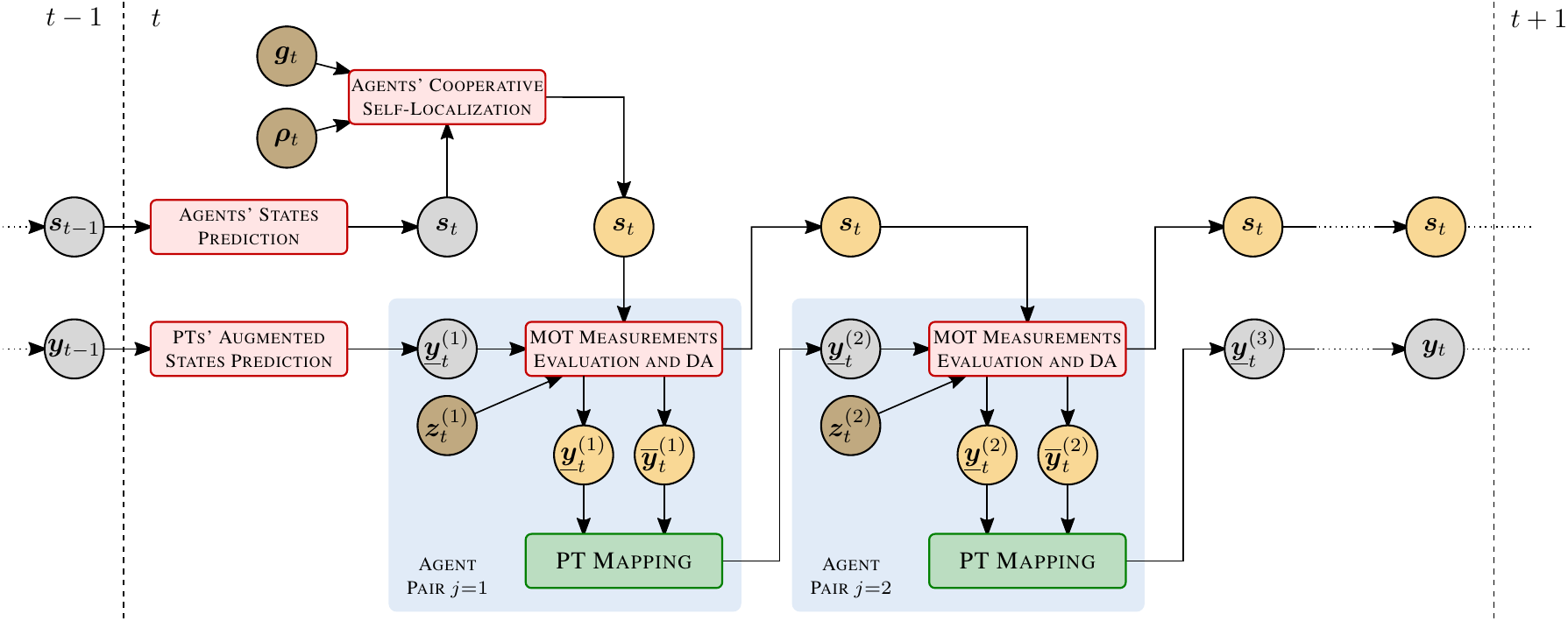}
	\vspace{-4mm}
	\caption{Block diagram providing the sequence of operations performed by the proposed method at time $t$. The red boxes represent the operations performed by the SPA-based algorithm as stated by the factorization of the joint posterior pdf in \eqref{eq:final-factorization}; the green boxes represent the PT mapping operations carried out between two consecutive agent pairs. The
		nodes represent random vectors: in particular, brown nodes are observations (i.e., navigation data, inter-agent and MOT measurements), and
		yellow nodes indicate random vectors whose \textit{beliefs} are updated through observations, following either the ``agents' cooperative self-localization'' or the ``MOT measurements evaluation and DA'' operations.
		The arrows link the random vectors to the operations that involve them.
		}
	\label{fig:figblockscheme}
	\vspace{-2mm}
\end{figure*}

Direct marginalization of the joint posterior pdf $f (\V{y}_{1:t}, \V{s}_{0:t}, \linebreak \VG{\alpha}_{1:t}, \VG{\beta}_{1:t} | \V{g}_{1:t}, \VG{\rho}_{1:t}, \V{z}_{1:t})$ for the computation of the marginal posterior pdfs $f(\V{y}_{k,t} | \V{g}_{1:t}, \VG{\rho}_{1:t}, \V{z}_{1:t} )$ and $f(\V{s}_{a,t} | \V{g}_{1:t}, \VG{\rho}_{1:t}, \V{z}_{1:t} )$ is generally infeasible, as it requires high-dimensional integration and summation.
Approximations of these marginal posterior pdfs, called \textit{beliefs}, can be efficiently obtained by applying the SPA on a factor graph \cite{KscFreLoe:01, LoeDauHuKorPinKsc:J07} carefully devised from the factorization in \eqref{eq:final-factorization}.

\subsection{Sum-product algorithm: overview  and notation}
\label{sec:review_factorGraphs}
Here, we briefly review factor graphs and the SPA.
Let us consider the generic problem of estimating $D$ parameter vectors $\VG{\lambda}_d$, $d \rmv\in \{1, \ldots, D\}$,
from a vector of observations $\VG{\pi}$. In the Bayesian setting, 
these vectors are random, and the estimation of $\VG{\lambda}_d$ is based on the posterior pdf $f(\VG{\lambda}_d|\VG{\pi})$.
This
pdf is a marginal pdf of the joint posterior pdf $f(\VG{\lambda} | \VG{\pi})$, where $\VG{\lambda} \deq \big[ \VG{\lambda}_1^{\T}, \ldots, \VG{\lambda}^{\T}_{D} \big]^{\T}\rmv\rmv$. 
The joint posterior pdf is assumed to be 
the product of certain lower-dimensional factors, i.e.,
\begin{align}
	f(\VG{\lambda} | \VG{\pi}) \propto \prod_{\ell} \kappa_{\ell} \big( \VG{\lambda}^{(\ell)}; \VG{\pi} \big) \ist ,
	\label{eq:general_factorization} 
\end{align}
where each argument $\VG{\lambda}^{(\ell)}$ comprises certain parameter vectors $\VG{\lambda}_d$,
and each $\VG{\lambda}_d$ can appear in several $\VG{\lambda}^{(\ell)}$. 
The factorization \eqref{eq:general_factorization} can be represented by a factor graph, which is constructed as follows: 
each parameter variable $\VG{\lambda}_d$ is represented by a variable node; 
each factor $\kappa_{\ell}(\cdot)$ is represented by a factor node; and variable node ``$\VG{\lambda}_d$'' and factor node ``$\kappa_{\ell}$'' are adjacent, i.e., connected by an edge, 
if $\VG{\lambda}_d$ is an argument of $\kappa_{\ell}(\cdot)$.

The SPA algorithm aims at computing the marginal posterior pdfs $f(\VG{\lambda}_d | \VG{\pi})$ in an efficient way, and is
based on the factor graph representing the factorization of $f(\VG{\lambda} | \VG{\pi})$ in \eqref{eq:general_factorization}.
For each node in the factor graph, certain messages are calculated, each of which is then passed to one of the adjacent nodes.
Let $\Set{V}_{\ell}$ denote the set of indices $d$ of all those variable nodes ``$\VG{\lambda}_d$'' that are adjacent to factor node ``$\kappa_{\ell}$''.
Then, factor node ``$\kappa_{\ell}$'' passes the following message to variable node ``$\VG{\lambda}_d$'' with $d \! \in \! \Set{V}_{\ell}$:
\begin{align}
	\hspace{-2mm}\zeta_{\kappa_{\ell} \to \VG{\lambda}_d}(\VG{\lambda}_d) = \!\! \int \!\! \kappa_{\ell} \big( \VG{\lambda}^{(\ell)}; \VG{\pi} \big) \hspace{-2mm} \prod_{\substack{d' \in \Set{V}_{\ell} \\ d' \neq d}} \hspace{-2mm} \eta_{\ist \VG{\lambda}_{d'} \to \kappa_{\ell}}(\VG{\lambda}_{d'}) \, 
\mathrm{d}\VG{\lambda}_{-d} \, .
	\label{eq:incomingMessage}
\end{align}
Here, $\int \ldots\, \mathrm{d}\VG{\lambda}_{-d}$ denotes integration with respect to all $\VG{\lambda}_{d'}$, $d' \! \in \! \Set{V}_{\ell}$, except $\VG{\lambda}_{d}$, and the messages $\eta_{\ist \VG{\lambda}_{d} \to \kappa_{\ell}}(\VG{\lambda}_{d})$ are calculated as described later. If the factorization \eqref{eq:general_factorization} involves (also) discrete variables, 
then the respective integrations in \eqref{eq:incomingMessage} have to be replaced with summations.
Furthermore, let ${\Set{F}}_{d}$ be the set of the indices $\ell$ of all those factors 
nodes ``$\kappa_{\ell}$'' that are adjacent to variable node ``$\VG{\lambda}_d$''.
Then, variable node ``$\VG{\lambda}_d$'' passes the following message to factor node ``$\kappa_{\ell}$'' with 
$\ell \! \in \! {\Set{F}}_{d}$:
\begin{align}
	\eta_{\ist \VG{\lambda}_d \to \kappa_{\ell}}(\VG{\lambda}_d) = \! \prod_{ \substack{\ell' \in {\Set{F}}_d \\ \ell' \neq \ell} } \!\! \zeta_{\kappa_{\ell'} \to \VG{\lambda}_d}(\VG{\lambda}_{d}) \ist.
	\label{eq:outgoingMessage}
\end{align}

For a factor graph with loops, the calculation of the messages is usually repeated in an iterative manner. 
There is no unique order --- or \textit{schedule} --- of message calculation, and different orders may lead to
different results. Finally, for each variable node ``$\VG{\lambda}_d$'', a belief $\tilde{f}(\VG{\lambda}_d)$ is calculated 
by multiplying all the incoming messages (passed from all the adjacent factor nodes) and normalizing the resulting product function such that 
$\int \rmv\tilde{f}(\VG{\lambda}_d) \ist \mathrm{d}\VG{\lambda}_d = 1$. The belief $\tilde{f}(\VG{\lambda}_d)$ provides the desired approximation of the marginal posterior pdf 
$f(\VG{\lambda}_d |\VG{\pi})$.

\subsection{SPA-based joint localization and tracking}
\label{subsec:SPA_Message_Passing_alg}

The factor graph derived from the factorization \eqref{eq:final-factorization} contains loops; therefore,
a message calculation schedule needs to be selected.
The proposed algorithm is based of the following rules:
\textit{(i)} messages are not sent backward in time;
\textit{(ii)} MOT measurements produced by the agent pairs are processed sequentially according to the arbitrary order established by the indexing function $\phi(\cdot)$, from agent pair $j = 1$
through agent pair $j = J$; and
\textit{(iii)} iterative message passing is only performed for agents' cooperative self-localization, and MOT measurements data association within each agent pair $j$.
Note that, similarly to the order of the SPA messages, different orders of the agent pairs may lead to different outcomes of the SPA-based algorithm. Within the proposed framework, the order selection is further complicated by the bistatic geometry of some agent pairs; indeed, a bistatic agent pair might have a favorable geometry to observe a particular target, but not necessarily \textit{all} the targets.
The selection of the optimal agent pair order is still an open problem and needs to be tailored according to the specific application and, in particular, to the system's architecture and specifications. These aspects are not addressed in this paper.
Finally, we observe that a parallel implementation, suitable for a distributed approach, is however possible as similarly done in \cite[Sec. IX-B]{MeyKroWilLauHlaBraWin:J18}.

Fig.~\ref{fig:figblockscheme} shows a block
diagram that provides an intuitive representation of the proposed
method.
First, the predictions of the joint agent state vector $\V{s}_{t-1}$
and the joint PT augmented state vector $\V{y}_{t-1}$
from $t-1$ to $t$ are performed; these operations are described by the red boxes ``agents' states prediction'' and ``PTs' augmented states prediction'', respectively.
The predicted agent states are then updated by using the navigation data $\V{g}_{t}$ and the inter-agent measurements $\VG{\rho}_{t}$
through the ``agents' cooperative self-localization''.
Next, the MOT measurements $\V{z}_{t}^{(j)}$ produced
at agent pair $j$ are processed through the ``MOT measurements evaluation and DA'' box using the updated agent states $\V{s}_{t}$ and the legacy PT augmented states $\underline{\V{y}}_{t}^{(j)}$;
this is performed sequentially, from agent pair $j = 1$ to agent pair $j = J$.
Once the MOT measurements at agent pair $j$ are processed, the beliefs of the PT states (both legacy and new) and the agent states are updated before the MOT measurements of the next agent pair $j+1$ are processed.
All these operations, represented by
the red boxes in Fig.~\ref{fig:figblockscheme}, correspond to macro-factors in the factorization of the joint posterior pdf in \eqref{eq:final-factorization}.
The
green boxes refer to the PT mapping operation (cf. Section~\ref{subsec:target_state_space_model}) that is carried out between any two consecutive agent pairs, i.e., $j-1$ and $j$, at current time $t$.

Combining the rules for the message
schedule stated above, and the generic SPA rules for calculating messages and beliefs described in Section~\ref{sec:review_factorGraphs}, we provide the expressions of the SPA messages for each of these operations in what follows; for clarity, the titles of the next four subsections recall the operations described by the red boxes in Fig.~\ref{fig:figblockscheme}.
The SPA messages are exchanged on the factor graphs in Fig.~\ref{fig:graph-self-loc}, Fig.~\ref{fig:graph-pt-pred}, and Fig.~\ref{fig:graph-tracking}; the first one relates to the agents' states prediction and cooperative self-localization, the second one to the PTs' augmented states prediction, and the last one to the MOT measurements evaluation and data association.
We observe that the agent state variable nodes at time $t$ in Fig.~\ref{fig:graph-self-loc}, i.e., variable nodes ``$\V{s}_{a}$'', coincide with the agent state variable nodes in Fig.~\ref{fig:graph-tracking}; analogously, the legacy PT augmented state variable nodes at time $t$ in Fig.~\ref{fig:graph-pt-pred}, i.e., variable nodes ``$\underline{\V{y}}_{\ell}$'', coincide with the legacy PT augmented state variable nodes in Fig.~\ref{fig:graph-tracking} when $j = 1$.
The structure of the factor graph in Fig.~\ref{fig:graph-self-loc} changes according to the availability, at time $t$, of the navigation data and the inter-agent measurements; a general case is there illustrated, in which all the agents have navigation data, and inter-agent measurements between any two agents are available.
Similarly, the structure of the factor graph in Fig.~\ref{fig:graph-tracking} changes, at each time $t$ and agent pair $j$, according to the number of legacy PTs and number of MOT measurements, as well as to the kind of MOT measurements, that is, monostatic, if $j_{1} = j_{2}$, or bistatic, if $j_{1} \neq j_{2}$; the latter is there illustrated, with two separate variable nodes for the Rx-agent, i.e., ``$\V{s}_{j_{1}}$'', and the Tx-agent, i.e., ``$\V{s}_{j_{2}}$''.

\subsubsection{Agents' states prediction}
\label{subsubsec:agent_states_pred}

\begin{figure}[!t]
	\centering
	\includegraphics{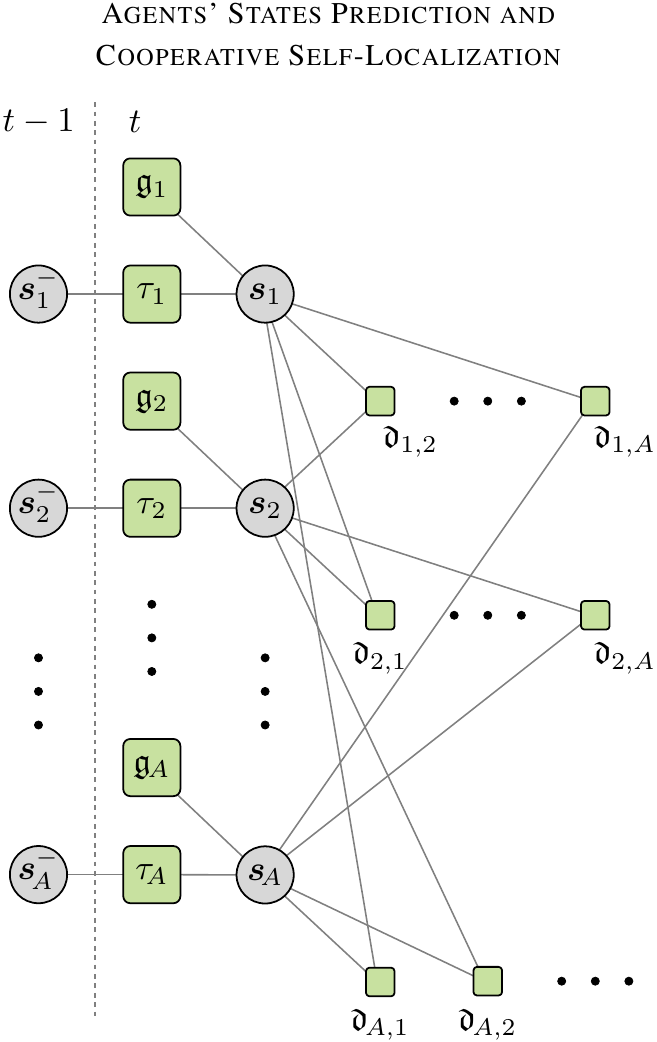}
	\caption{Factor graph representing the agents' states prediction and agents' cooperative self-localization portions of the factorization in \eqref{eq:final-factorization} for one time step $t$.
	Grey circles are variable nodes, and green squares are factor nodes.
	The following short notations are used:
	$\V{s}_{a}^{-} \deq \V{s}_{a,t-1}$; 
	$\V{s}_{a} \deq \V{s}_{a,t}$; 
	$\mathfrak{g}_{a} \deq \mathfrak{g}_{a}(\V{g}_{a,t} | \linebreak \V{s}_{a,t})$;
	$\tau_{a} \deq \tau_{a}(\V{s}_{a,t} | \V{s}_{a,t-1})$; 
	and $\mathfrak{d}_{a,a'} \deq \mathfrak{d}(\VG{\rho}_{t}^{(a,a')} | \V{s}_{a,t}, \V{s}_{a'\rmv,t})$.}
	\label{fig:graph-self-loc}
	\vspace{-2mm}
\end{figure}

The prediction of the state of agent $a \in \Set{A}$ is performed by computing the message $\zeta_{ \tau_{a} \to \V{s}_{a} } (\V{s}_{a,t}) $ from factor node ``$\tau_{a}$'' to variable node ``$\V{s}_{a}$'' in Fig.~\ref{fig:graph-self-loc}.
The expression of this message is as follows
\begin{align}
	\zeta_{ \tau_{a} \to \V{s}_{a} } (\V{s}_{a,t}) = \int \tau_{a}(\V{s}_{a,t} | \V{s}_{a,t-1}) \tilde{f}_{J} (\V{s}_{a,t-1})  \, \text{d} \V{s}_{a,t-1} \ist ,
\end{align}
where $\tilde{f}_{J}(\V{s}_{a,t-1})$ is the belief of the agent state at previous time $t-1$, computed at the last agent pair $J$; its expression is provided later in Section~\ref{subsubsec:meas_update}.

\subsubsection{Agents' cooperative self-localization}
\label{subsubsec:agents_self_localization}
The variable nodes ``$\V{s}_{a}$'', and
factor nodes ``$\mathfrak{d}_{a,a'}$'' and ``$\mathfrak{d}_{a'\rmv,a}$'' in Fig.~\ref{fig:graph-self-loc} define a loopy graph.
Therefore, the messages related to the agents' cooperative self-localization are iteratively computed as follows.
At each iteration $n = 1, \ldots, N_{\text{SL}}$ of the agents' cooperative self-localization loop, 
the messages $\eta^{(n)}_{\V{s}_{a} \to \mathfrak{d}_{a,a'} }(\V{s}_{a,t})$
and $\eta^{(n)}_{\V{s}_{a} \to \mathfrak{d}_{a'\rmv,a}}(\V{s}_{a,t})$
are calculated as 
\begin{align}
	\eta^{(n)}_{\V{s}_{a} \to \mathfrak{d}_{a,a'}} \big( \V{s}_{a,t} \big) &= \zeta_{\tau_{a} \to \V{s}_{a} } \big( \V{s}_{a,t} \big) \ist \zeta_{ \mathfrak{g}_{a} \to \V{s}_{a}} \big( \V{s}_{a,t} \big) \nonumber \\[0mm]
	&\hspace{5mm} \times \Bigg( \prod_{ \substack{a'' \in \Set{T}^{(a)}_{t} \\ a'' \neq a'} } \zeta^{(n-1)}_{ \mathfrak{d}_{a,a''} \to \V{s}_{a} } \big( \V{s}_{a,t} \big) \Bigg) \nonumber \\[0mm]
	&\hspace{5mm} \times \Bigg( \prod_{a'' \in \Set{R}^{(a)}_{t} } \zeta^{(n-1)}_{ \mathfrak{d}_{a'',a} \to \V{s}_{a} } \big( \V{s}_{a,t} \big) \Bigg) \ist , \label{eg:self-loc-messag-from-s-1}
	\intertext{and}
	\eta^{(n)}_{\V{s}_{a} \to \mathfrak{d}_{a'\rmv,a} } \big( \V{s}_{a,t} \big) &= \zeta_{\tau_{a} \to \V{s}_{a} } \big( \V{s}_{a,t} \big) \ist \zeta_{ \mathfrak{g}_{a} \to \V{s}_{a}} \big( \V{s}_{a,t} \big) \nonumber \\[0mm]
	&\hspace{5mm} \times \Bigg( \prod_{a'' \in \Set{T}^{(a)}_{t} } \zeta^{(n-1)}_{ \mathfrak{d}_{a,a''} \to \V{s}_{a} } \big( \V{s}_{a,t} \big) \Bigg) \nonumber \\[0mm]
	&\hspace{5mm} \times \Bigg( \prod_{ \substack{a'' \in \Set{R}^{(a)}_{t} \\ a'' \neq a'} }  \zeta^{(n-1)}_{ \mathfrak{d}_{a'',a} \to \V{s}_{a} }\big( \V{s}_{a,t} \big) \Bigg) \ist . \label{eg:self-loc-messag-from-s-2}
\end{align}
We observe that the factor node $\mathfrak{d}_{a,a'}$ refers to the likelihood of the inter-agent measurement $\VG{\rho}^{(a,a')}_{t}$
produced by Rx-agent $a$
using the signal transmitted by Tx-agent $a'$; vice versa, the factor node $\mathfrak{d}_{a'\rmv,a}$ refers to the likelihood of the inter-agent measurement $\VG{\rho}^{(a'\rmv,a)}_{t}$
produced by Rx-agent $a'$
using the signal transmitted by Tx-agent $a$.
Therefore, the agent state $\V{s}_{a,t}$ acts as Rx-agent for the message $\eta^{(n)}_{\V{s}_{a} \to \mathfrak{d}_{a,a'}}$, and as Tx-agent for the message $\eta^{(n)}_{\V{s}_{a} \to \mathfrak{d}_{a'\rmv,a}}$.
In both \eqref{eg:self-loc-messag-from-s-1} and \eqref{eg:self-loc-messag-from-s-2}, the message $\zeta_{ \mathfrak{g}_{a} \to \V{s}_{a}} ( \V{s}_{a,t} )$, that is passed from factor node ``$\mathfrak{g}_{a}$'' to variable node ``$\V{s}_{a}$'', is computed as
\begin{align}
	\zeta_{ \mathfrak{g}_{a} \to \V{s}_{a}} \big( \V{s}_{a,t} \big) =
	\begin{dcases}
		\mathfrak{g}_{a} \big( \V{g}_{a,t}  \big| \V{s}_{a,t} \big)	&	a \in \Set{A}^{\V{g}}_{t} \ist , \\[0mm]
		1	&	a \notin \Set{A}^{\V{g}}_{t} \ist ,
	\end{dcases}
\end{align}
and messages $\zeta^{(n)}_{ \mathfrak{d}_{a,a'} \to \V{s}_{a} }(\V{s}_{a,t})$ and $\zeta^{(n)}_{ \mathfrak{d}_{a'\rmv,a} \to \V{s}_{a}}(\V{s}_{a,t})$
are calculated as 
\begin{align}
	&\hspace{-5mm}\zeta^{(n)}_{ \mathfrak{d}_{a,a'} \to \V{s}_{a} } \big( \V{s}_{a,t} \big) \nonumber \\[0mm]
	&\hspace{0mm} = \int \mathfrak{d} \big( \VG{\rho}^{(a,a')}_{t} \big| \V{s}_{a,t}, \V{s}_{a'\rmv,t} \big) \eta^{(n)}_{\V{s}_{a'} \to \mathfrak{d}_{a,a'} } \big( \V{s}_{a'\rmv,t} \big) \,\text{d} \V{s}_{a'\rmv,t} \ist , \label{eg:self-loc-messag-to-s-1}
	\intertext{and}
	&\hspace{-5mm}\zeta^{(n)}_{ \mathfrak{d}_{a'\rmv,a} \to \V{s}_{a} } \big( \V{s}_{a,t} \big) \nonumber \\[0mm]
	&\hspace{0mm} = \int \mathfrak{d} \big( \VG{\rho}^{(a'\rmv,a)}_{t} \big| \V{s}_{a,t}, \V{s}_{a'\rmv,t} \big) \eta^{(n)}_{\V{s}_{a'} \to \mathfrak{d}_{a'\rmv,a} } \big( \V{s}_{a'\rmv,t} \big) \,\text{d} \V{s}_{a'\rmv,t} \ist . \label{eg:self-loc-messag-to-s-2}
\end{align}
The iteration constituted by
\eqref{eg:self-loc-messag-from-s-1}--\eqref{eg:self-loc-messag-to-s-2} is initialized by $\zeta^{(0)}_{ \mathfrak{d}_{a,a'} \to \V{s}_{a} } ( \V{s}_{a,t} ) = 1$ and $\zeta^{(0)}_{ \mathfrak{d}_{a'\rmv,a} \to \V{s}_{a} } ( \V{s}_{a,t} ) = 1$.\noeqref{eg:self-loc-messag-to-s-1}
Eventually, at iteration $n = N_{\text{SL}}$, the
belief of each agent state $a \in \Set{A}$ after cooperative self-localization is calculated as:
\begin{align}
	&\tilde{f}_{0}(\V{s}_{a,t}) = \dfrac{1}{C_{a,t}^{(0)}} \zeta_{ \tau_{a} \to \V{s}_{a} } \big( \V{s}_{a,t} \big) \ist \zeta_{ \mathfrak{g}_{a} \to \V{s}_{a}} \big( \V{s}_{a,t} \big) \nonumber \\[0mm]
	&\hspace{2mm} \times \Bigg( \prod_{a' \in \Set{R}^{(a)}_{t} } \zeta^{(N_{\text{SL}})}_{ \mathfrak{d}_{a'\rmv,a} \to \V{s}_{a} } \big( \V{s}_{a,t} \big) \Bigg) \Bigg( \prod_{a' \in \Set{T}^{(a)}_{t} } \zeta^{(N_{\text{SL}})}_{ \mathfrak{d}_{a,a'} \to \V{s}_{a} } \big( \V{s}_{a,t} \big) \Bigg) \ist , \nonumber \\[-2mm]
	\label{eq:belief-zero-agent}
\end{align}
where $C_{a,t}^{(0)}$ is a normalization constant defined such that $\int \tilde{f}_{0}(\V{s}_{a,t}) \, \mathrm{d}\V{s}_{a,t} = 1$.

\subsubsection{PTs' augmented states prediction}
\label{subsubsec:pred}

\begin{figure}[!t]
	\centering
	\includegraphics{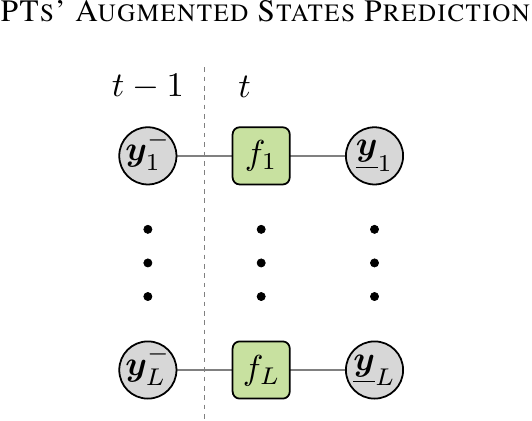}
	\caption{Factor graph representing the PTs' augmented states prediction portion of the factorization in \eqref{eq:final-factorization} for one time step $t$.
	Grey circles are variable nodes, and green squares are factor nodes.
	The following short notations are used:
	$L \deq L_{t}^{(1)}$;
	$\V{y}_{\ell}^{-} \deq \V{y}_{\ell,t-1}$; 
	$\protect\underline{\V{y}}_{\ell} \deq \protect\underline{\V{y}}_{\ell,t}^{(1)}$; 
	and $f_{\ell} \deq f (\protect\underline{\V{y}}_{\ell,t}^{(1)} | \V{y}_{\ell,t-1} )$.}
	\label{fig:graph-pt-pred}
	\vspace{-4mm}
\end{figure}

Let us recall that for each PT augmented state $\V{y}_{\ell,t-1}$, $\ell \in \Set{K}_{t-1}$, at time $t-1$, there is one legacy PT augmented state $\underline{\V{y}}_{\ell,t}^{(1)}$, $\ell \in \Set{L}_{t}^{(1)}$, at the first agent pair at time $t$.
The prediction of PT $\ell$ is performed by computing the message $\zeta_{f_{\ell} \to \underline{\V{y}}_{\ell}} (\underline{\V{y}}_{\ell,t}^{(1)})$ from factor node ``$f_{\ell}$'' to variable node ``$\underline{\V{y}}_{\ell}$'' in Fig.~\ref{fig:graph-pt-pred} as 
\begin{align}
	&\zeta_{ f_{\ell} \to \underline{\V{y}}_{\ell}} \big( \underline{\V{y}}_{\ell,t}^{(1)} \big) = \zeta_{ f_{\ell} \to \underline{\V{y}}_{\ell}} \big( \underline{\V{x}}_{\ell,t}^{(1)}, \underline{r}_{\ell,t}^{(1)} \big) \nonumber \\[0mm]
	&\hspace{5mm} = \!\! \sum_{r_{\ell,t-1} \in \{0,1\} } \int f \big( \underline{\V{x}}_{\ell,t}^{(1)}, \underline{r}_{\ell,t}^{(1)} \big| \V{x}_{\ell,t-1}, r_{\ell,t-1}) \nonumber \\[-2mm]
	& \hspace{30mm} \times \tilde{f}_{J} \big( \V{x}_{\ell,t-1}, r_{\ell,t-1} \big) \,\text{d} \V{x}_{\ell,t-1} \ist ,
	 \label{eq:message-PT-pred}
\end{align}
where $ \tilde{f}_{J} ( \V{x}_{\ell,t-1}, r_{\ell,t-1} )$ is the belief of the PT augmented state at previous time $t - 1$ computed at the last agent pair $J$; the computation of this belief is detailed in the Section~\ref{subsubsec:meas_update}.
Note that, from \eqref{eq:message-PT-pred} and the fact that $\tilde{f}_{J} ( \V{x}_{\ell,t-1}, r_{\ell,t-1} )$ is normalized, it follows that $\zeta_{ f_{\ell} \to \underline{\V{y}}_{\ell}} ( \underline{\V{x}}_{\ell,t}^{(1)}, \underline{r}_{\ell,t}^{(1)} )$ is normalized too, i.e., $\sum_{\underline{r}_{\ell,t}^{(1)} \in \{ 0,1 \}} \int \zeta_{ f_{\ell} \to \underline{\V{y}}_{\ell}} ( \underline{\V{x}}_{\ell,t}^{(1)}, \underline{r}_{\ell,t}^{(1)} ) \text{d}\underline{\V{x}}_{\ell,t}^{(1)} = 1$.

\begin{figure*}
	\centering
	\includegraphics{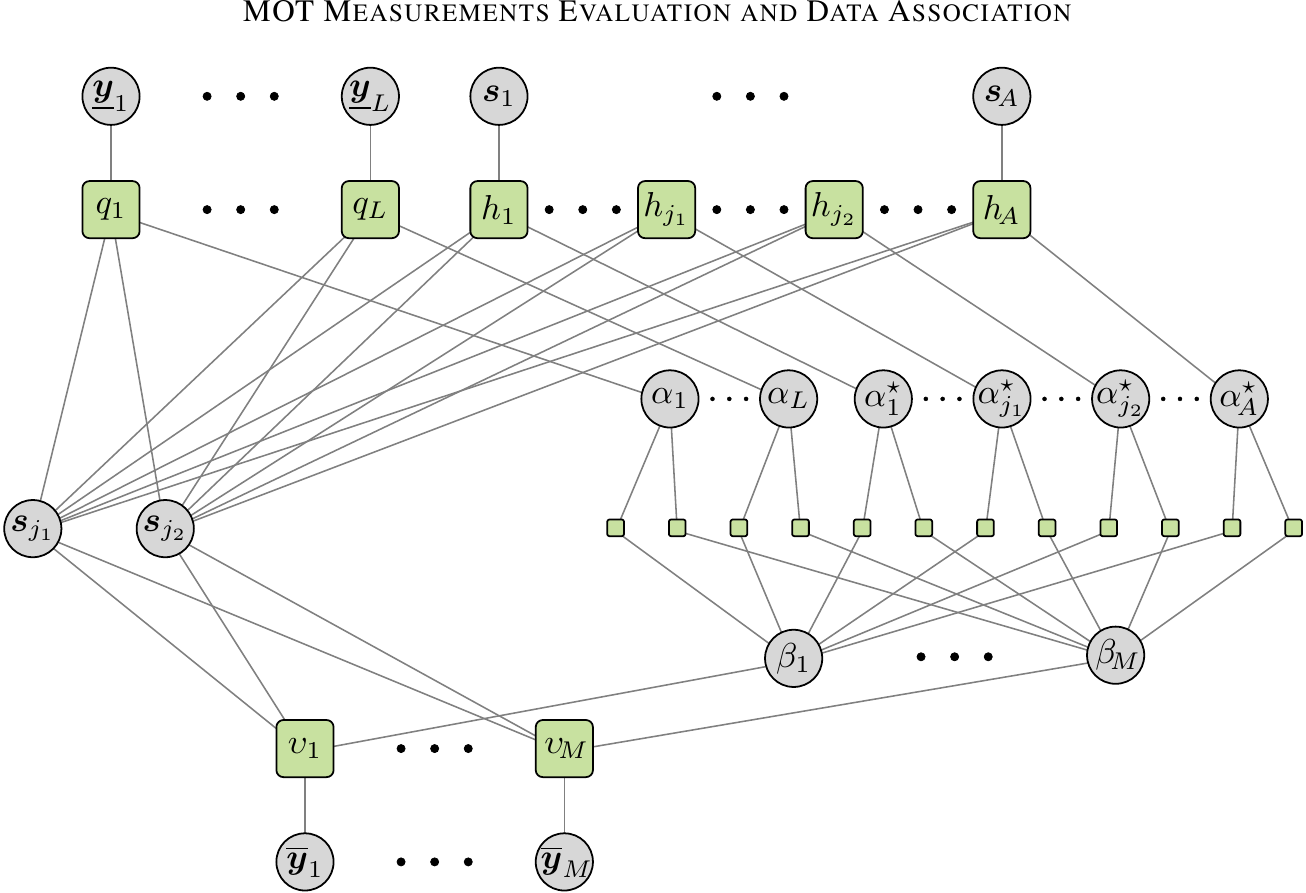}
	\caption{Factor graph representing the MOT measurements evaluation and data association portion of the factorization in \eqref{eq:final-factorization} for one time step $t$ and agent pair $j$.
	Grey circles are variable nodes, and green squares are factor nodes.
	The following short notations are used:
	$L \deq L_{t}^{(j)}$;
	$\protect\underline{\V{y}}_{\ell} \deq \protect\underline{\V{y}}_{\ell,t}^{(j)}$; 	
	$\V{s}_{a} \deq \V{s}_{a,t}$;
	$q_{\ell} \deq q ( \protect\underline{\V{y}}_{\ell,t}^{(j)}, \alpha_{\ell,t}^{(j)}, \V{s}_{j_{1},t}, \V{s}_{j_{2},t}; \V{z}_{t}^{(j)} )$;
	$h_{a} \deq h ( \V{s}_{a,t}, \alpha_{L + a,t}^{(j)}, \V{s}_{j_{1},t}, \V{s}_{j_{2},t}; \V{z}_{t}^{(j)} )$; 
	$\alpha_{\ell} \deq \alpha_{\ell,t}^{(j)}$; 
	$\alpha^{\star}_{a} \deq \alpha_{L + a,t}^{(j)}$; 
	$M \deq M_{t}^{(j)}$;
	$\beta_{m} \deq \beta_{m,t}^{(j)}$; 
	$\upsilon_{m} \deq \upsilon ( \overline{\V{y}}_{m,t}^{(j)}, \beta_{m,t}^{(j)}, \linebreak \V{s}_{j_{1},t}, \V{s}_{j_{2},t}; \V{z}_{m,t}^{(j)} )$;
	and $\overline{\V{y}}_{m} \deq \overline{\V{y}}_{m,t}^{(j)}$.}
	\label{fig:graph-tracking}
	\vspace{-2mm}
\end{figure*}

\subsubsection{MOT measurements evaluation and data association}
\label{subsubsec:meas_eval}
The MOT measurements evaluation and data association steps are performed at each agent pair $j$, sequentially from $j = 1$ to $j = J$.
With reference to Fig.~\ref{fig:graph-tracking}, the MOT measurements evaluation step consists in computing the following messages:
from factor nodes ``$q_{\ell}$'' to variable nodes ``$\alpha_{\ell}$'', i.e., $\zeta_{q_{\ell} \to \alpha_{\ell}} (\alpha_{\ell,t}^{(j)})$;
from factor nodes ``$h_{a}$'' to variable nodes ``$\alpha^{\star}_{a}$'', i.e., $\zeta_{h_{a} \to \alpha^{\star}_{a}} (\alpha_{L + a}^{(j)} )$; and
from factor nodes ``$\upsilon_{m}$'' to variable nodes ``$\beta_{m}$'', i.e., $\zeta_{\upsilon_{m} \to \beta_{m}} (\beta_{m,t}^{(j)})$.
We recall that the MOT measurements can be either monostatic, if $j_{1} = j_{2}$, or bistatic, if $j_{1} \neq j_{2}$;
next, we provide expressions of these messages for both cases by using the Dirac delta and the Kronecker delta.

The message $\zeta_{q_{\ell} \to \alpha_{\ell}} (\alpha_{\ell,t}^{(j)})$ is computed as
\begin{align}
	&\zeta_{q_{\ell} \to  \alpha_{\ell} } \big( \alpha^{(j)}_{\ell,t} \big) \nonumber \\[0mm]
	&\hspace{4.5mm} = \sum_{\underline{r}^{(j)}_{\ell,t} \in \{0,1\} } \iiint q \big( \underline{\V{x}}^{(j)}_{\ell,t}, \underline{r}^{(j)}_{\ell,t}, \alpha^{(j)}_{\ell,t}, \V{s}_{j_{1},t}, \V{s}_{j_{2},t}; \V{z}^{(j)}_{t} \big) \nonumber \\[0mm]
	&\hspace{9mm} \times \tilde{f}_{j-1} \big( \underline{\V{x}}^{(j)}_{\ell,t}, \underline{r}^{(j)}_{\ell,t} \big) \tilde{f}_{j-1} \big( \V{s}_{j_{1},t} \big) \Big( \tilde{f}_{j-1} \big( \V{s}_{j_{2},t} \big) \Big)^{1 - \delta_{j_{1},j_{2}}} \nonumber \\[0mm]
	&\hspace{9mm} \times \Big( \delta \big( \V{s}_{j_2,t} \big) \Big)^{\delta_{j_1,j_2}}
	\, \text{d}\underline{\V{x}}^{(j)}_{\ell,t} \, \text{d}\V{s}_{j_1,t} \, \text{d}\V{s}_{j_2,t} \ist .
	\label{eq:mot-measurements-message-q-to-alpha}
\end{align}
For monostatic MOT measurements, i.e., for $j_{1} = j_{2}$, the Kronecker delta $\delta_{j_{1},j_{2}}$ is $1$;
moreover, according to \eqref{eq:def-q1-r1}--\eqref{eq:def-q1-r0} and \eqref{eq:def-pd1}--\eqref{eq:def-pd2}, and to \eqref{eq:def-q2-r1}--\eqref{eq:def-q2-r0} and \eqref{eq:def-likelihood-f0}, the function $q(\cdot) = q_{1}(\cdot) \, q_{2}(\cdot)$ does not depend on the vector $\V{s}_{j_{2},t}$ (note that this also applies to function $h(\cdot) = h_{1}(\cdot) \, h_{2}(\cdot)$ according to \eqref{eq:def-h1} and \eqref{eq:def-h2-1}--\eqref{eq:def-h2-2}, and function $\upsilon(\cdot)$ according to
\eqref{eq:def-upsilon-2} and \eqref{eq:def-likelihood-f0}).
Therefore, the message in \eqref{eq:mot-measurements-message-q-to-alpha} in the case of monostatic MOT measurements particularizes as
\begin{align}
	&\zeta_{q_{\ell} \to  \alpha_{\ell} } \big( \alpha^{(j)}_{\ell,t} \big) = \!\! \sum_{\underline{r}^{(j)}_{\ell,t} \in \{0,1\} } \!\! \iiint q \big( \underline{\V{x}}^{(j)}_{\ell,t}, \underline{r}^{(j)}_{\ell,t}, \alpha^{(j)}_{\ell,t}, \V{s}_{j_{1},t}; \V{z}^{(j)}_{t} \big) \nonumber \\[0mm]
	&\hspace{7mm} \times \tilde{f}_{j-1} \big( \underline{\V{x}}^{(j)}_{\ell,t}, \underline{r}^{(j)}_{\ell,t} \big) \tilde{f}_{j-1} \big( \V{s}_{j_{1},t} \big)  \nonumber \\[0mm]
	&\hspace{7mm} \times \delta \big( \V{s}_{j_2,t} \big)  \, \text{d}\underline{\V{x}}^{(j)}_{\ell,t} \,\text{d}\V{s}_{j_1,t} \text{d}\V{s}_{j_2,t} \ist \nonumber \\[0mm]
	&\hspace{1.5mm} = \! \int \! \delta \big( \V{s}_{j_2,t} \big) \, \text{d}\V{s}_{j_2,t} \hspace{-2mm} \! \sum_{\underline{r}^{(j)}_{\ell,t} \in \{0,1\} } \!\! \iint \! q \big( \underline{\V{x}}^{(j)}_{\ell,t}, \underline{r}^{(j)}_{\ell,t}, \alpha^{(j)}_{\ell,t}, \V{s}_{j_{1},t}; \V{z}^{(j)}_{t} \big) \nonumber \\[0mm]
	&\hspace{7mm} \times \tilde{f}_{j-1} \big( \underline{\V{x}}^{(j)}_{\ell,t}, \underline{r}^{(j)}_{\ell,t} \big) \tilde{f}_{j-1} \big( \V{s}_{j_{1},t} \big) \, \text{d}\underline{\V{x}}^{(j)}_{\ell,t} \,\text{d}\V{s}_{j_1,t} \nonumber \\[0mm]
	&\hspace{1.5mm} = \!\! \sum_{\underline{r}^{(j)}_{\ell,t} \in \{0,1\} } \!\! \iint q \big( \underline{\V{x}}^{(j)}_{\ell,t}, \underline{r}^{(j)}_{\ell,t}, \alpha^{(j)}_{\ell,t}, \V{s}_{j_{1},t}; \V{z}^{(j)}_{t} \big) \nonumber \\[0mm]
	&\hspace{7mm} \times \tilde{f}_{j-1} \big( \underline{\V{x}}^{(j)}_{\ell,t}, \underline{r}^{(j)}_{\ell,t} \big) \tilde{f}_{j-1} \big( \V{s}_{j_{1},t} \big) \, \text{d}\underline{\V{x}}^{(j)}_{\ell,t} \,\text{d}\V{s}_{j_1,t} \ist ,
\end{align}
where we used the fact that $\int \delta (\V{s}_{a,t}) \, \text{d}\V{s}_{a,t} = 1$.
For bistatic MOT measurements, i.e., for $j_{1} \neq j_{2}$, instead, the Kronecker delta $\delta_{j_{1},j_{2}}$ is $0$, and the message in \eqref{eq:mot-measurements-message-q-to-alpha} becomes
\begin{align}
	&\zeta_{q_{\ell} \to  \alpha_{\ell} } \big( \alpha^{(j)}_{\ell,t} \big) \nonumber \\[0mm]
	&\hspace{5mm} = \sum_{\underline{r}^{(j)}_{\ell,t} \in \{0,1\} } \iiint q \big( \underline{\V{x}}^{(j)}_{\ell,t}, \underline{r}^{(j)}_{\ell,t}, \alpha^{(j)}_{\ell,t}, \V{s}_{j_{1},t}, \V{s}_{j_{2},t}; \V{z}^{(j)}_{t} \big) \nonumber \\[0mm]
	&\hspace{10mm} \times \tilde{f}_{j-1} \big( \underline{\V{x}}^{(j)}_{\ell,t}, \underline{r}^{(j)}_{\ell,t} \big) \tilde{f}_{j-1} \big( \V{s}_{j_{1},t} \big) \nonumber \\[0mm]
	&\hspace{10mm} \times \tilde{f}_{j-1} \big( \V{s}_{j_{2},t} \big) \, \text{d}\underline{\V{x}}^{(j)}_{\ell,t} \,\text{d}\V{s}_{j_1,t} \, \text{d}\V{s}_{j_2,t} \ist .
\end{align}
We note that the message in \eqref{eq:mot-measurements-message-q-to-alpha} depends on the beliefs at the previous agent pair $j - 1$ of the Rx-agent and Tx-agent states, i.e., $\tilde{f}_{j-1}(\V{s}_{j_{1},t})$  and $\tilde{f}_{j-1}(\V{s}_{j_{2},t})$, respectively, and of the legacy PTs augmented states, i.e., $\tilde{f}_{j-1}(\underline{\V{x}}_{\ell,t}^{(j)},\underline{r}_{\ell,t}^{(j)})$.
These beliefs are initialized at $j = 1$ with the prediction messages of the agent states and PT augmented states described in Section~\ref{subsubsec:agents_self_localization} and Section~\ref{subsubsec:pred}, respectively.
Specifically, the agent states beliefs are initialized with $\tilde{f}_{0} ( \V{s}_{j_{1},t} )$ and $\tilde{f}_{0} ( \V{s}_{j_{2},t} )$ as computed in \eqref{eq:belief-zero-agent}, and the PT augmented states beliefs are initialized with $\tilde{f}_{0} ( \underline{\V{x}}^{(1)}_{\ell,t}, \underline{r}^{(1)}_{\ell,t} ) = \zeta_{ f_{\ell} \to \underline{\V{y}}_{\ell}} ( \underline{\V{x}}_{\ell,t}^{(1)}, \underline{r}_{\ell,t}^{(1)} )$ as computed in \eqref{eq:message-PT-pred}.
Finally, we recall that before processing the MOT measurements at any agent pair $j > 1$, legacy and new PTs at agent pair $j - 1$ are
mapped into legacy PTs at agent pair $j$.
The PT mapping is formally described by the following expression
\begin{align}
	\tilde{f}_{j-1} & \big( \underline{\V{x}}_{\ell,t}^{(j)}, \underline{r}_{\ell,t}^{(j)} \big) = \hspace{-2mm} \sum_{r_{k,t}^{(j-1)} \in \{ 0,1 \}} \hspace{-2mm} \delta_{r_{k,t}^{(j-1)} \rmv , \ist \underline{r}_{\ell,t}^{(j)}} \\[0mm]
	&\times \int \tilde{f}_{j-1} \big( \V{x}_{k,t}^{(j-1)}, r_{k,t}^{(j-1)} \big) \delta \big( \V{x}_{k,t}^{(j-1)} - \underline{\V{x}}_{\ell,t}^{(j)} \big) \, \text{d}\V{x}_{k,t}^{(j-1)} \ist ,
\end{align}
where, with an abuse of notation, $\V{x}_{k,t}^{(j-1)}$ and $r_{k,t}^{(j-1)}$ represent the state and the existence variable of either a legacy or a new PT at agent pair $j-1$.

The message $\zeta_{h_{a} \to \alpha^{\star}_{a}} (\alpha_{L + a}^{(j)} )$ is expressed as
\begin{align}
	&\zeta_{h_{a} \to \alpha_{a}^{\star} } \big( \alpha^{(j)}_{L + a,t} \big) \nonumber \\[0mm]
	&\hspace{5mm} = \iiint h \big( \V{s}_{a,t} , \alpha^{(j)}_{L + a,t} , \V{s}_{j_{1},t}, \V{s}_{j_{2},t} ; \V{z}^{(j)}_{t} \big) \nonumber \\[0mm]
	&\hspace{10mm} \times \tilde{f}_{j-1} \big( \V{s}_{a,t} \big) \tilde{f}_{j-1} \big( \V{s}_{j_{1},t} \big) \Big( \tilde{f}_{j-1} \big( \V{s}_{j_{2},t} \big) \Big)^{1 - \delta_{j_{1}, j_{2}}} \nonumber \\[0mm]
	&\hspace{10mm} \times \Big( \delta(\V{s}_{j_2,t}) \Big)^{\delta_{j_1,j_2}} \, \text{d}\V{s}_{a,t} \, \text{d}\V{s}_{j_1,t} \, \text{d}\V{s}_{j_2,t} \ist .
\end{align}
We observe that, according to \eqref{eq:def-h1}, \eqref{eq:def-h2-1}, and \eqref{eq:def-h2-2}, if agent $a$ is either the Rx-agent $j_{1}$, or the Tx-agent $j_{2}$, the function $h (\cdot)$ is non-zero --- specifically, equal to $1$ --- if and only if $\alpha^{(j)}_{L + a,t} = 0$.
Thus, if $a = j_{1}$ or $a = j_{2}$, then $\zeta_{h_{a} \to \alpha_{a}^{\star} } ( \alpha^{(j)}_{L + a,t} ) = \linebreak \delta_{\alpha^{(j)}_{L + a,t},0}$; this follows from the
fact that Rx-agent $j_{1}$ and Tx-agent $j_{2}$ cannot produce MOT measurements at agent pair $j$.

Finally, the message $\zeta_{\upsilon_{m} \to \beta_{m}} (\beta_{m,t}^{(j)})$ is calculated as
\begin{align}
	&\zeta_{\upsilon_{m} \to  \beta_{m} } \big( \beta^{(j)}_{m,t} \big) \nonumber \\[0mm]
	&\hspace{5mm} = \! \! \sum_{\overline{r}^{(j)}_{m,t}\in \{0,1\}} \iiint \upsilon \big(\overline{\V{x}}^{(j)}_{m,t}, \overline{r}^{(j)}_{m,t}, \beta^{(j)}_{m,t}, \V{s}_{j_{1},t}, \V{s}_{j_{2},t}; \V{z}^{(j)}_{m,t}  \big) \nonumber \\[0mm]
	&\hspace{10mm} \times \tilde{f}_{j-1} \big( \V{s}_{j_1,t} \big) \Big( \tilde{f}_{j-1} \big( \V{s}_{j_2,t} \big) \Big)^{1-\delta_{j_1,j_2}} \nonumber \\[0mm]
	&\hspace{10mm} \times \Big( \delta \big(\V{s}_{j_2,t} \big) \Big)^{\delta_{j_1,j_2}}
\text{d}\overline{\V{x}}^{(j)}_{m,t} \, \text{d}\V{s}_{j_1,t} \, \text{d}\V{s}_{j_2,t} \ist .
\end{align}

The data association step is an iterative procedure that converts the messages $\zeta_{q_{\ell} \to  \alpha_{\ell} } ( \alpha^{(j)}_{\ell,t} )$, $\zeta_{h_{a} \to \alpha_{a}^{\star} } ( \alpha^{(j)}_{L + a,t} )$, and \linebreak $\zeta_{\upsilon_{m} \to  \beta_{m} } ( \beta^{(j)}_{m,t} )$, into the following messages:
$\eta_{\alpha_{\ell} \to q_{\ell}} ( \alpha^{(j)}_{\ell,t} )$, from variable nodes ``$\alpha_{\ell}$'' to factor nodes ``$q_{\ell}$''; 
$\eta_{\alpha_{a}^{\star} \to h_{a}} ( \alpha^{(j)}_{L + a,t} )$, from variable nodes ``$\alpha_{a}^{\star}$'' to factor nodes ``$h_{a}$''; and
$\eta_{\beta_{m} \to \upsilon_{m}} ( \beta^{(j)}_{m,t} )$, from variable nodes ``$\beta_{m}$'' to factor nodes ``$\upsilon_{m}$''.
This iterative procedure is described in \cite[Sec. IX-A3]{MeyKroWilLauHlaBraWin:J18}, and expressions of the messages are provided therein.

\subsubsection{Beliefs calculation}
\label{subsubsec:meas_update}
Once the MOT measurements produced
at agent pair $j$ are incorporated, the information they provide is used to eventually update the agent states
and
PT augmented states beliefs.
For the legacy PT augmented states, the following messages are computed from factor nodes ``$q_{\ell}$'' to variable nodes ``$\underline{\V{y}}_{\ell}$'', $\ell \in \Set{L}_{t}^{(j)}$, that is,
\begin{align}
	\nonumber \\[-8mm]
	&\zeta_{q_{\ell} \to \underline{\V{y}}_{\ell}} \big( \underline{\V{y}}^{(j)}_{\ell,t} \big) = \zeta_{q_{\ell} \to \underline{\V{y}}_{\ell}} \big( \underline{\V{x}}^{(j)}_{\ell,t}, \underline{r}^{(j)}_{\ell,t} \big) = \sum_{\alpha^{(j)}_{\ell,t} = 0}^{M^{(j)}_{t}} \eta_{\alpha_{\ell} \to q_{\ell}} \big( \alpha^{(j)}_{\ell,t} \big) \nonumber \\[0mm]
	&\hspace{5mm} \times \iint q( \underline{\V{x}}^{(j)}_{\ell,t} ,  \underline{r}^{(j)}_{\ell,t} , \alpha^{(j)}_{\ell,t} , \V{s}_{j_{1},t}, \V{s}_{j_{2},t} ; \V{z}^{(j)}_{t} ) \nonumber \\[0mm]
	&\hspace{5.5mm} \times \tilde{f}_{j-1} \big( \V{s}_{j_1,t} \big) \Big( \tilde{f}_{j-1} \big( \V{s}_{j_2,t} \big) \Big)^{1-\delta_{j_1,j_2}} \nonumber \\[0mm]
	&\hspace{5.5mm} \times \Big( \V{s}_{j_2,t}) \Big)^{\delta_{j_1,j_2}} \text{d}\V{s}_{j_1,t} \, \text{d}\V{s}_{j_2,t} \ist .
\end{align}
Then, the updated beliefs
are obtained as
\begin{align}
	&\tilde{f}_{j} \big( \underline{\V{x}}^{(j)}_{\ell,t},\underline{r}^{(j)}_{\ell,t} \big) = \frac{1}{\underline{C}_{\ell,t}^{(j)}} \tilde{f}_{j-1} \big( \underline{\V{x}}^{(j)}_{\ell,t},\underline{r}^{(j)}_{\ell,t} \big) \, \zeta_{q_{\ell} \to \underline{\V{y}}_{\ell}} \big( \underline{\V{x}}^{(j)}_{\ell,t},\underline{r}^{(j)}_{\ell,t} \big) \ist ,
\end{align}
where $\underline{C}_{\ell,t}^{(j)}$ is a normalization constant defined such that
$\sum_{\underline{r}_{\ell,t}^{(j)} \in \{ 0,1 \}} \int \tilde{f}_{j} ( \underline{\V{x}}^{(j)}_{\ell,t},\underline{r}^{(j)}_{\ell,t} ) \, \text{d} \underline{\V{x}}^{(j)}_{\ell,t} = 1$.
(For clarity, we recall that $\tilde{f}_{0} ( \underline{\V{x}}^{(1)}_{\ell,t}, \underline{r}^{(1)}_{\ell,t} ) = \zeta_{ f_{\ell} \to \underline{\V{y}}_{\ell}} ( \underline{\V{x}}_{\ell,t}^{(1)}, \underline{r}_{\ell,t}^{(1)} )$.)
Similarly, for the new PT augmented states, the messages from factor nodes ``$\upsilon_{m}$'' to variable nodes ``$\overline{\V{y}}_{m}$'' are computed as
\begin{align}
	&\zeta_{\upsilon_{m} \to \overline{\V{y}}_{m}} \big( \overline{\V{y}}^{(j)}_{m,t} \big) = \zeta_{\upsilon_{m} \to \overline{\V{y}}_{m}} \big( \overline{\V{x}}^{(j)}_{m,t},\overline{r}^{(j)}_{m,t} \big) \nonumber \\[0mm]
	&\hspace{5mm} = \sum_{\beta^{(j)}_{m,t} = 0}^{L^{(j)}_{t} + A} \eta_{\beta_{m} \to \upsilon_{m}} \big( \beta^{(j)}_{m,t} \big) \nonumber \\[0mm]
	&\hspace{10mm} \times \iint \upsilon \big( \overline{\V{x}}^{(j)}_{m,t} ,  \overline{r}^{(j)}_{m,t} , \beta^{(j)}_{m,t},  \V{s}_{j_{1},t}, \V{s}_{j_{2},t} ; \V{z}^{(j)}_{m,t} \big) \nonumber \\[0mm]
	&\hspace{10mm} \times \tilde{f}_{j-1} \big( \V{s}_{j_1,t} \big) \Big( \tilde{f}_{j-1} \big( \V{s}_{j_2,t} \big) \Big)^{1-\delta_{j_1,j_2}} \nonumber \\[0mm]
	&\hspace{10mm} \times \Big( \delta \big( \V{s}_{j_2,t} \big) \Big)^{\delta_{j_1,j_2}}
\text{d}\V{s}_{j_1,t} \, \text{d}\V{s}_{j_2,t} \ist
\end{align}
and the updated beliefs
are obtained as
\begin{align}
	\tilde{f}_{j} \big( \overline{\V{x}}^{(j)}_{m,t}, \overline{r}^{(j)}_{m,t} \big) = \frac{1}{\overline{C}_{m,t}^{(j)}} \zeta_{\upsilon_{m} \to \overline{\V{y}}_{m}} \big( \overline{\V{x}}^{(j)}_{m,t},\overline{r}^{(j)}_{m,t} \big) \ist,
\end{align}
where the normalization constant $\overline{C}_{m,t}^{(j)}$ is defined
such that $\sum_{\overline{r}_{m,t}^{(j)} \in \{ 0,1 \}}\int \tilde{f}_{j} ( \overline{\V{x}}^{(j)}_{m,t}, \overline{r}^{(j)}_{m,t} ) \, \text{d} \overline{\V{x}}^{(j)}_{m,t} = 1$.

For the agent states $a \in \Set{A}$, the messages $\zeta_{  h_{a} \to \V{s}_{a}}(\V{s}_{a,t})$ passed from the factor nodes ``$h_{a}$'' to the variable nodes ``$\V{s}_{a}$'' are calculated according to
\begin{align}
	&\zeta_{  h_{a} \to \V{s}_{a}} \big( \V{s}_{a,t} \big) = \sum_{\alpha^{(j)}_{L + a,t} = 0}^{M^{(j)}_{t}} \eta_{\alpha_{a}^{\star} \to h_{a}} \big( \alpha^{(j)}_{L + a,t} \big) \nonumber \\[0mm]
	&\hspace{10mm} \times \iint h \big( \V{s}_{a,t} ,  \alpha^{(j)}_{L + a,t} , \V{s}_{j_{1},t}, \V{s}_{j_{2},t} ; \V{z}^{(j)}_{t} \big) \nonumber \\[0mm]
	&\hspace{10mm} \times \tilde{f}_{j-1} \big( \V{s}_{j_1,t} \big) \Big( \tilde{f}_{j-1} \big( \V{s}_{j_2,t} \big) \Big)^{1-\delta_{j_1,j_2}} \nonumber \\[0mm]
	&\hspace{10mm} \times \Big( \delta \big( \V{s}_{j_2,t} \big) \Big)^{\delta_{j_1,j_2}} \, \text{d}\V{s}_{j_1,t} \, \text{d}\V{s}_{j_2,t} \ist .
	\label{eq:message-h-to-agent}
\end{align}
As before, we observe that if agent $a$ is either the Rx-agent $j_{1}$, or the Tx-agent $j_{2}$, i.e., $a = j_{1}$ or $a = j_{2}$, the function $h (\cdot)$ is non-zero if and only if $\alpha^{(j)}_{L + a,t} = 0$, from which it follows that $\zeta_{  h_{a} \to \V{s}_{a}} ( \V{s}_{a,t} ) = \eta_{\alpha_{a}^{\star} \to h_{a}} ( \alpha^{(j)}_{L + a,t} = 0)$.
Additionally, for
Rx-agent $j_{1}$ and
Tx-agent $j_{2}$ further computations are needed, as the following messages have to be calculated: $\zeta_{q_{\ell} \to \V{s}_{j_{1}}} (\V{s}_{j_{1},t})$ and $\zeta_{q_{\ell} \to \V{s}_{j_{2}}} (\V{s}_{j_{2},t})$, from factor nodes ``$q_{\ell}$'' to variable nodes ``$\V{s}_{j_{1}}$'' and
``$\V{s}_{j_{2}}$'', respectively; $\zeta_{\upsilon_{m} \to \V{s}_{j_{1}}} (\V{s}_{j_{1},t})$ and $\zeta_{\upsilon_{m} \to \V{s}_{j_{2}}} (\V{s}_{j_{2},t})$, from factor nodes ``$\upsilon_{m}$'' to variable nodes ``$\V{s}_{j_{1}}$'' and
``$\V{s}_{j_{2}}$'', respectively; and $\zeta_{h_{a} \to \V{s}_{j_{1}}} (\V{s}_{j_{1},t})$ and $\zeta_{h_{a} \to \V{s}_{j_{2}}} (\V{s}_{j_{2},t})$, from factor nodes ``$h_{a}$'' to variable nodes ``$\V{s}_{j_{1}}$'' and
``$\V{s}_{j_{2}}$'', respectively.
Hereafter, we provide the expressions of these messages for the Rx-agent $j_{1}$; the messages related to the Tx-agent $j_{2}$ can be derived similarly by substituting $j_{1}$ with $j_{2}$ and vice versa.
The messages $\zeta_{q_{\ell} \to \V{s}_{j_{1}}} (\V{s}_{j_{1},t})$, $\ell \in \Set{L}_{t}^{(j)}$, are defined as
\begin{align}
	&\zeta_{q_{\ell} \to \V{s}_{j_1}} \big( \V{s}_{j_1,t} \big) = \sum_{\alpha^{(j)}_{\ell,t} = 0}^{M^{(j)}_{t}} \sum_{\underline{r}^{(j)}_{\ell,t}\in \{0,1\}} \eta_{\alpha_{\ell} \to q_{\ell}} \big( \alpha^{(j)}_{\ell,t} \big) \nonumber \\[0mm]
	&\hspace{5mm}  \times  \iint q \big( \underline{\V{x}}^{(j)}_{\ell,t}, \underline{r}^{(j)}_{\ell,t}, \alpha^{(j)}_{\ell,t}, \V{s}_{j_{1},t}, \V{s}_{j_{2},t}; \V{z}^{(j)}_{t} \big) \nonumber \\[0mm]
	&\hspace{5mm} \times \tilde{f}_{j-1} \big( \underline{\V{x}}_{\ell,t}^{(j)}, \underline{r}_{\ell,t}^{(j)} \big) \Big( \tilde{f}_{j-1} \big( \V{s}_{j_{2},t} \big) \Big)^{1-\delta_{j_{1},j_{2}}} \nonumber \\[0mm]
	&\hspace{5mm} \times \Big( \delta \big( \V{s}_{j_{2},t} \big) \Big)^{\delta_{j_{1},j_{2}}} \, \text{d}\underline{\V{x}}^{(j)}_{\ell,t} \, \text{d}\V{s}_{j_{2},t} \ist ;
\end{align}
the messages $\zeta_{\upsilon_{m} \to \V{s}_{j_{1}}} (\V{s}_{j_{1},t})$, $m \in \Set{M}_{t}^{(j)}$, are defined as
\begin{align}
	&\zeta_{\upsilon_{m} \to  \V{s}_{j_{1}}} \big( \V{s}_{j_{1},t} \big) = \sum_{\beta^{(j)}_{m,t} = 0}^{L^{(j)}_{t} + A} \sum_{\overline{r}^{(j)}_{m,t}\in \{0,1\}} \eta_{\beta_{m} \to \upsilon_{m}} \big( \beta^{(j)}_{m,t} \big) \nonumber \\[0mm]
	&\hspace{5mm} \times \iint \upsilon \big( \overline{\V{x}}^{(j)}_{m,t}, \overline{r}^{(j)}_{m,t}, \beta^{(j)}_{m,t}, \V{s}_{j_{1},t}, \V{s}_{j_{2},t}; \V{z}^{(j)}_{m,t} ) \nonumber \\[0mm]
	&\hspace{5mm} \times \Big( \tilde{f}_{j-1} \big( \V{s}_{j_{2},t} \big) \Big)^{1-\delta_{j_{1},j_{2}}} \nonumber \\[0mm]
	&\hspace{5mm} \times \Big( \delta \big( \V{s}_{j_{2},t} \big) \Big)^{\delta_{j_{1},j_{2}}} \, \text{d}\overline{\V{x}}^{(j)}_{m,t} \, \text{d}\V{s}_{j_{2},t} \ist ;
\end{align}
finally, the messages $\zeta_{h_{a} \to \V{s}_{j_{1}}} (\V{s}_{j_{1},t})$, $a \in \Set{A}$, are defined as
\begin{align}
	&\zeta_{  h_{a} \to \V{s}_{j_{1}}} \big( \V{s}_{j_{1},t} \big) = \sum_{\alpha^{(j)}_{L + a,t} = 0}^{M^{(j)}_{t}} \eta_{\alpha_{a}^{\star} \to h_{a}} \big( \alpha^{(j)}_{L + a,t} \big) \nonumber \\[0mm]
	&\hspace{10mm} \times \iint h \big( \V{s}_{a,t} ,  \alpha^{(j)}_{L + a,t} , \V{s}_{j_{1},t}, \V{s}_{j_{2},t} ; \V{z}^{(j)}_{t} \big) \nonumber \\[0mm]
	&\hspace{10mm} \times \tilde{f}_{j-1} \big( \V{s}_{a,t} \big) \Big( \tilde{f}_{j-1} \big( \V{s}_{j_2,t} \big) \Big)^{1-\delta_{j_1,j_2}} \nonumber \\[0mm]
	&\hspace{10mm} \times \Big( \delta \big( \V{s}_{j_2,t} \big) \Big)^{\delta_{j_1,j_2}} \, \text{d}\V{s}_{a,t} \, \text{d}\V{s}_{j_2,t} \ist .
	\label{eq:message-h-to-rx-agent}
\end{align}
We observe that the expression of the message in \eqref{eq:message-h-to-rx-agent} is consistent with the expression in \eqref{eq:message-h-to-agent} in that, if $a = j_{1}$, then $\zeta_{h_{a} \to \V{s}_{j_{1}}} (\V{s}_{j_{1},t}) = \eta_{\alpha_{a}^{\star} \to h_{a}} ( \alpha^{(j)}_{L + a,t} = 0)$.
Eventually, the updated belief of the Rx-agent state is computed as
\begin{align}
	\tilde{f}_{j} \big( \V{s}_{j_{1},t} \big) &= \dfrac{1}{C_{j_{1},t}^{(j)}} \tilde{f}_{j-1} \big( \V{s}_{j_{1},t} \big) \hspace{-.5mm} \prod_{\ell \in \Set{L}_{t}^{(j)}} \hspace{-.5mm} \zeta_{q_{\ell} \to \V{s}_{j_{1}}} \big( \V{s}_{j_{1},t} \big) \nonumber \\[0mm]
	&\hspace{5mm}\times \hspace{-1mm} \prod_{m \in \Set{M}_{t}^{(j)}} \hspace{-1mm} \zeta_{\upsilon_{m} \to \V{s}_{j_{1}}} \big( \V{s}_{j_{1},t} \big) \prod_{a \in \Set{A}} \zeta_{h_{a} \to \V{s}_{j_{1}}} \big( \V{s}_{j_{1},t} \big) \ist ,
\end{align}
while the belief for any other agent state, i.e., $a \in \Set{A} \setminus \{ j_{1}, j_{2} \}$, is computed as
\begin{align}
	\tilde{f}_{j} \big( \V{s}_{a,t} \big) = \dfrac{1}{C_{a,t}^{(j)}} \tilde{f}_{j - 1} \big( \V{s}_{a,t} \big) \, \zeta_{h_{a} \to \V{s}_{a}} \big( \V{s}_{a,t} \big) \ist ,
\end{align}
where the normalization constant $C_{a,t}^{(j)}$ is defined such that $\int \tilde{f}_{j} ( \V{s}_{a,t} ) \, \mathrm{d}\V{s}_{a,t} \! = \! 1$;
we recall that $\tilde{f}_{0} (\V{s}_{a,t})$ is given in \eqref{eq:belief-zero-agent}.

\subsubsection{Implementation details}
\label{subsubsec:particle_impl}
The SPA-based joint localization and tracking algorithm detailed in this section is implemented following a particle-based approach, where each pdf is described by a set of $N_{\text{P}}$ particles.
This choice is particularly appropriate for non-Gaussian settings and pdf, which intrinsically appear in data association problems	 \cite{MenBraAllNicKocBauLepBra:c20,SolMeyBraHla:J19_67,MeyWin:C18}.
Although both agent states and PT states are here unknown --- unlike other SPA-based approaches that assume perfect knowledge of the sensors' position~\cite{SolMeyBraHla:J19_67,MeyKroWilLauHlaBraWin:J18} --- the proposed algorithm scales quadratically with $N_{\text{P}}$ by virtue of a proper stacking of the particles.
Furthermore, it scales linearly with the number of MOT measurements, number of agent pairs, and number of iterations of the agents' cooperative self-localization loop and the data association loop, and quadratically with the number of legacy PTs.

Moreover, as mentioned in Section \ref{subsec:target_state_space_model}, in order to keep a tractable number of PTs over time, a pruning step is performed.
Specifically, once all the MOT measurements at time $t$ are processed, any PT $k \in \Set{K}_{t}$ with existence probability $f(r_{k,t} = 1 | \V{g}_{1:t}, \VG{\rho}_{1:t}, \V{z}_{1:t})$ smaller than a threshold $P_{\text{pr}}$, is removed from the set $\Set{K}_{t}$ and it is not carried over to the next time $t + 1$ as legacy PT.
Besides this pruning step, the number $K_t$ of PTs at time $t$ is unbounded.

\section{Experimental Results}
\label{sec:sim_res}
In this section, performance results of the proposed joint cooperative self-localization and multitarget tracking approach are provided.
The maritime domain is considered as application scenario, thus the kinematics of agents and targets, the accuracy of measurements, and other parameters are chosen accordingly.
First, in Section~\ref{subsec:exp_set}, a simulated scenario is used to show how to take advantage of target information for the localization of agents.
Then, in Section~\ref{sec:exp_res_real_data}, an application to real maritime data is presented.

\subsection{Simulated scenario}
\label{subsec:exp_set}

\subsubsection{Set-up}
\label{subsubsec:set-up}

\begin{figure}[!t]
	\centering
	
	\includegraphics{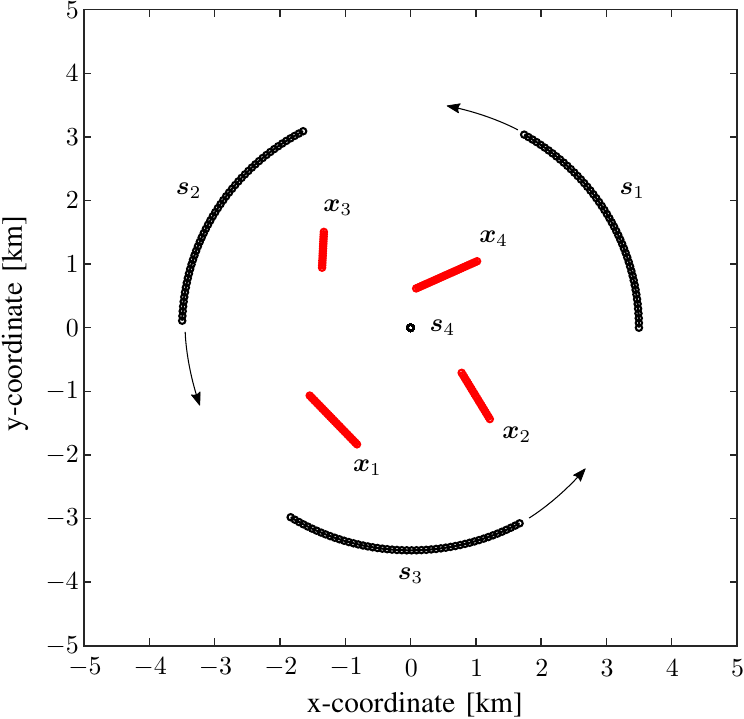}

	\caption{Illustration of the simulated scenario (the time index $t$ is omitted).
	The black circles
		are the agents' positions over time: the
		arrows
		represent the counterclockwise
		directions of agents
		$a = 1$, $2$, and $3$, while agent $a = 4$ is anchored.
	The red circles
		are the targets's positions over time.
		}
	\label{fig:figsimscenario}
	
	\vspace{-3mm}
	
\end{figure}

The simulated scenario is shown in Fig.~\ref{fig:figsimscenario}.
An area of 10$\ist\times$10 km is surveyed by $A = 4$ agents over 50 time steps;
the time step duration is $T_{\text{s}} = 30$ s. Agents $a = 1$, $2$, and $3$ move counterclockwise along a circle of radius 3.5 km and center $(0,0)$ at a constant radial velocity of $0.69$ m/s, while agent $a = 4$ is anchored at the center. The static agent is the only Tx-agent of the considered scenario, while the moving ones are all Rx-agents, leading to the following sets: $\Set{R}= \{1 , 2, 3 \}$ and $\Set{T}= \{ 4 \}$. Agents can communicate and sense over the whole area, i.e., there is no limitation on the sensing/communication range. The agent state
$\V{s}_{a,t} = [ \check{\V{s}}_{a,t}^{\T} , \dot{\check{\V{s}}}_{a,t}^{\T} ]^{\T} \in \mathbb{R}^4$
comprises both position, i.e., $\check{\V{s}}_{a,t}$, and velocity, i.e., $\dot{\check{\V{s}}}_{a,t}$, over a 2D space, and its dynamics is modeled with a nearly constant velocity (NCV) model, that is (cf. \eqref{eq:agent-dynamic-model}), $\V{s}_{a,t} = \VG{\varepsilon}_{a} ( \V{s}_{a,t-1}, \V{u}_{a,t} ) = \V{A} \V{s}_{a,t-1} + \linebreak \V{W} \V{u}_{a,t}$, where $\V{A} \rmv\in \mathbb{R}^{4 \times 4}$ and $\V{W} \!\in\rmv \mathbb{R}^{4 \times 2}$ are as in \cite[Sec.\ 6.3.2]{BarRonKir:01}, and the
process noise term $\V{u}_{a,t}$ is Gaussian distributed with mean $\V{0}$ and time-invariant covariance matrix $\omega_{\text{A}}^{2} \ist \mathbf{I}_{2}$, with per-component standard deviation (std) $\omega_{\text{A}} = 0.1$ m/s$^2$.

The scenario also includes four mobile targets, each moving at a constant speed randomly drawn from $[ -1.54, 1.54 ]$ m/s.
They appear and disappear at different times, thus are detectable respectively in the following time intervals:
$t \in [5,  35]$, $t \in [10,40]$, $t \in [20,40]$, and $t \in [30,45]$. The PT state $\V{x}_{k,t} = [ \check{\V{x}}_{k,t}^{\T}, \dot{\check{\V{x}}}_{k,t}^{\T} ]^{\T} \in \mathbb{R}^4$ comprises both position, i.e., $\check{\V{x}}_{k,t}$, and velocity, i.e., $\dot{\check{\V{x}}}_{k,t}$, over a 2D space, and its dynamics (cf. \eqref{eq:PT-dynamic-model}) follows an NCV model similar to the one adopted for the agent states, with a per-component
process noise standard deviation of $\omega_{\text{T}} = 0.1$ m/s$^2$.

The prior pdf $f (\V{s}_{0})$, assuming that the states of the agents at time $t = 0$ are independent,
can be written as
\begin{align}
	f \big( \V{s}_{0} \big) = \prod_{a \in \Set{A}} f \big( \V{s}_{a,0} \big) = \prod_{a \in \Set{A}} f \big( \check{\V{s}}_{a,0} \big) f \big( \dot{\check{\V{s}}}_{a,0} \big) \ist ,
\end{align}
where the pdf $f(\check{\V{s}}_{a,0})$ is uniform over a circle of radius 150 m around the true position, and $f(\dot{\check{\V{s}}}_{a,0})$ is uniform
on the 2D interval $[-2.57 \, , 2.57 ] \times [-2.57 \, , 2.57] \, \text{m/s}$.

Agents cooperatively localize themselves by combining
navigation data $\V{g}_{a,t}$ and inter-agent measurements $\VG{\rho}^{(a,a')}_{t}$.
The former are available at agents $a = 3$ and $4$ at all times, i.e., $\Set{A}^{\V{g}}_{t} = \Set{A}^{\V{g}} = \{ 3 , 4 \}$, 
and provide only position information, that is (cf. \eqref{eq:nav-data}), $\V{g}_{a,t} = \VG{\theta}_{a} ( \V{s}_{a,t}, \V{n}_{a,t} ) = \check{\V{s}}_{a,t} + \V{n}_{a,t}$, $a \in \Set{A}^{\V{g}}$.
The noise term $\V{n}_{a,t}$ has a Gaussian distribution with mean $\V{0}$ and time-invariant covariance matrix $\sigma_{a}^{2} \ist \mathbf{I}_{2}$, with
\begin{align}
	\sigma_{a} =
	\begin{cases}
		20 \, \text{m}	& a = 3 \,, \\[0mm]
		5 \, \text{m}	& a = 4 \,.
	\end{cases} 
\end{align} 
The latter are range-bearing measurements, available at all times and among all agents, defined as (cf. \eqref{eq:inter-agent-meas}):
\begin{align}
	\VG{\rho}_{t}^{(a,a')} &= \VG{\vartheta} \big( \V{s}_{a,t}, \V{s}_{a'\rmv,t}, \V{w}_{t}^{(a,a')} \big) \\[1mm]
	&= \begin{bmatrix}
		2 \norm{\check{\V{s}}_{a,t} - \check{\V{s}}_{a'\rmv,t}} \\[1mm]
		\angle(\check{\V{s}}_{a'\rmv,t} - \check{\V{s}}_{a,t})
	\end{bmatrix}
	+ \V{w}_{t}^{(a,a')} \ist ,
\end{align}
where the noise term $\V{w}_{t}^{(a,a')}$ is Gaussian distributed with mean $\V{0}$ and time-invariant covariance matrix $\diag(\varsigma_{\rho,\text{r}}^{2},\varsigma_{\rho,\text{b}}^{2})$, equal for all pairs $(a,a')$; the standard deviations are set to $\varsigma_{\rho,\text{r}} = 20$ m and $\varsigma_{\rho,\text{b}} = 1$ deg.
The number of iterations of the agents' cooperative self-localization loop
is set to $N_{\text{SL}} = 5$.

At time $t$ and agent pair $j$, agents and PTs give rise to MOT measurements $\V{z}^{(j)}_{m,t}$, $m \in \Set{M}_{t}^{(j)}$, as described in Section~\ref{subsubsec:MOT}.
We recall that an MOT measurement can be monostatic ($j_{1} = j_{2}$) or bistatic ($j_{1} \neq j_{2}$), and can derive --- unless it is a false alarm --- from a reflection from a PT or from an agent (except that from the Rx-agent $j_{1}$ and the Tx-agent $j_{2}$).
We thus have four possible MOT measurement models. A monostatic MOT measurement $\V{z}_{m,t}^{(j)}$ comprises range and bearing information and, assuming that it rises from PT $k$,
is modeled as (cf. \eqref{eq:inter-agent-meas1})
\begin{align}
	\V{z}^{(j)}_{m,t} &= \V{\gamma}_{\texttt{mono}} \big(\V{x}_{k,t}, \V{s}_{j_1,t},\V{v}^{(j)}_{m,t} \big) \nonumber \\[0mm]
	&=\begin{bmatrix}
	    2 \norm{\check{\V{s}}_{j_1,t} - \check{\V{x}}_{k,t}}  \\[1mm]
		\angle(\check{\V{x}}_{k,t} - \check{\V{s}}_{j_1,t})
	\end{bmatrix} + \V{v}^{(j)}_{m,t} \ist .
	\label{eq:mot-meas-model-mono}
\end{align}
Similarly, a bistatic MOT measurement has a bistatic range and bearing information, with the latter representing the
AoA of the reflected signal at the Rx-agent; therefore, assuming that it is generated by PT $k$, it is modeled as (cf. \eqref{eq:inter-agent-meas2})
\begin{align}
	\V{z}^{(j)}_{m,t} &= \V{\gamma}_{\texttt{bi}} \big(\V{x}_{k,t}, \V{s}_{j_1,t}, \V{s}_{j_2,t}, \V{v}^{(j)}_{m,t} \big) \nonumber \\[0mm]
	&=\begin{bmatrix}
		\norm{\check{\V{s}}_{j_1,t} - \check{\V{x}}_{k,t}} + \norm{\check{\V{s}}_{j_2,t} - \check{\V{x}}_{k,t}} \\[1mm]
		\angle(\check{\V{x}}_{k,t} - \check{\V{s}}_{j_1,t})
	\end{bmatrix} + \V{v}^{(j)}_{m,t} \ist .
	\label{eq:mot-meas-model-bi}
\end{align}
The noise term $\V{v}^{(j)}_{m,t}$ in both \eqref{eq:mot-meas-model-mono} and \eqref{eq:mot-meas-model-bi} 
is Gaussian distributed with mean $\V{0}$ and time-invariant covariance matrix $\diag(\varsigma_{\text{z,r}}^{2}, \varsigma_{\text{z,b}}^{2})$, equal for all agent 
pairs $j$; the standard deviations are set to $\varsigma_{\text{z,r}} = 20$ m and $\varsigma_{\text{z,b}} = 1$ deg.
The analogous cases for monostatic/bistatic MOT measurements originating from agent's reflection can be easily derived, thus omitted.
Note that all the observations, sent to a fusion centre, refer to a common spatial reference system.
The detection probability is assumed constant among each agent pair, regardless of the type of MOT measurement (monostatic or bistatic);
moreover, it is independent of the legacy object
state, Rx-agent state, and Tx-agent state, that is, (cf. \eqref{eq:def-pd1}) $P_{\text{d},\texttt{mono}}^{(j)} ( \V{o}_{i,t}^{(j)}, \V{s}_{j_{1}\rmv,t} ) = P_{\text{d},\texttt{bi}}^{(j)} ( \V{o}_{i,t}^{(j)}, \linebreak \V{s}_{j_{1}\rmv,t}, \V{s}_{j_{2}\rmv,t} ) = P_{\text{d}} = 0.7$.
Furthermore, the mean number of false alarms is $\mu_{\text{c}}^{(j)} = \mu_{\text{c}} = 3$, and the false alarm distribution $f_{\text{c}}^{(j)} ( \V{z}^{(j)}_{m,t} ) = f_{\text{c}} ( \V{z}^{(j)}_{m,t} )$ is uniform over the entire surveillance area.
Finally, the new PT state $\overline{\V{x}}_{m,t}^{(j)}$ related to MOT measurement $\V{z}_{m,t}^{(j)}$ as described in Section~\ref{subsec:target_state_space_model}, is distributed according to $f_{\text{n}} (\overline{\V{x}}_{m,t}^{(j)}) = f (\check{\overline{\V{x}}}_{m,t}^{(j)}) f (\dot{\check{\overline{\V{x}}}}_{m,t}^{(j)})$, with Gaussian pdf $f (\check{\overline{\V{x}}}_{m,t}^{(j)})$ centered around the Cartesian MOT measurement (converted from the range-bearing space) and with
covariance matrix $\varsigma_{\text{n}}^{2} \mathbf{I}$, with per-component standard deviation $\varsigma_{\text{n}} = 500$ m, and uniform
pdf $f (\dot{\check{\overline{\V{x}}}}_{m,t}^{(j)})$
on the 2D interval 
$[-1.54 \, , \linebreak 1.54] \times [-1.54 \, , 1.54] \, \text{m/s}$;
the mean number of new PTs is $\mu_{\text{n}}^{(j)} = \mu_{\text{n}} = 0.1$.

All agent and PT state pdfs are described by sets of $N_{\text{P}} = 1000$ particles.
Furthermore, the pruning threshold is set to $P_{\text{pr}} = 0.01$. 
Table~\ref{tab:simulationParameters} summarizes the main parameters used for the performance evaluation in the simulated scenario. 

\begin{table}[!b]

	\vspace{-3mm}

	\renewcommand{\arraystretch}{1.2}
	\small	
		
	\caption{Simulation parameters.}
	\label{tab:simulationParameters}
	\centering
	\begin{tabular}{ l | c | c }
		\textsc{Parameter}	&	\textsc{Symbol}	&	\textsc{Value}	\\[.5mm]
		\hline \hline
		Number of agents & $A$ & 4\\
		Set of Rx-agents & $\Set{R}$ & $\{ 1, 2, 3 \}$ \\
		Set of Tx-agents & $\Set{T}$ & $\{ 4 \}$ \\
		Set of agents with navigation data & $\Set{A}^{\V{g}}$ & $\{ 3, 4 \}$  \\
		Process noise std	& $\omega_{\text{A}}$, $\omega_{\text{T}}$	& 0.1 m/s$^2$\\
		Navigation data std, agent $a = 3$ & $\sigma_{3}$ & 20 m\\
		Navigation data std, agent $a = 4$ & $\sigma_{4}$ & 5 m\\
		Range information std  & $\varsigma_{\rho,\text{r}}$, $\varsigma_{\text{z,r}}$	& 20 m\\
		Bearing information std & $\varsigma_{\rho,\text{b}}$, $\varsigma_{\text{z,b}}$	& 1 deg\\
		Detection probability & $P_\text{d}$ & 0.7 \\
		Mean number of false alarms &  $\mu_{\text{c}}$ & $3$  \\
		New PT, prior position std	&  $\varsigma_{\text{n}}$ & 500 m  \\
		Mean number of new PTs & $\mu_{\text{n}}$ & 0.1 \\
		Survival probability & $p_{\text{s}}$ & 0.99 \\
		Pruning threshold & $P_{\text{pr}}$ & 0.01 \\
		PT existence threshold & $P_{\text{ex}}$& 0.75\\
		Number of particles & $N_{\text{P}}$ & 1000 \\
		\hline
	\end{tabular}
\end{table}

\subsubsection{Discussion}
\label{subsubsec:discussion}
In this settings, we compare the performance of the proposed
algorithm that jointly performs cooperative self-localization and multitarget tracking, with respect to the case in which the two tasks are performed independently.
To ease the notation and
for the reader's convenience, we refer to
the proposed method as joint localization and tracking (JLT), and
to the alternative approach as separate localization and tracking (SLT). The difference between SLT and JLT is that the former estimates the agent states at time $t$ only once by running the cooperative self-localization;
then, the multitarget tracking task is performed considering the estimated agent states as \textit{true} states.
This is equivalent to running the classical SPA-based multitarget tracking algorithm described in \cite{MeyKroWilLauHlaBraWin:J18} in which the states (i.e., positions) of the sensors are assumed known.
The proposed JLT algorithm, instead, repeatedly estimates the agent states after the MOT measurements from each agent pair $j$ are processed, as described in Section \ref{subsec:SPA_Message_Passing_alg} and shown in Fig. \ref{fig:figblockscheme}.
On the figures, JLT-related quantities are reported with
dotted lines, while SLT-related quantities with dashed lines. 

It is important to
emphasize that in the JLT approach targets are not only unknown objects to be localized, but they represent  a valuable information correspondingly used by the agents to refine their own localization through the proposed cooperative mechanism. To highlight this, in the simulation we induce an outage condition to agents $a = 1$, and $2$ in the time interval $t \in [ 10, 40 ]$, meaning that they do not have availability of inter-agent measurements from the Tx-agent $a = 4$.
In practical use cases, outage can be related to perturbations on the communication channel due to hardware malfunctioning, interference, or hacking. Results are averaged over 250 Monte Carlo iterations.

\begin{figure}[!t]
	\centering
	\hspace{0mm} \includegraphics{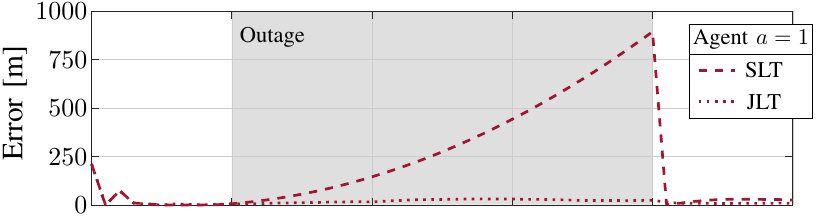} \\[1mm]
	\hspace{0mm} \includegraphics{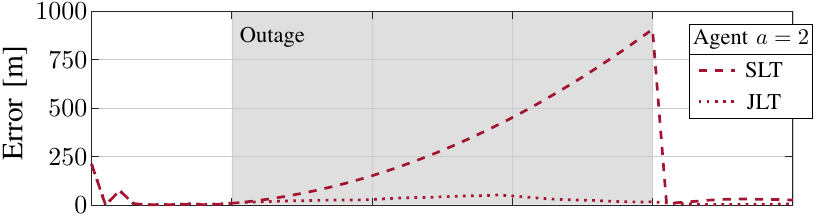} \\[1mm]
	\hspace{0mm} \includegraphics{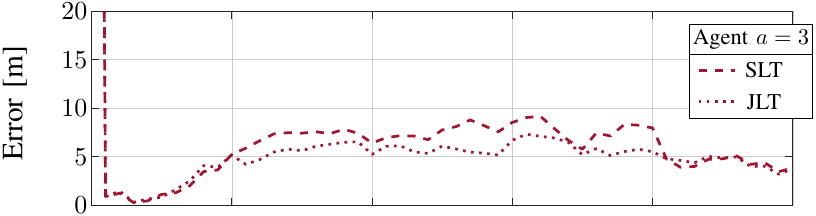} \\[1mm]
	\hspace{0mm} \includegraphics{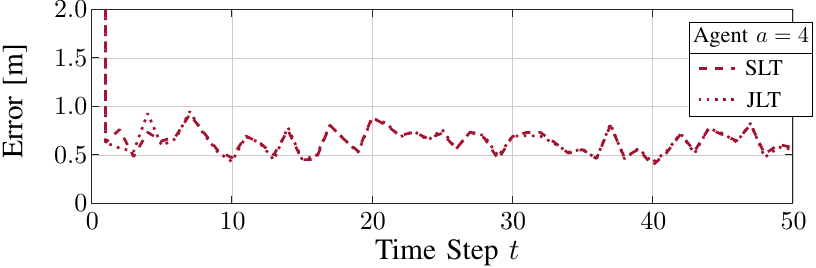}
	\vspace{-2mm}
	\caption{Performance comparison between JLT and	SLT in terms of agents localization error over time.}
	\label{fig:figagentaccuracy}
	\vspace{-2mm}
\end{figure}

In Fig.~\ref{fig:figagentaccuracy} we provide the position error over time for each agent individually, for both JLT and SLT. The most interesting result is related to the agents in outage condition, i.e., $a = 1$, and $2$: we observe that
the JLT is still able to localize them by exploiting the target information. The SLT, instead, can only rely on motion prediction,
that leads to high position errors.
Considering practical implementation, the use of JLT might allow, for instance, the recovery of the agent, which would be hardly achieved in case of SLT.
As a second comment, we note that JLT outperforms SLT in the localization of agent $a = 3$, while there are no benefits for the localization of agent  $a = 4$, since it is anchored and navigation data are extremely precise by default.
\begin{figure}[!t]
	\centering
	
	\includegraphics{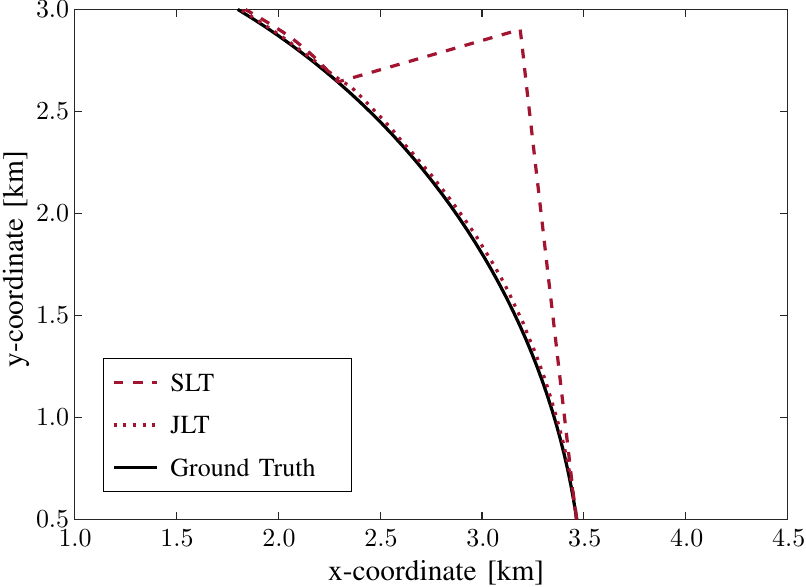}
	\vspace{-1mm}
	\caption{Zoomed reconstructed trajectory of agent $a = 1$. The black solid line is the ground truth trajectory, the red dashed line indicates the SLT estimated trajectory, and the red dotted line refers to the JLT estimated trajectory.}
	\label{fig:figagentzoom}
	\vspace{-2mm}
\end{figure}
To show the effect of relying on motion prediction only (SLT) with respect to a profitable use of target implicit information (JLT), in Fig.~\ref{fig:figagentzoom} we zoom in on the reconstructed trajectory of agent $a = 1$,
highlighting the huge difference between the two methodologies.
Note that the effect that the target location information supports the localization of the agents has been
previously demonstrated in \cite{MeyHliWye:J16}. However, as mentioned in Section \ref{subsec:background}, the algorithm proposed in \cite{MeyHliWye:J16} is limited by the fact that the number of targets that can be tracked is time-invariant and has to be known in advance, and it assumes a perfect knowledge of the association between targets and measurements.

\begin{figure}[!t]
	\centering
	
	\subfloat[MOSPA error.]{
		\includegraphics{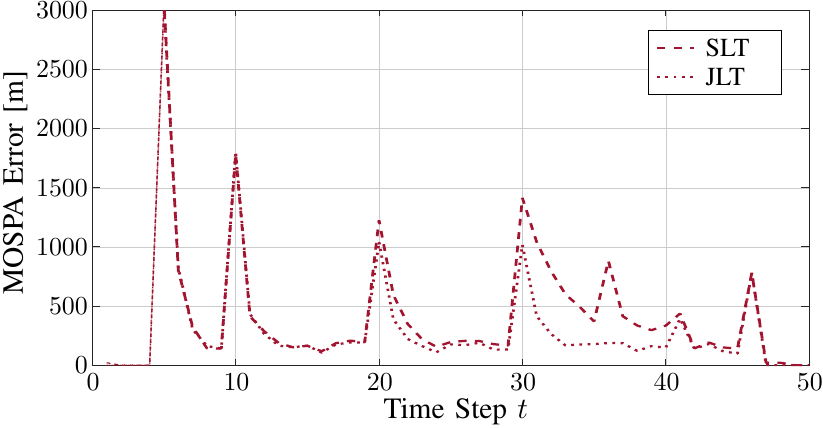}
		\label{fig:mospa} 
	}  
      
	\centering

	\subfloat[Mean number of detected targets.]{
		\hspace{1.6mm}		
		\includegraphics{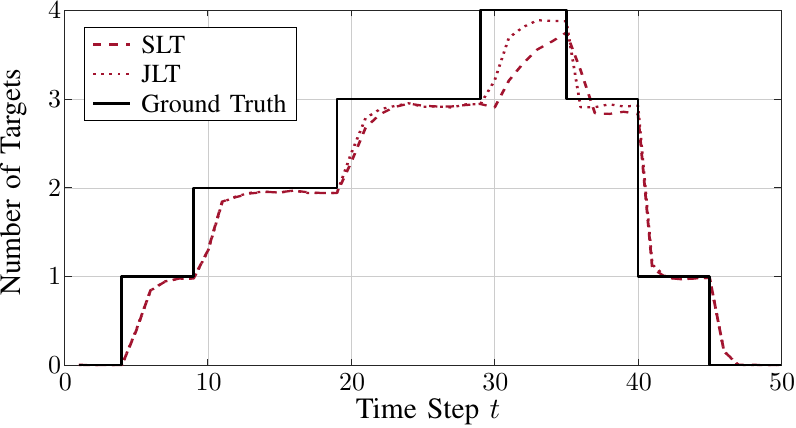}
		\label{fig:numberoftargets}
	}
	
	\caption{Performance comparison between the multitarget tracking capabilities of JLT and SLT in the simulated scenario 
		in terms of (a) MOSPA error and (b) estimated mean number of detected targets over time.
		}
	\vspace{-2mm}
	\label{fig:targetMetric}
\end{figure}

After the analysis on agent localization, we now focus on the capability of the proposed technique to perform multitarget tracking.
Results are given in terms of mean optimal subpattern assignment (MOSPA) error \cite{SchVoVo:J08} of order 1 and cut off parameter of 5000 m in Fig.~\ref{fig:mospa}, and in terms of estimated mean number of detected targets in Fig.~\ref{fig:numberoftargets};
the MOSPA error accounts for localization errors for correctly detected targets, and errors for missed targets and false targets.
Results in Fig.~\ref{fig:mospa} indicate superior tracking capabilities of the JLT over the SLT, in particular
after the outage that starts at $t = 10$ (300 s).
Indeed, the outage does not allow the SLT to accurately estimate
the positions of agents $a = 1$, and $2$; this directly affects the multitarget tracking task, since the MOT measurements they produce cannot be effectively used.
After $t = 30$ (900 s), the difference in MOSPA error between JLT and SLT increases, because of the appearance of the last target that is not promptly detected by the SLT.
The tardiness in target detection is also observable in Fig.~\ref{fig:numberoftargets}, where we report the mean number of detected PTs over time.

Results demonstrate how target information is of high importance for practical application, e.g., in maritime surveillance. We proved the advantages of considering a joint
framework for cooperative self-localization and multitarget tracking rather than perform the two tasks independently. 

\subsection{Application to real maritime data}
\label{sec:exp_res_real_data}
\begin{figure}[!b]
	\centering
	\vspace{-2mm}
	\includegraphics{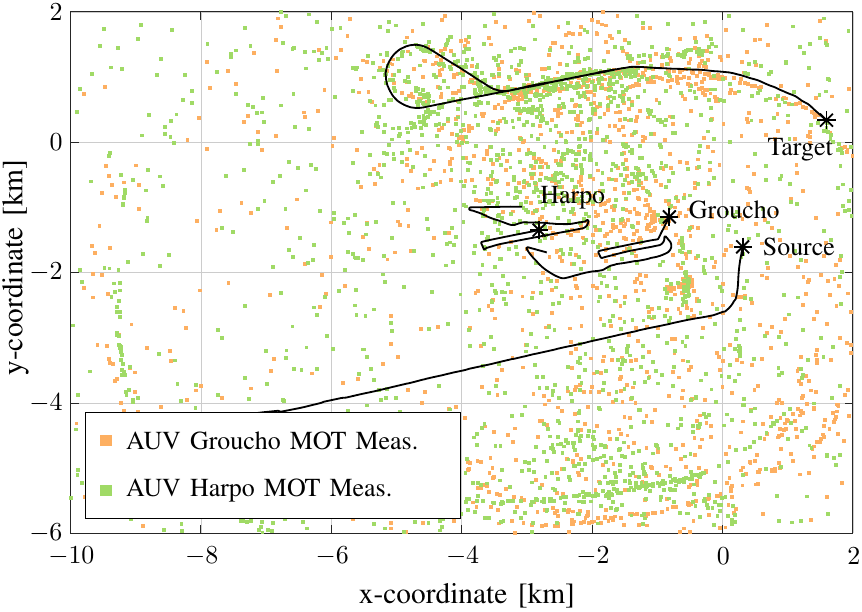}
	
	\caption{MOT measurements (converted in x-y coordinates) produced by AUVs Groucho and Harpo, respectively in orange and green, during the LCAS18 campaign. The black solid lines are the ground truth trajectories of the sonar source, the AUVs Groucho and Harpo, and the target; the asterisks mark the initial positions of these trajectories.}
	\label{fig:mot-measurements}
	
\end{figure}

This section presents the performance of JLT and SLT assessed in a real
maritime application. We consider a hybrid, autonomous, robotic network developed by NATO Centre for Maritime Research and Experimentation (CMRE) for surveillance applications.
The network consists of mobile and fixed gateways that form the communication infrastructure, and of autonomous underwater vehicles (AUVs) capable of detecting and tracking possible threats, and communicating the acquired data to the command and control center \cite{FerMunTesBra:J17,FerMunLeP:J18}.
The data we use
was gathered during the littoral continuous active sonar trial conducted off the coast of Piombino, near Livorno, Italy, in November 2018 (LCAS18) \cite{FerPetDeMMorMic:C19}.
During this trial the network consisted of $A = 6$ agents, classified as follows:
\begin{itemize}
	\item surface agents: a towed sonar source ($a = 1$), two stationary and co-located acoustic modems ($a = 2$, and $3$), and a
		 nearly-stationary waveglider ($a = 4$);
		\item underwater agents: two AUVs ($a = 5$, and $6$), named as Groucho and Harpo, each towing a uniform linear array of microphones and equipped with an acoustic modem for communications.
\end{itemize}
The classification of Rx-agents and Tx-agents is chosen as follows:
the set of Tx-agents is constituted by all surface and underwater agents, i.e., $\Set{T} = \{ 1,2, \ldots, 6 \}$, while the set of Rx-agents comprises all agents but the towed sonar source, i.e., $\Set{R} = \{ 2,3, \ldots, 6 \}$. 
Practically, inter-agent measurements are available between all the Tx-agents and Rx-agents, 
while MOT measurements are only produced by the AUVs, i.e., $a = 5$, and $6$,
using the signal
transmitted by the towed sonar source, i.e., $a = 1$; this means that the agent pairs involved in the sequential processing of the MOT measurements are $(j_{1},j_{2}) = (5,1)$ and $(j_{1},j_{2}) = (6,1)$, from which it then follows that $J = \{ 1,2 \}$.
Fig.~\ref{fig:mot-measurements} shows all the MOT measurements produced by Groucho and Harpo during the trial, respectively in orange and green, and the ground-truth trajectories of the sonar source, the NATO research vessel Leonardo acting as target, and the AUVs. The ground-truth position of the former two is provided by their on-board GNSS receivers, whereas the ground-truth of Groucho and Harpo is provided by their INS.
We
note that in this real application, the observations (navigation data, inter-agent and MOT measurements) are not guaranteed to be acquired at the same time, nor they are available at all times because of the challenging propagation conditions posed by the underwater environment.
Nevertheless, the flexibility of the proposed algorithm allows
to perform each single task (e.g., the agents' cooperative self-localization or the processing of the MOT measurements at a specific agent pair) only when the relevant observations are available.

\begin{figure}[!t]
	\centering
	
	\subfloat[Snapshot at minute 37.]{
		\includegraphics{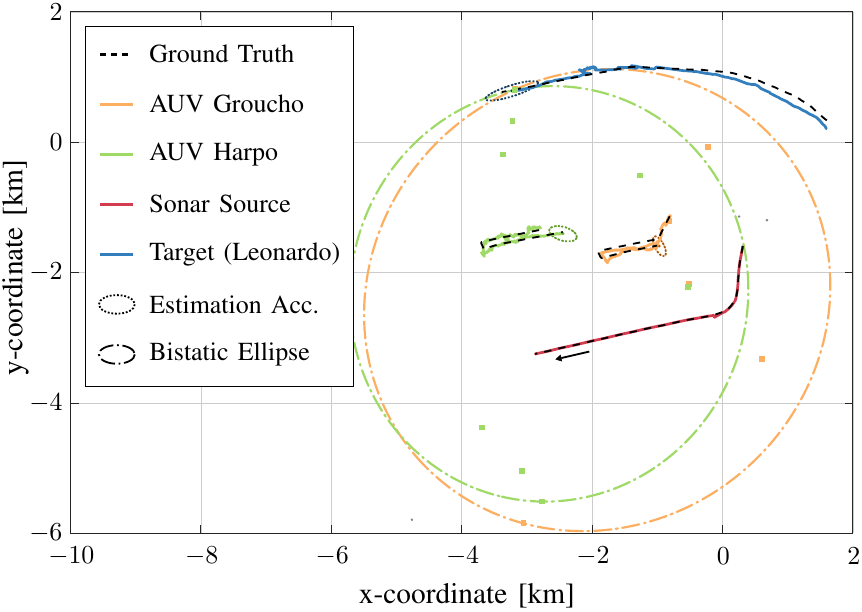}
		\label{fig:algorithm-in-action-1} 
	}
	
	\centering

	\subfloat[Snapshot at minute 87.]{
		\includegraphics{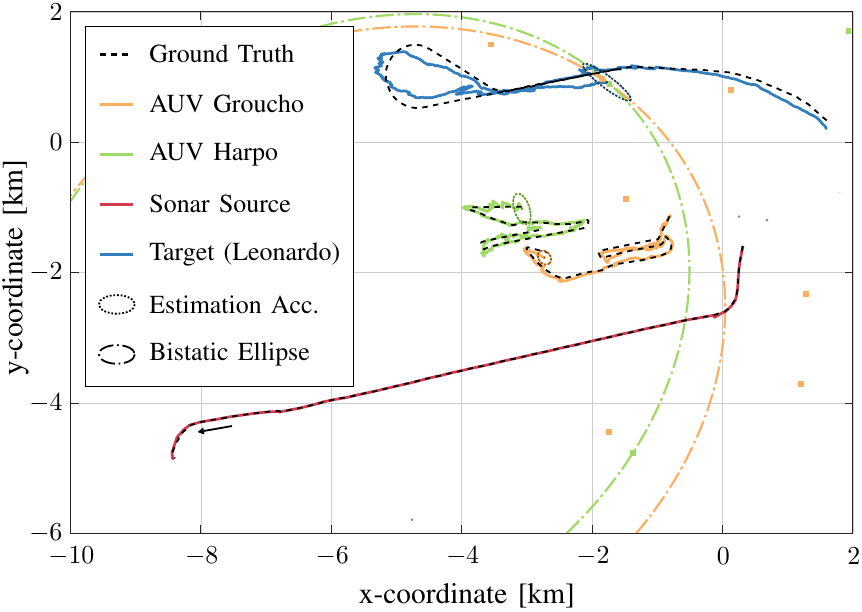}
		\label{fig:algorithm-in-action-2}
	}
	
	\caption{Behavior of the JLT using the LCAS18 data set. The figures are snapshots of the
		output produced by JLT after (a) 37 minutes and (b) 87 minutes since the beginning of the trial, and include MOT measurements, estimated trajectories, and estimated current positions of agents and target.
	Solid lines represent the estimated trajectory up to current time.
	Dotted ellipses indicate the estimate and accuracy of agents and target current location (note that the ellipse associated to the sonar source is hardly visible, as its estimated position is very accurate due to the availability of navigation data; the black arrow indicates the direction of movement).
	Dotted-dashed ellipses are examples of bistatic ellipses associated to Groucho's and Harpo's MOT measurements closest to the target.
	}
	\vspace{-2mm}
	\label{fig:algorithm-in-action}
\end{figure}

Navigation data are available for all the surface agents by means of on-board GNSS receivers; the standard deviation of the position information is set to 5 m, and the standard deviation of the velocity information (used by the towed sonar source only) is set to 0.1 m/s.
Range and bearing characterize both inter-agent and MOT measurements.
Standard deviations of range and bearing information are set to 70 m and 7 deg, respectively. 
Additionally, inter-agent and MOT measurements are duplicated because of the port-starboard ambiguity, i.e., the inability of the AUVs to discriminate if a signal comes from the port side or from the starboard side due to the intrinsic cylindrical symmetry of the towed array \cite{BraWilLePMarMat:J14}.
The other parameters are as
reported in Table~\ref{tab:simulationParameters}, with the exception of the mean number of false alarms set to $\mu_{\text{c}} = 9$, and the number of particles set to $N_{\text{P}} = 250$.

\begin{figure}[!t]

	\centering

	\includegraphics{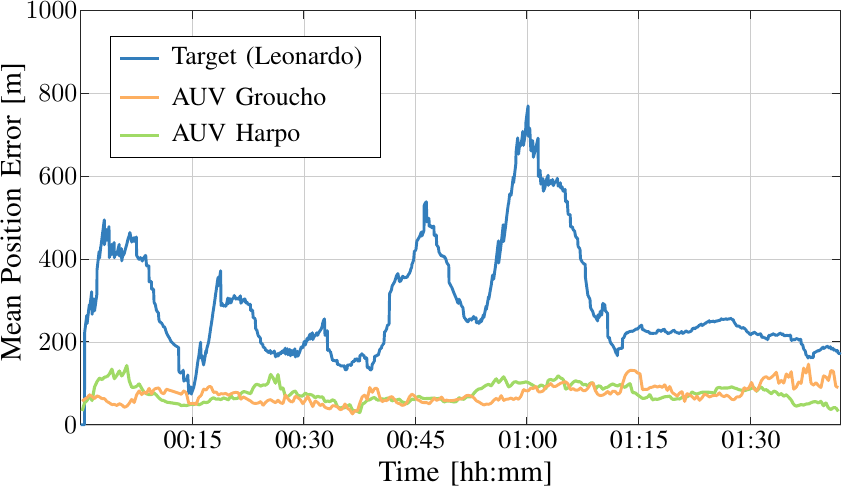}
	
	\caption{Real data experiment. Mean error on the JLT estimated positions of the target (in blue), AUV Groucho (in orange) and AUV Harpo (in green).}
	\label{fig:figrealdataauvpos}
	
	\vspace{-2mm}	
	
\end{figure}

Fig.~\ref{fig:algorithm-in-action-1} and Fig.~\ref{fig:algorithm-in-action-2} are snapshots of the output produced by JLT after 37 and 87 minutes, respectively, since the beginning of the
trial.
The images show the estimated trajectories of the non-stationary agents and the target, and estimate and accuracy (i.e., 95\% confidence interval) of their current positions.
Examples of bistatic ellipses associated to Groucho's and Harpo's MOT measurements closest to the target are also provided.
A bistatic ellipse is the locus of points in which the sum of the distances from the sonar source and the AUV (i.e., the foci) is constant and equal to the bistatic range component of the AUV's MOT measurement (cf. \eqref{eq:mot-meas-model-bi}) \cite{Wil:B11}.
Because of the port-starboard ambiguity
two actual contacts are visible on each bistatic ellipse, one close to the target and the other along the specular direction.
These snapshots show that it is possible to joint localize agents (towed sonar source and AUVs) and target through a fusion of navigation
data, inter-agent and MOT measurements.
Fig.~\ref{fig:figrealdataauvpos} shows the position errors over time averaged over 20 Monte Carlo iterations obtained with JLT for the target\footnote{For each Monte Carlo iteration, the position error for the target is computed as the distance between its ground-truth trajectory and the PT that is consistently closest to it over time.}, Groucho, and Harpo (the Monte Carlo iterations only differ for the drawing of the particles).
The higher
mean position error for the target is clearly due to the fact that only MOT measurements are used to estimate its state; moreover, the peaks occur
when either the target itself is maneuvering and/or one or both the AUVs are maneuvering.
Table~\ref{tab:real-data-comparison} reports a comparison between
JLT and SLT in terms of time-averaged and maximum mean position errors, and time-averaged MOSPA error with order 1 and cutoff parameter of 1000 m.
JLT outperforms SLT in terms of time-averaged and maximum mean position errors for both the AUVs.
These results confirm the benefit of exploiting the target location information for estimating the agents' states.
Regarding the tracking of the target, JLT provides a slightly higher time-averaged mean position error, but a lower maximum mean position error and a lower time-averaged MOSPA error, that accounts for both missed targets and false targets.
\begin{table}[!b]
\centering
    	\renewcommand{\arraystretch}{1.2}
	\small	
	\caption{Real data experiment. Performance comparison \\ between JLT and SLT.}
	\label{tab:real-data-comparison}
	\centering
	\begin{tabular}{ l | c || c | c }
		\multicolumn{2}{c ||}{\textsc{Mean Position Error}}	&	\textsc{JLT}	&	\textsc{SLT}\\[.5mm]
		\hline \hline
		\rule{0mm}{3.5mm}\multirow{2}{*}{Target (Leonardo)}	&	Time-averaged	&	$285.4$ m	&	$267.7$ m	\\
			&	Max	&	$770.0$ m	&	$842.0$ m	\\[.5mm] \hline
		\rule{0mm}{3.5mm}\multirow{2}{*}{AUV Groucho}	&	Time-averaged	&	$75.9$ m	&	$117.2$ m	\\
			&	Max	&	$143.4$ m	&	$189.8$ m	\\[.5mm] \hline
		\rule{0mm}{3.5mm}\multirow{2}{*}{AUV Harpo}	&	Time-averaged	&	$74.2$ m	&	$79.6$ m	\\
			&	Max	&	$146.6$ m	&	$165.9$ m	\\[.5mm] \hline \hline
		\multicolumn{2}{c ||}{\rule{0mm}{3.5mm}\textsc{Time-averaged MOSPA Error}}	&	$348.2$ m	&	$405.1$ m	\\[.5mm]	\hline		
	\end{tabular}
\end{table}\begin{figure}[!t]
	\centering

	\includegraphics{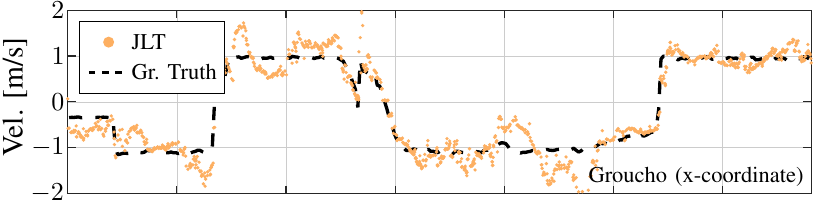} \\[1mm]
	\includegraphics{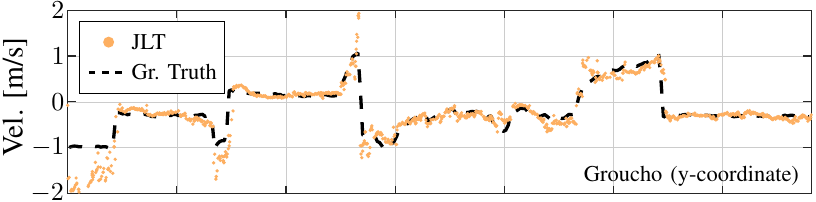} \\[1mm]
	\includegraphics{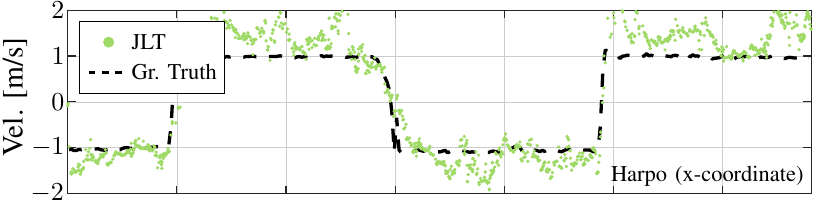} \\[1mm]
	\includegraphics{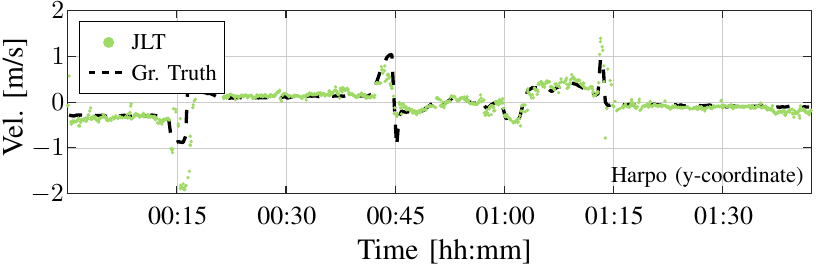}
	
	\caption{Real data experiment. Comparison between the Cartesian components of the JLT estimated velocities of
		Groucho (top figures) and
		Harpo (bottom figures),
		and their ground truth velocities
		provided by the on-board INS.}
	\label{fig:figrealdataauvvel}
\end{figure}\begin{figure*}[!t]

\setcounter{MYtempeqncnt}{\value{equation}}
\setcounter{equation}{45}

\begin{align}
	f \big( \V{y}_{1:t}, & \V{s}_{0:t}, \VG{\alpha}_{1:t}, \VG{\beta}_{1:t} \big| \V{g}_{1:t}, \VG{\rho}_{1:t}, \V{z}_{1:t}, \V{m}_{1:t} \big) \nonumber \\[0mm]
	&\propto f \big( \V{y}_{1:t}, \V{s}_{0:t}, \VG{\alpha}_{1:t}, \VG{\beta}_{1:t} , \V{g}_{1:t}, \VG{\rho}_{1:t}, \V{z}_{1:t}, \V{m}_{1:t} \big) \nonumber \\[0mm]
	&= f \big( \V{y}_{t}, \V{s}_{t}, \VG{\alpha}_{t}, \VG{\beta}_{t}, \V{g}_{t}, \VG{\rho}_{t}, \V{z}_{t}, \V{m}_{t} \big| \V{y}_{1:t-1}, \V{s}_{0:t-1}, \VG{\alpha}_{1:t-1}, \VG{\beta}_{1:t-1}, \V{g}_{1:t-1}, \VG{\rho}_{1:t-1}, \V{z}_{1:t-1}, \V{m}_{1:t-1} \big) \nonumber \\[0mm]
	& \hspace{10mm} \times f \big( \V{y}_{1:t-1}, \V{s}_{0:t-1}, \VG{\alpha}_{1:t-1}, \VG{\beta}_{1:t-1}, \V{g}_{1:t-1}, \VG{\rho}_{1:t-1}, \V{z}_{1:t-1}, \V{m}_{1:t-1} \big) \nonumber \\[0mm]
	&= f \big( \V{y}_{t}, \V{s}_{t}, \VG{\alpha}_{t}, \VG{\beta}_{t}, \V{g}_{t}, \VG{\rho}_{t}, \V{z}_{t}, \V{m}_{t} \big| \V{y}_{t-1}, \V{s}_{t-1} \big)  f \big( \V{y}_{1:t-1}, \V{s}_{0:t-1}, \VG{\alpha}_{1:t-1}, \VG{\beta}_{1:t-1}, \V{g}_{1:t-1}, \VG{\rho}_{1:t-1}, \V{z}_{1:t-1}, \V{m}_{1:t-1} \big) \nonumber \\[0mm]
	&= f \big(
		\V{s}_{0} \big) \prod_{t' = 1}^{t} f \big( \V{y}_{t'}, \V{s}_{t'}, \VG{\alpha}_{t'}, \VG{\beta}_{t'}, \V{g}_{t'}, \VG{\rho}_{t'}, \V{z}_{t'}, \V{m}_{t'} \big| \V{y}_{t'-1}, \V{s}_{t'-1} \big)
	\label{eq:app-factorization-1}
\end{align}
\hrulefill
\end{figure*}\begin{figure*}[!t]
\begin{align}
	&f \big( \overline{\V{y}}_{t}, \VG{\alpha}_{t}, \VG{\beta}_{t}, \V{z}_{t}, \V{m}_{t} \big| \underline{\V{y}}_{t}^{(1)}, \V{s}_{t} \big) \nonumber \\[0mm]
	&= f \big( \overline{\V{y}}_{t}^{(J)}, \VG{\alpha}_{t}^{(J)}, \VG{\beta}_{t}^{(J)}, \V{z}_{t}^{(J)}, M_{t}^{(J)} \big| \underline{\V{y}}_{t}^{(1)}, \overline{\V{y}}_{t}^{(1)} \rmv\rmv\rmv\rmv, \ldots, \overline{\V{y}}_{t}^{(J-1)}, \VG{\alpha}_{t}^{(1)} \rmv\rmv\rmv\rmv, \ldots, \VG{\alpha}_{t}^{(J-1)}, \VG{\beta}_{t}^{(1)} \rmv\rmv\rmv\rmv, \ldots, \VG{\beta}_{t}^{(J-1)}, \V{z}_{t}^{(1)} \rmv\rmv\rmv\rmv, \ldots, \V{z}_{t}^{(J-1)}, M_{t}^{(1)} \rmv\rmv\rmv\rmv, \ldots, M_{t}^{(J-1)}, \V{s}_{t} \big) \nonumber \\[0mm]
	&\hspace{10mm} \times f \big( \overline{\V{y}}_{t}^{(1)} \rmv\rmv\rmv\rmv, \ldots, \overline{\V{y}}_{t}^{(J-1)}, \VG{\alpha}_{t}^{(1)} \rmv\rmv\rmv\rmv, \ldots, \VG{\alpha}_{t}^{(J-1)}, \VG{\beta}_{t}^{(1)} \rmv\rmv\rmv\rmv, \ldots, \VG{\beta}_{t}^{(J-1)}, \V{z}_{t}^{(1)} \rmv\rmv\rmv\rmv, \ldots, \V{z}_{t}^{(J-1)}, M_{t}^{(1)} \rmv\rmv\rmv\rmv, \ldots, M_{t}^{(J-1)} \big) \nonumber \\[0mm]
	&= f \big( \overline{\V{y}}_{t}^{(J)}, \VG{\alpha}_{t}^{(J)}, \VG{\beta}_{t}^{(J)}, \V{z}_{t}^{(J)}, M_{t}^{(J)} \big| \underline{\V{y}}_{t}^{(J)}, \VG{\alpha}_{t}^{(1)} \rmv\rmv\rmv\rmv, \ldots, \VG{\alpha}_{t}^{(J-1)}, \VG{\beta}_{t}^{(1)} \rmv\rmv\rmv\rmv, \ldots, \VG{\beta}_{t}^{(J-1)}, \V{z}_{t}^{(1)} \rmv\rmv\rmv\rmv, \ldots, \V{z}_{t}^{(J-1)}, M_{t}^{(1)} \rmv\rmv\rmv\rmv, \ldots, M_{t}^{(J-1)}, \V{s}_{t} \big) \nonumber \\[0mm]
	&\hspace{10mm} \times f \big( \overline{\V{y}}_{t}^{(1)} \rmv\rmv\rmv\rmv, \ldots, \overline{\V{y}}_{t}^{(J-1)}, \VG{\alpha}_{t}^{(1)} \rmv\rmv\rmv\rmv, \ldots, \VG{\alpha}_{t}^{(J-1)}, \VG{\beta}_{t}^{(1)} \rmv\rmv\rmv\rmv, \ldots, \VG{\beta}_{t}^{(J-1)}, \V{z}_{t}^{(1)} \rmv\rmv\rmv\rmv, \ldots, \V{z}_{t}^{(J-1)}, M_{t}^{(1)} \rmv\rmv\rmv\rmv, \ldots, M_{t}^{(J-1)} \big) \nonumber \\[0mm]
	&= f \big( \overline{\V{y}}_{t}^{(J)}, \VG{\alpha}_{t}^{(J)}, \VG{\beta}_{t}^{(J)}, \V{z}_{t}^{(J)}, M_{t}^{(J)} \big| \underline{\V{y}}_{t}^{(J)}, \V{s}_{t} \big) \nonumber \\[0mm]
	&\hspace{10mm} \times f \big( \overline{\V{y}}_{t}^{(1)} \rmv\rmv\rmv\rmv, \ldots, \overline{\V{y}}_{t}^{(J-1)}, \VG{\alpha}_{t}^{(1)} \rmv\rmv\rmv\rmv, \ldots, \VG{\alpha}_{t}^{(J-1)}, \VG{\beta}_{t}^{(1)} \rmv\rmv\rmv\rmv, \ldots, \VG{\beta}_{t}^{(J-1)}, \V{z}_{t}^{(1)} \rmv\rmv\rmv\rmv, \ldots, \V{z}_{t}^{(J-1)}, M_{t}^{(1)} \rmv\rmv\rmv\rmv, \ldots, M_{t}^{(J-1)} \big) \nonumber \\[0mm]
	&= \prod_{j = 1}^{J} f \big( \overline{\V{y}}_{t}^{(j)}, \VG{\alpha}_{t}^{(j)}, \VG{\beta}_{t}^{(j)}, \V{z}_{t}^{(j)}, M_{t}^{(j)} \big| \underline{\V{y}}_{t}^{(j)}, \V{s}_{t} \big) \ist .
	\label{eq:joint-pdf-prod-over-agent-pair}
\end{align}
\hrulefill

\setcounter{equation}{\value{MYtempeqncnt}}

\end{figure*}Lastly, we compare in Fig.~\ref{fig:figrealdataauvvel} the JLT estimated velocities of the AUVs --- along the two Cartesian coordinates --- and the velocities provided by the INS:
results show a good flexibility to abrupt heading variations and quite  accurate velocity estimation overall.

The assessment on real data  proves the ability of the proposed SPA-based 
algorithm to jointly perform cooperative self-localization and multitarget tracking. The results
provides evidence of how to profitably take advantage of intrinsic target information to perform agents localization.

\section{Concluding Remarks}
\label{sec:conclusions}

In this work, we developed a joint technique for cooperative self-localization and multitarget tracking in
agent networks.
The proposed algorithm is general enough to be tailored to any
multi-agent system, where agents are equipped with diverse perception sensors and communication devices. The proposed
method performs the mandatory tasks of self-determining the network topology  (i.e., the agent network localization)  and detecting and
tracking an unknown and arbitrary number of targets, where existence probabilities are used to declare their actual presence or to opt for their removal (pruning). The developed
technique takes advantage of target information to update and refine the agent
positions, assigning an opportunistic role to targets.
This latter benefit might not be so relevant in case of a large availability of navigation data and/or inter-agent measurements,
but it has been proven through simulations to be of utmost importance in case of malfunctioning or outage conditions.
An important aspect of the proposed
algorithm is the flexibility: the algorithm intrinsically handles time-variant properties of agents and network topology (such as a connection of a new agent, or the disappearance of an extant one) and it also admits the
coexistence of different types of observations (navigation data, inter-agent and MOT measurements).
Lastly, the extension of the data association problem to include
agents as well (and not only targets) allows us to consider realistic conditions of signal propagation, where reflections from both agents and targets are unavoidably present and affect the signal processing chain.

The
joint cooperative self-localization and multitarget tracking method proposed in this article assumes a centralized architecture of information exchange 
in a complex
network made by several agents.
Future work will include the study of distributed/decentralized architectures where the exchange of local target/agent states among the agents, rather than of observations with a centralized node, is more convenient, e.g., in terms of robustness.
Promising paradigms for distributed architectures are the consensus networks
\cite{BraMarMatWil:J10,PapaRepMeyBraHla:J18,ShaSauBucVar:J19,OlfFaxMux:J07}  and the adaptive networks \cite{MatBraMarSay:J16,Say:J14}.

\appendix

\section{Derivation of \eqref{eq:joint-posterior-pdf}}
\label{sec:app-joint-post-pdf}
Here we derive the factorization in \eqref{eq:joint-posterior-pdf} of the joint posterior pdf $f ( \V{y}_{1:t}, \V{s}_{0:t}, \VG{\alpha}_{1:t}, \VG{\beta}_{1:t} | \V{g}_{1:t}, \VG{\rho}_{1:t}, \V{z}_{1:t} )$.
Since the MOT measurements $\V{z}_{1:t}$ are observed, hence known, the joint vector of numbers of MOT measurements $\V{m}_{1:t}$ is also known, that is $f ( \V{y}_{1:t}, \V{s}_{0:t}, \VG{\alpha}_{1:t}, \VG{\beta}_{1:t} | \V{g}_{1:t}, \VG{\rho}_{1:t}, \V{z}_{1:t} ) = f ( \V{y}_{1:t}, \V{s}_{0:t}, \VG{\alpha}_{1:t}, \linebreak \VG{\beta}_{1:t} | \V{g}_{1:t}, \VG{\rho}_{1:t}, \V{z}_{1:t}, \V{m}_{1:t} )$.
Then, we obtain the factorization in \eqref{eq:app-factorization-1} by using assumption (A6) in the third step.
Recalling from Section~\ref{subsec:target_state_space_model} that $\V{y}_{t}$ is the vector stacking the legacy PT augmented states at the first agent pair and all the new PT augmented states introduced at time $t$, that is, $\V{y}_{t} = [ \underline{\V{y}}_{t}^{(1)\T}, \overline{\V{y}}_{t}^{\T} ]^{\T}$, each factor $f ( \V{y}_{t}, \V{s}_{t}, \VG{\alpha}_{t}, \VG{\beta}_{t}, \V{g}_{t}, \VG{\rho}_{t}, \V{z}_{t}, \V{m}_{t} | \V{y}_{t-1}, \V{s}_{t-1})$ of the product in \eqref{eq:app-factorization-1} can be further expressed as
\begin{align}
	f \big( \V{y}_{t}, & \V{s}_{t}, \VG{\alpha}_{t}, \VG{\beta}_{t}, \V{g}_{t}, \VG{\rho}_{t}, \V{z}_{t}, \V{m}_{t} \big| \V{y}_{t-1}, \V{s}_{t-1} \big) \nonumber \\[0mm]
	&= f \big( \overline{\V{y}}_{t}, \VG{\alpha}_{t}, \VG{\beta}_{t}, \V{g}_{t}, \VG{\rho}_{t}, \V{z}_{t}, \V{m}_{t} \big| \underline{\V{y}}_{t}^{(1)}, \V{s}_{t}, \V{y}_{t-1}, \V{s}_{t-1} \big) \nonumber \\[0mm]
	&\hspace{10mm} \times f \big( \underline{\V{y}}_{t}^{(1)}, \V{s}_{t} \big| \V{y}_{t-1}, \V{s}_{t-1} \big) \nonumber \\[0mm]
	&= f \big( \overline{\V{y}}_{t}, \VG{\alpha}_{t}, \VG{\beta}_{t}, \V{g}_{t}, \VG{\rho}_{t}, \V{z}_{t}, \V{m}_{t} \big| \underline{\V{y}}_{t}^{(1)}, \V{s}_{t} \big) \nonumber \\[0mm]
	&\hspace{10mm} \times f \big( \underline{\V{y}}_{t}^{(1)}, \V{s}_{t} \big| \V{y}_{t-1}, \V{s}_{t-1} \big) \label{eq:app-factorization-2-step1} \\[0mm]
	&= f ( \overline{\V{y}}_{t}, \VG{\alpha}_{t}, \VG{\beta}_{t}, \V{g}_{t}, \VG{\rho}_{t}, \V{z}_{t}, \V{m}_{t} | \underline{\V{y}}_{t}^{(1)}, \V{s}_{t}) \nonumber \\[0mm]
	&\hspace{10mm} \times f (\underline{\V{y}}_{t}^{(1)} | \V{y}_{t-1}) f (\V{s}_{t} | \V{s}_{t-1}) \label{eq:app-factorization-2-step2} \\[0mm]
	&= f ( \overline{\V{y}}_{t}, \VG{\alpha}_{t}, \VG{\beta}_{t}, \V{z}_{t}, \V{m}_{t} | \underline{\V{y}}_{t}^{(1)}, \V{s}_{t}) f ( \V{g}_{t} | \V{s}_{t}) f ( \VG{\rho}_{t} | \V{s}_{t} ) \nonumber \\[0mm]
	&\hspace{10mm} \times f (\underline{\V{y}}_{t}^{(1)} | \V{y}_{t-1}) f (\V{s}_{t} | \V{s}_{t-1}) \label{eq:app-factorization-2-step3} \ist ,
\end{align}
where assumption (A7) is used in \eqref{eq:app-factorization-2-step1},
(A3)
in \eqref{eq:app-factorization-2-step2}, and
(A9)
in \eqref{eq:app-factorization-2-step3}.
Finally, by using
assumption
(A8) and the fact that new PTs at agent pairs $1, \ldots, j - 1$ become legacy PTs at agent pairs $j, \ldots, J$, 
the joint pdf $f (\overline{\V{y}}_{t}, \VG{\alpha}_{t}, \VG{\beta}_{t}, \V{z}_{t}, \V{m}_{t} | \underline{\V{y}}_{t}^{(1)}, \V{s}_{t})$ in \eqref{eq:app-factorization-2-step3} can be factorized as in \eqref{eq:joint-pdf-prod-over-agent-pair}.
Eventually, by inserting \eqref{eq:joint-pdf-prod-over-agent-pair} into \eqref{eq:app-factorization-2-step3}, and \eqref{eq:app-factorization-2-step3} into \eqref{eq:app-factorization-1}, we obtain the factorization in \eqref{eq:joint-posterior-pdf}.

\vspace{-2mm}
\section*{Acknowledgment}
The collection of LCAS18 sea trial data used in this paper was made possible by the LCAS Multi National - Joint Research Project, with participants CMRE (NATO), DSTG (AUS), DRDC (CAN), DSTL (GBR), CSSN (ITA), FFI (NOR), DTA (NZL), and ONR (USA).

\balance
\renewcommand{\baselinestretch}{1}
\selectfont
\bibliographystyle{IEEEtran}
\bibliography{IEEEabrv,CompleteBibliography}

\end{document}